\documentclass[12pt,preprint]{aastex}

\received{22~August~2005}
\accepted{27~March~2006}
\slugcomment{accepted by ApJS}

\shorttitle{Solar CO}
\shortauthors{T.\ R.\ Ayres et al.}

\begin{document}

\title{Solar Carbon Monoxide, Thermal Profiling,\\
and the Abundances of C, O, and Their Isotopes}

\author{Thomas R.\ Ayres\footnote{Visiting Astronomer, 
National Solar Observatory, operated by the
Association of Universities for Research in Astronomy, Inc. (AURA), 
under cooperative agreement with the National Science Foundation.}}
\affil{Center for Astrophysics and Space Astronomy,\\
University of Colorado, 389 UCB [CASA],\\
Boulder, CO 80309-0389; ayres@casa.colorado.edu }

\author{Claude Plymate and Christoph U.\ Keller}
\affil{National Solar Observatory, PO Box
                26732,\\ Tucson, AZ 85726-6732;
 plymate@noao.edu,  ckeller@noao.edu \\[5mm] 
 {\em received:}~~22~August~2005; {\em accepted:}~~27~March~2006}

\begin{abstract}

A solar photospheric ``thermal profiling'' analysis is presented, exploiting 
the infrared (2.3--4.6~\micron) rovibrational bands of carbon monoxide (CO) 
as observed with the McMath-Pierce Fourier transform spectrometer (FTS) at
Kitt Peak, and from above the Earth's atmosphere by the Shuttle-borne 
ATMOS experiment.  Visible continuum intensities and center-limb behavior 
constrained the temperature profile of the deep photosphere, while CO 
center-limb behavior defined the thermal structure at higher altitudes.  
The oxygen abundance was self consistently determined from weak CO absorptions 
(for C/O$\equiv\,0.5$). Our analysis was meant to complement recent studies 
based on 3-D convection models which, among other things, have revised the 
historical solar oxygen (and carbon) abundance downward by a factor of nearly 
two; although in fact our conclusions do not support such a revision.  Based 
on various considerations, an $\epsilon_{\rm~O}=\,700{\pm}100$~ppm (parts per 
million relative to hydrogen) is recommended; the large uncertainty reflects 
the model sensitivity of CO.  New solar isotopic ratios also are reported: 
$^{12}$C/$^{13}$C=$80{\pm}1$, $^{16}$O/$^{17}$O=$1700{\pm}220$, and 
$^{16}$O/$^{18}$O=$440{\pm}6$; all significantly lower than terrestrial.  
CO synthesis experiments utilizing a stripped down version of the 3-D 
model---which has large temperature fluctuations in the middle photosphere, 
possibly inconsistent with CO ``movies'' from the Infrared Imaging 
Spectrometer (IRIS), and a steeper mean temperature gradient than matches 
visible continuum center-limb measurements---point to a lower oxygen abundance 
($\sim$500~ppm) and isotopic ratios closer to terrestrial.  A low oxygen 
abundance from CO---and other molecules like OH---thus hinges 
on the reality of the theoretically predicted mid-photospheric convective 
properties.     

\end{abstract}


\section{Introduction}

For nearly three decades, the ``solar neutrino problem'' haunted Solar Physics.
The apparent dearth of the elusive subnuclear byproducts of proton--proton cycle
fusion presented severe challenges
for theoretical descriptions of interior conditions in the Sun; models that heretofore 
had been regarded as
ironclad and unassailable (Bahcall, Basu, \& Pinsonneault 2003,
and references to previous work therein).  The neutrino problem thankfully was resolved in recent 
years, ironically from the particle-physics rather than astrophysics side 
(e.g., Bahcall, Krastev, \& Smirnov 1999), and solar (and stellar)
interior models once again seemed secure. 

A new challenge, however, has confronted
the even more refined contemporary generation of solar
interior models crafted for---and to a large extent
by---helioseismology; what one might call the ``solar oxygen crisis.''
Sophisticated analyses, employing time dependent
3-D simulations of convection driven solar surface velocity fields and thermal inhomogeneities, 
have pointed to a surprisingly low oxygen abundance, based on
an impressive collection of tracers including forbidden and weak
permitted \ion{O}{1} lines in the visible spectrum, and hydroxyl (OH) rovibrational
and pure rotational bands in the 3--10~\micron\ thermal infrared (Asplund et al.\ 2004,
and references to previous work therein).

The new low O abundance ($\epsilon_{\rm O}=$\,460$\pm 50$ 
parts per million [ppm] relative to 
hydrogen versus the most recent
previously recommended value 650$\pm 100$ ppm [Grevesse \& Sauval 1998, hereafter GS98]: see Figure~1) 
plays an important role
in interior models of the Sun because oxygen ranks third after hydrogen and helium 
by number in a solar chemical mixture, and contributes very substantially to the interior 
radiative opacity.  Furthermore, the next most abundant species---C, N, and 
Ne---are very difficult to measure (as is O itself) in the visible spectrum
owing to lack of suitable absorption lines, and changes in
their abundances often follow closely any alterations to $\epsilon_{\rm O}$.  In other
words, abundance ratios such as C/O ($\sim 0.5$ [Allende Prieto, Lambert, \& Asplund 2002]) 
or Ne/O (derived from solar energetic particles [e.g., Meyer 1985] or 
high resolution spectra of Ne and O ions in the solar
transition zone [Warren 2005]) 
often are regarded as better determined than the absolute abundances themselves.
Thus, a revision in $\epsilon_{\rm O}$ could have a domino-like effect on the overall solar
chemical composition and heavy element mass fraction $Z$, one of the crucial parameters
governing the interior structure of a star and its evolution.

The revised low oxygen abundance is said to resolve outstanding issues in interstellar 
medium and young
stellar population studies (Asplund et al.\ 2004), which previously had pegged the Sun as an ``oxygen-rich''
dwarf; at odds with its middle-aged status in a galaxy undergoing steady
metal enrichment owing to successive generations of stellar birth and death (Timmes, Woosley, \& Weaver 1995).  
At the same time,
the solar oxygen crisis has provoked the helioseismology
community by casting doubt on the previous spectacularly good 
agreement between simulated interior
sound speed profiles, and the depth of the convection zone, with values
deduced from detailed, high precision measurements 
of surface $p-$mode
oscillations from the ground and space (Bahcall et al.\ 2005).  Ironically enough, the
new oxygen crisis also is having an impact on understanding solar neutrino
fluxes (Bahcall \& Serenelli 2005). 

There appears to be
no easy way out of the dilemma from the helioseismology side: the upward revisions
in interior opacities needed to accommodate the new low $\epsilon_{\rm O}$ are larger
than permitted by uncertainties in the best contemporary theoretical and laboratory
atomic physics data  (Antia \& Basu 2005).  
On the low O side, the latest generation of time dependent 3-D
solar convection models have become quite sophisticated, and are said to allow the
precise matching of intensity profiles of weak velocity-sensitive photospheric
lines without the {\em ad hoc}\/ specification of extraneous dynamical parameters (such as the micro- or macroturbulent
velocity fields familiar from classical 1-D abundance modeling [e.g.,
Gray 1976]).  This is an 
extremely important advance in the {\em ab initio}\/ modeling---and understanding---of the structure
of the solar 
photosphere, and by extension of all late-type stars dominated by strong
convective heat fluxes in their surface layers.

Furthermore, the analyses by Asplund and collaborators
of the fundamental atomic physics of the various oxygen abundance diagnostics, of blending
issues, and of departures from Local Thermodynamic Equilibrium (LTE) in the permitted \ion{O}{1} lines
are persuasive. In fact, the three major facets of the
low O problem---3-D model, atomic physics (including non-LTE effects), and 
blending---play roughly equal, positively reinforcing roles in explaining the nearly
0.3~dex (factor of $\sim 2$) decrease of the revised oxygen abundance relative to the
value recommended little more than a decade ago (e.g., $850{\pm}70$ ppm
[Grevesse \&  Anders 1991, hereafter GA91]).  Thus the solar low O problem truly is vexing.

The purpose of the present paper is to add, in our opinion, two key perspectives:
(1) the importance of accurately ``calibrating'' the reference photospheric model,
be it 3-D or 1-D, against, say, absolute visible continuum intensities and center-limb
behavior; and (2) a complementary oxygen abundance (and thermal profile)
diagnostic overlooked in the
recent work, namely
the rovibrational bands of solar carbon monoxide with its
fundamental ($\Delta{v}= 1$) near 4.6~\micron\ and first overtone ($\Delta{v}=2$) near
2.3~\micron.  The former issue---properly scaling the reference model in temperature---was
discussed specifically
in a pioneering comprehensive study of the solar carbon, nitrogen, and oxygen abundances by Lambert
(1978, hereafter L78), and more generally by Ayres (1977, hereafter A77; 1978a).  
The solar CO first overtone spectrum,
and its significance for the carbon abundance (and thus for $\epsilon_{\rm O}$ if C/O
is known) was described by Tsuji
(1977) and Ayres (1978b), and an extension to the fundamental bands by
Ayres \& Testerman (1981, hereafter AT81).

To preview our conclusions, the solar CO analysis does not support the
recently proposed substantial lowering of the oxygen (and carbon) abundances, but
instead favors a higher value for oxygen (and carbon), closer to the recent recommendations of
GS98, and the earlier work of L78 and GA91.  
In addition,
we derive new accurate abundance ratios for 
$^{13}$C, $^{17}$O, and $^{18}$O.  The stable isotopes of C and O are signatures of
galactic chemical evolution, 
and should increase over time as the galaxy becomes progressively enriched in the
nuclear detritus of successive generations of star formation.  
The oxygen isotopes, in particular, can trace fractionation processes in the solar
nebula associated with the establishment (or not) of dust-gas equilibrium.  It is
commonly lamented in the solar system literature, however, that solar astronomy has failed 
to provide definitive
ratios for either $^{16}$O/$^{17}$O or $^{16}$O/$^{18}$O (Wiens et al.\ 2004). 
This was one of the 
major goals, in fact, of NASA's ill-fated {\em Genesis}\/ Discovery mission (Burnett et al.\ 2003).  
Nevertheless, we demonstrate that
it is straightforward
to derive accurate oxygen (and carbon) isotopic abundance ratios from CO lines, and further
that the
 per~mil (parts per thousand) 
differences---e.g., ${\delta}^{18}$O---are an order of magnitude, or more,
larger than anticipated
by theoretical models of the gas-dust chemistry in primitive solar system material.

\section{Observations}

The CO rovibrational spectrum is a valuable tracer for the mid-photospheric
thermal profile, and the oxygen abundance,
for several reasons.  First, the weak lines of the $\Delta{v}=1$,\,2 bands form deep
in the warmer layers of the photosphere where the CO concentration is unsaturated, 
in the sense that
$n_{\rm CO}\sim \epsilon_{\rm C}\,\epsilon_{\rm O}$.  If one knows the C/O ratio
to better accuracy than $\epsilon_{\rm C}$ itself, then the CO column density becomes
{\em quadratically sensitive}\/ to $\epsilon_{\rm O}$.  

Second, the CO-rich  
zone nevertheless is shallow enough to lie above the layers where the strongest 
convective overshooting occurs,
in fact where the thermal fluctuation pattern is thought to experience a reversal (e.g., Uitenbroek
2000a), and measured temperature inhomogeneities are mild
($\Delta{T}_{\rm rms}\sim 40$~K: e.g., Ayres \& Brault 1990, hereafter AB90).

Third, the $\Delta{v}=1$,\,2 bands contain literally thousands of individual
transitions covering a wide range of $gf$-values and lower level excitation energies,
compared with the mere handful of useful \ion{O}{1} lines in the visible (L78).  Also,
potential blends with atomic absorption features are rare in the infrared, particularly
in the 4.6~\micron\ interval, contrasted to the
heavy atomic line blanketing in the visible.  (A key issue in the current 
low O debate
is the strength of a weak \ion{Ni}{1} blend in the important [\ion{O}{1}] 
0.630~\micron\ feature.)

Fourth, the CO $\Delta{v}=1$ bands closely trace the photospheric thermal profile: the
strongest
lines respond preferentially to the conditions in the higher layers where they 
become optically thick, whereas the weaker lines reflect conditions at lower
altitudes, closer to the infrared continuum formation horizon.
Observations of the center-limb behavior of representative CO lines can serve the same
``thermal profiling''
purpose, in fact redundantly.
A subsidiary issue involves the thermal structure of the very 
high layers, above the
nominal base of the warm chromosphere, 
where the anomalous limb darkening of the strongest
CO fundamental lines (Noyes \& Hall 1972a, AT81) and their off-limb
emissions (Solanki, Livingston, \& Ayres 1994) reveal
cold gas at altitudes where---in traditional 1-D models like the FAL\,C of
Fontenla, Avrett, \& Loeser (1993, hereafter FAL93)---temperatures should 
be too hot ($T\sim 7000$~K) for molecules to survive.  
The origin and spatial scope of this counterintuitive ``COmosphere'' is
controversial (Ayres 2002, and references to previous work therein; hereafter A02), but---as we describe
below (\S{3.3.4})---any high altitude cool material has only a minor influence on
deriving an oxygen abundance from the weak $\Delta{v}=1$,\,2 lines.

Fifth, the CO rovibrational bands form very close to Local Thermodynamic
Equilibrium (LTE), owing to the large
atomic hydrogen inelastic collision rates that quench the pure rotation and 
rotation-vibration transitions compared with their slow radiative decays 
(see Ayres \& Wiedemann 1989, and references to previous work therein, hereafter AW89).  
LTE formation applies, as well, to the background continuum in the infrared longward of
1.64~\micron, where the opacity sources thankfully are simple
and well understood: mainly
free-free (f--f) transitions of the negative hydrogen ion, H$^{-}$, with a very small contribution
from hydrogen bound-free (b--f) and f--f
(Vernazza, Avrett, \& Loeser 1976).  
The CO gas-phase chemistry also is very close to ``instantaneous
chemical equilibrium'' (ICE) in the middle photosphere, even in the face of
time dependent molecular
formation and destruction, and advection processes,
as demonstrated in recent detailed
2-D hydrodynamical simulations by Wedemeyer-B\"{o}hm et al.\ (2005), and earlier in
1-D time dependent models by Asensio Ramos et al.\ (2003).

Sixth, the CO molecular parameters including the all important dissociation
energy ($D_0$= 11.108~eV: Morton \& Noreau 1994, hereafter MN94), line positions and transition
strengths (Goorvitch 1994, hereafter G94) are thought to be accurately known.  Owing to the importance of
CO as a terrestrial pollutant, these values have been vetted thoroughly,
at least for the lowest $vJ$ transitions that are prevalent in a
cold planetary atmosphere like Earth's.  

The final practical advantage of the CO IR bands is the availability of very high quality
measurements from the ground and space.  The large Fourier transform spectrometer (FTS)
on the National Solar Observatory's
McMath-Pierce telescope at Kitt Peak (Brault 1978), for example,
can record the fundamental region at resolutions of
$R\equiv \omega/\Delta\omega\sim 2\times10^{5}$, or more, ($\omega$ is
the frequency measured in wavenumbers [cm$^{-1}$]),
fully resolving the narrow solar CO lines, and with signal-to-noise (S/N) in
excess of several thousand.  The McMath-Pierce also is home to the
Infrared Imaging Spectrometer (IRIS), a grating-based instrument with lower resolving power
($\sim 6\times10^4$), but a long slit stigmatic imaging capability, and ten times higher
time resolution than the FTS can muster (Uitenbroek, Noyes, \& Rabin 1994, hereafter UNR94;
Ayres \& Rabin 1996, hereafter AR96).  Furthermore, the
Shuttle-borne ATMOS experiment (also an FTS) 
obtained excellent solar disk center reference spectra in the thermal infrared during
its several flights (Farmer 1994), achieving $2\times10^5$ resolution and 
S/N$>$\,1000 in the best quality scans provided to the community.

Despite the many positive attributes of CO as an abundance diagnostic, a specific
downside---shared by molecules in general---is the strong sensitivity of
the dissociative equilibrium to temperature.  For this reason it is imperative
to accurately characterize the thermal conditions in the mid-photosphere where
the abundance sensitive weak CO lines arise.  

The following sections summarize the several types of observational
material used in the subsequent thermal profiling---and oxygen abundance---analysis; 
including besides 2--6~\micron\ CO,
calibrated intensity highpoints in the 0.4--0.7~\micron\ visible continuum region,
and continuum center-limb behavior at visible and infrared wavelengths.

\subsection{ATMOS/ATLAS-3}

The Shuttle-borne ATMOS
instrument was designed to study trace molecular
species in the Earth's atmosphere backlighted
by the rising or setting Sun as viewed from low-Earth orbit.  The baseline solar reference spectra,
obtained from zenith pointings, were mainly free of terrestrial contamination, and consequently are
extremely valuable for a study like the present one.  The best results
were obtained from the final ATMOS flight in 1994 November as part of the ATLAS-3 payload
(Abrams et al.\ 1999).
For the particular scan
ranges used here, isolated by one of three passband filters, the instrument accepted
a 1~mrad-diameter (200\arcsec) circular region at disk center, which 
corresponds to $\mu= 1$ for all intents
and purposes ($\mu\equiv \cos{\theta}$, where $\theta$ is the heliocentric angle:
0\arcdeg\ at disk center, 90\arcdeg\ at the extreme limb). 

Scans were downloaded from the ATMOS public archive and calibrated according to the
specifications in the 
associated documentation.  For the CO first overtone interval, 3200--4800~cm$^{-1}$, 
only scan \#4 had suitable
coverage.  For the CO fundamental region,
1600--2400 cm$^{-1}$, scans \#3 and \#9 provided overlapping coverage.  Each
was independently normalized to a continuum level by sequentially fitting low order
polynomials to intensity highpoints in partially overlapping 5~cm$^{-1}$ intervals:
the ATMOS scans are dominated by the ubiquitous---but mainly isolated---absorption 
features of the CO bands, and generally there are smooth intervals
between the lines where the pristine continuum---free of
solar or telluric absorptions---shines through.  The
two separate continuum-normalized scans then were cross-correlated to determine any
global frequency offsets.  The derived 
$\sim 0.5$~pixel shift was applied
to scan \#9; then the two datasets were combined by {\em interleaving}\/ the 
intensity points, rather than interpolating and coadding which would have
involved a smoothing of at least one of the scans, perhaps compromising the
intrinsic high resolution of the ATMOS spectra.  Thus, the tracings presented here
for the CO fundamental region consist of two entirely independent datasets: any deviations
represent mainly differences in the assigned local continuum levels or 
scan-specific distortions, and therefore are a fair
measure of systematic intensity uncertainties beyond the purely photometric noise component.

Figure~2 illustrates selected intervals of the fundamental and first overtone CO bands
from the ATMOS spectra.  Figure~3 illustrates selected transitions from the 1--0 R-branch 
($\Delta{v}\equiv v_{\rm upper} - v_{lower}= +1$, $\Delta{J}\equiv J_{\rm upper} - J_{lower}=+1$; 
P-branch transitions have $\Delta{J}= -1$) of the CO fundamental 
from ATMOS, together with line profiles calculated with an ``optimum'' 1-D thermal profile described later
(\S{3.3}).

\subsection{McMath-Pierce FTS}

The large (1~m path difference) FTS on the 
McMath-Pierce telescope achieves superb
spectral resolution and excellent signal-to-noise in the thermal infrared, with broad frequency coverage
and negligible scattered light.  
The McMath-Pierce telescope itself, with its 1.58~m primary mirror
and all reflecting unobscured design, not only can
observe beyond the $\sim$2.5~$\mu$m
cutoff of typical window materials used in evacuated solar telescopes, but also
delivers sub-arcsecond diffraction limited
performance ($\sim$~0\farcs8 at 5~\micron) during periods of good seeing.

These capabilities are crucial for center-limb
work in the CO bands, where core depths of a wide range of narrow
$\Delta v=1$ lines must be measured accurately at each $\mu$ value,
and one wants to record as close as a few arcseconds from the
limb in order to obtain the best slant-angle advantage for observing
the highest atmospheric layers.  Errors of only a few percent in line depths, due to 
under resolving line profiles or the presence of scattered light, 
can translate to an overestimate of hundreds of degrees in 
core brightness temperatures (an indirect diagnostic of
kinetic temperatures near $\tau\sim 1$: AT81).  

The main disadvantages of the FTS are that
it can record only a single spatial point at a time (defined by an entrance aperture
which can have an arbitrary shape and size within certain limits), and it requires
several minutes to accumulate a full interferogram at highest resolution
and S/N.  By its nature, the FTS also benefits from
a highly stable source region; a particular concern, for example, in observations very close to the 
edge of the disk, where seeing fluctuations or wind shake
could transiently alter the intensity field.  For that
reason, the highest quality previous FTS work on the anomalous limb darkening
of the CO bands had been done at $\mu\gtrsim 0.2$ (AT81, AB90), corresponding
to 8.1~mm or 19\arcsec\ inside the IR limb.  (The plate scale of the
McMath-Pierce is 2.37\arcsec\ mm$^{-1}$.)  

\subsubsection{Fast Tip/Tilt Image Stabilizer}

Recently the FTS was outfitted with
a novel fast tip/tilt image stabilization system, developed by C.~Keller and
C.~Plymate of NSO; a major improvement for limb observations.  The system mounts on the front plate of the
FTS, and delivers a stabilized image to the entrance port
of the interferometer.  The transfer
optics of the stabilizer pick off a portion of the solar image, collimate it onto a
piezioelectrically actuated tip/tilt mirror, and refocus the corrected
image onto a beamsplitter.  The 98\,\% reflected light is directed to
the FTS entrance aperture, while the 2\,\% transmitted image
is sensed by a CCD camera, which can operate in a variety of tracking
modes---limb, quad cell, or granular correlation---to complete the 
feedback loop.  The CCD accepts a $\sim$100~\AA\ passband centered
in the red at $\sim$~7000~\AA.  The system operates
at 500~Hz, fast enough to compensate for 
seeing fluctuations and image shake.  

\subsubsection{2002 April Observing Run}

The observations described here were acquired during a 5-day run in 2002 April (13--17),
mainly on the 14th.  A 98~cm full
path difference of the FTS was utilized, and the interferograms were accumulated in
double-sided mode.  The theoretical resolving power was
$R\sim$~2$\times 10^5$ in the 2145~cm$^{-1}$ (4.66~$\mu$m)
atmospheric ``window'' where most CO $\Delta v=1$
measurements are made.
Each interferogram required nearly four minutes
to record, including fly-back.  Twin L--N$_{2}$-cooled InSb detectors (``A'' and ``B'')
were sampled at 2500~Hz. (The two outputs view 
opposite sides of the FTS recombining optic and thus are
out of phase with one another.  The 
interferogram signal is the difference of the two, which 
has the advantage of canceling noise [such as 60 Hz] common 
to both detectors.)  A 
broad-band filter limited the spectral
coverage to 1800--2900~cm$^{-1}$ (3.4--5.6~$\mu$m), 
to suppress out-of-band photometric noise.

On the 14th, we took advantage of superb
image quality and lack of (wind-induced) image shake to work close to the limb,
exploiting the new image stabilizer.  
We chose a 10\,mm$\times$0.25\,mm rectangular entrance
aperture (24\arcsec$\times$0{\farcs}6; the radius of the solar image
was 957\arcsec).  The width of the
slit was comparable to the diffraction limit of the telescope, and small
enough to impose a negligible $\delta\mu/\mu$ on observations very close to 
the limb.  The slit was as long as could be accommodated in good
focus with the FTS, but short enough that $\delta\mu/\mu$ due to curvature
of the limb was negligible. 

We zeroed the coordinate frame of the image stabilizer, in limb tracking mode,
at the half power point of the A+B central fringe signal (representing source 
intensity) as it rolled off across the edge of the disk.
Owing to the passband limiting filter, we were measuring the true infrared
limb, which can be displaced a few arcseconds from the visible limb 
by atmospheric
refraction.  We then backed the limb away from the aperture a set number
of steps on the image stabilization system (the plate scale at the CCD
was 0\farcs14~pix$^{-1}$), to position a specific
$\mu$-value over the entrance slot.  We frequently checked the null position
at the IR limb to ensure that no drifts had occurred.  During that
morning, we obtained 21 separate useful observations at a range of
$\mu$-values: from $\mu= 0.169$ (5.83~mm inside the limb) down to as close
as $\mu= 0.076$ (1.17~mm inside the limb).  We concluded with a pair
of measurements at disk center ($\mu= 1$).  

Since the image stabilizer had to be mounted
in a fixed orientation on the vertical circular ``table'' of the FTS,
we were not able to actively compensate for the diurnal rotation of the solar image.
Accordingly, our observations effectively covered an arc at
the limb as the solar disk rotated over the fixed slit position;  
providing, in fact, a broad, unbiased spatial average at the limb.

A full account of the FTS program will be provided in a future paper, including
details of the data reduction, scan averaging, and compensation for atmospheric absorption.  
Figure~4 compares a
disk center tracing from the McMath-Pierce FTS with the ATMOS/ATLAS-3 spectrum
described previously.  
Center-limb lineshapes of selected 2145~cm$^{-1}$
CO transitions will be presented later (\S{3.3.3}).

\subsection{McMath-Pierce IRIS}

The Infrared Imaging Spectrograph (IRIS), commissioned in 1993, utilizes the
Main (``vertical'') spectrograph of the McMath-Pierce, a $256\times256$ Amber
Engineering InSb camera cooled by solid nitrogen, and the
control system of the Near Infrared Magnetograph (NIM: Rabin 1994).
We made use of two general types of IRIS observations in the present
study: (1) short exposure
single slit measurements of the CO translimb (emission) spectrum; and (2) rapid cadence sequences 
of raster scans to record 2-D
time dependent behavior of the CO bands at disk center.

\subsubsection{Translimb CO Measurements}

Translimb CO measurements are a diagnostic for the coolest parts of the high altitude ``COmosphere,''
especially the height of the temperature minimum (``$T_{\rm min}$'').  
The strong saturated low excitation CO features, in particular, can
have significant contributions from this cold ($T< 4000$~K)
region, thereby affecting the determination of the 
temperature stratification and potentially also the oxygen abundance.

The present study utilized sequences of translimb image bursts obtained 1996 May~9,  
representative of the quiet Sun near sunspot minimum, as described fully by A02.  
Although not stated explicitly
in that study in these terms, the single component thermal profile 
that best matches the off-limb CO emissions has a
$T_{\rm min}$ at a column mass density $m\sim 2.5\times10^{-3}$ gr~cm$^{-2}$,
compared with the $m\sim 5\times10^{-2}$ of the FAL\,C reference
model (FAL93).  Unfortunately, the off-limb CO emissions---as recorded by
the relatively low dispersion IRIS instrument---are not useful for quantifying
$T_{\rm min}$ itself.  However, the new FTS extreme limb scans
can be used for that purpose, as described later (\S{3.3.4}).  (To preview our results, we obtain
$T_{\rm min}\sim 3500$~K, as compared with the 4400~K of the FAL\,C model.)

\subsubsection{Disk-Center CO Timeseries}

The second set of IRIS data comprised high spatial resolution, rapid cadence
surface temperature maps recorded during a coordinated multiwavelength campaign in
1999 May.  The program, which included simultaneous UV imaging spectroscopy
from {\em SoHO}~SUMER and narrow band UV imaging from {\em TRACE,}\/ has 
been described previously in
a qualitative way by Ayres (2000).  Here, we provide additional details
concerning the IRIS part of the program.

The objective on the IRIS side was to measure, with high time resolution, 
temperature and velocity fluctuations at several levels of the outer
photosphere, ultimately to compare with the response of the overlying
chromosphere imaged by SUMER and {\em TRACE.}  In the present work, however, the amplitude of thermal
fluctuations in the mid- and high photosphere play a direct and crucial role in backing out realistic
abundances from temporally and spatially averaged solar CO measurements.   

The IRIS spectra described here were obtained on the morning of 1999 May~14.  The
visible seeing was good at the start of observations (08:30 MST), but deteriorated later
in the morning.  The infrared seeing usually is much better than in the visible, and the
effective resolution often is limited
by telescope diffraction rather than the atmosphere.

We operated the vertical spectrograph in single pass, recording
a free spectral range of 2.3~cm$^{-1}$ centered at 2143~cm$^{-1}$; with a
``minifying'' lens in the transfer optics to achieve a spatial coverage
of 97\arcsec\ along the slit.  The slit was oriented N/S heliocentric, with a width of
160~\micron\ (0{\farcs}4).  In mapping mode, the
slit was stepped sequentially in the E/W direction, 8 steps per raster
with a step spacing of 1\farcs9; thus covering a swath 13\arcsec$\times$97\arcsec\
at Sun center.
The exposure time at each slit position 
was 300~ms, and with stepping and data writing overheads, a full area scan required
45~s to complete (well sampling the photospheric $p$-mode oscillations timescale [5 minutes]
and that associated with higher frequency chromospheric modes [3 minutes]).  We accumulated
40 rasters per observation: 320 spectral frames in 30~minutes.

In each individual spectrogram, the main parameters of interest were central depths
and frequency shifts of the CO lines, and the relative variation of continuum intensity, as
recorded along
the 1-D slit direction.  (Reduction and measurement techniques are described by AR96).
The depth of a strong saturated CO line is tied closely to the
kinetic temperature of the atmospheric
layer in which the line core becomes optically thick.  Similarly,
the Doppler shift records the line-of-sight component of the velocity field where the
feature forms.  Finally, the relative continuum intensity responds to temperature contrasts in the
layers where
the IR background continuum becomes optically thick.  In general, the 
line FWHM also would be a parameter
of interest, reflecting possible changes in the unresolved velocity field.  
Empirically, however,
we found very little spatial variation in the fitted FWHMs.

Assembling the 1-D measurements
for the 8 pointings in each area scan yielded a 2-D spatial map for that 45~s
time interval: velocity and intensity fluctuations for the CO lines
(which were coadded into two sets: ``strong'' [2--0~R6, 3--2~R14, 4--3~R23]
and ``weak'' [7--6~R67, 7--6~R68]), and
intensity for the background continuum.  The $\Delta{I}$ fluctuations were translated to 
equivalent $\Delta{T}$
according to a disk center absolute continuum intensity obtained from a model calculation.
An example of a 10 frame sequence is shown in Figure~5.  Figure~6 illustrates cross-correlations of several
combinations of
line and continuum parameters, imposing discrete temporal lags 
out to $\pm$5 steps, in multiples of the fundamental 45~s cadence, over one of the 40 frame sequences.  In
the full interval (450~s), 
solar rotation would carry the scene about 1\arcsec\ under the fixed area scan,
comparable to the diffraction-limited spatial resolution.

In the bottom row, one sees that the continuum thermal fluctuations maintain a noticeable self correlation
(tilted contours) over the full range of lags displayed ($\sim{\pm}$4~minutes).  
This likely is a signature of photospheric
convective granulation, which has a coherence time of order 10 minutes.  Notice, also,
that the temperature fluctuations are relatively mild: the rms value is only
about 22~K.  In contrast, while
the weak CO lines show a modest positive correlation with the continuum at zero lag, the
correlation rapidly dissipates for lags on either side of zero.  
At zero (or any other) 
lag, the weak CO lines (with a larger $\Delta{T}_{\rm rms}\sim 42$~K)
show no evidence for a {\em negative}\/ branch of the continuum 
correlation, as might have been anticipated as a signature of
``reversed granulation'' reported
previously in the strong CO 3--2~R14 line from
an IRIS CO time series utilizing a fixed slit (Uitenbroek 2000a), 
and in simulations of weak 7--6~R68
with a 3-D model in the same work.  

Here, the strong CO lines also show hardly any correlation
with the continuum, even at zero lag; and a
larger range of $\Delta{T}$, although the rms still is a relatively
modest 83~K.  Thus, the thermal 
variations at high altitudes probed
by the strong CO lines appear to be unconnected to
the inhomogeneities of the middle photosphere recorded in the IR continuum.  This point is emphasized
in Fig.~5 where the patches of highest and lowest $\Delta{T}$ are highlighted by contours in
the strong-line and continuum maps, and repeated in the other panels.  Few of the peak
temperature excursions in the continuum print through to the outer photosphere.  In fact,
the granular contrast in the thermal IR continuum is very low, probably as a consequence
of formation in the higher altitude zone where the granulation pattern is reversing
and is more chaotic than in deeper layers.  That would account for the
apparent lack of anti-correlation with the strong CO lines, but also suggests that
the convective penetration in these layers is more muted than predicted by the
3-D simulations.   

The important role of $p$--mode
oscillations in driving temperature fluctuations in the strong CO $\Delta{v}=1$ lines
was recognized at the time these
features first were observed with high resolution and sensitivity at the McMath-Pierce in
the early 1970's (Noyes \& Hall 1972b).  
Independently of how temperature (and velocity) perturbations in the CO and
continuum layers might be connected (or not), the high spatial resolution 
IRIS maps provide a direct probe of these excursions in the altitude range 
most relevant to the CO absorptions.  This information can be exploited
to test the response of simulated spectra to the fluctuating 
environment of the middle and upper photosphere.

\subsection{Absolute Continuum Intensities and Center-Limb Behavior}

The final piece of the observational puzzle, to set the stage for the
thermal profile (and abundance) analysis, are measurements of continuum highpoints
in the visible spectrum, and continuum center-limb behavior from the
visible out to the 5~\micron\ CO $\Delta{v}=1$ region.  Continuum highpoints
provide a fundamental calibration of the temperature scale of
a solar model in its deepest layers, while 
the center-limb behavior responds primarily to
temperature {\em gradients}\/ over the emission formation zone, 
$\tau_{\rm C}\sim~$0.1--1.  Since the IR continua beyond
3~\micron\ form {\em above}\/ the visible continuum, their
center-limb behavior can be exploited to transfer the visible-based temperature calibration
to the even higher layers where the CO lines arise.

\subsubsection{Visible and Infrared Center-limb Behavior}

We used two sources for continuum center-limb behavior: Neckel \& Labs (1994, hereafter NL94) for
$\lambda < 1.1~\micron$, and Spickler, Benner, \& Russell (1996, hereafter SBR96) for $\lambda > 2.4~\micron$.
The NL94 measurements were based on monochromatic ($\sim 2\times10^5$~resolution)
drift scans obtained
with the McMath-Pierce telescope and
vertical spectrograph, regularly during the period between solar minimum in 1986 and
maximum in 1990.  These data agree well with previously published work
by Pierce and colleagues (Pierce \& Slaughter 1977; Pierce, Slaughter, \& Weinberger 1977),
typically better than $\pm$0.5\,\% according to our own comparisons.  In the 
thermal infrared, SBR96 reported center-limb curves recorded by the HALOE
occultation experiment on the Upper Atmosphere Research Satellite, at eight wavelengths between
2.45~\micron\ and 9.85~\micron, with bandwidths corresponding to a resolution of $\sim 50$.
We considered only the six measurements $\le 6.25$~\micron, approximately the
upper wavelength limit of the CO $\Delta{v}=1$ bands.  The visible
and IR center-limb
curves will appear later (\S{3.3.1}), when we compare the empirical behavior with predictions
of different thermal models.

\subsubsection{Absolute Continuum Intensities}

We adopted the Thuillier et al.\ (2004, hereafter T04) solar irradiance reference spectra as a basis for
calibrating absolute disk center intensity highpoints.  These spectra cover the
wavelength range from kilovolt soft X-rays to the 2.4~\micron\ mid-IR, although 
we made use only of the visible
portion, 0.4--0.7~\micron.  The reference spectra were derived from a variety of sources, for two
levels of solar activity.  (The irradiance differences between sunspot minimum and maximum are entirely 
negligible in the visible, however).  
The mid-UV/visible portion (0.20--0.87~\micron)
was recorded at a resolution of $\sim 10$~\AA\ 
by the SOLSPEC instrument on the three flights of the
Shuttle ATLAS platform, mentioned
previously in the context of the ATMOS experiment.  Integrations of the reference spectra
yielded a total solar irradiance (TSI) within 1\,\% of TSI values measured 
independently by active cavity radiometer (``ACRIM'') instruments in the same epoch.  
A detailed error estimate (see Thuillier et al.\
[2003], their Table~IIIa--d) indicated a 1\,$\sigma$ uncertainty of $\pm$1\,\% for the 0.45--0.70~\micron\
range of interest here.  T04 further normalized their reference spectrum to
the estimated TSI of 1367 W~m$^{-2}$ at 1~AU during that epoch, to improve the absolute
accuracy to approximately that of the TSI measurements ($\lesssim 1$\,\%).

The reference spectra are disk average
irradiances $f_{\lambda}$ at 1 AU, at the coarse resolution of 10~\AA.  For
modeling purposes, it is desirable to extract specific intensity $I_{\lambda}$ highpoints 
in the visible that are as close as possible to the pristine continuum level at disk center.  
We accomplished this as follows. 

R.~L.~Kurucz kindly provided a high resolution irradiance atlas of the solar spectrum based on
groundbased FTS scans corrected very carefully for telluric absorptions, 
instrumental filter response, and so forth.  These spectra 
had been normalized to a continuum level determined by identifying narrow windows where
detailed spectral simulations indicated minimal solar (and telluric) line absorption, and adjusting for
such structure if necessary.  We smoothed these high resolution 
continuum normalized tracings to the $\sim 10$~\AA\
bins of the T04 reference spectrum.  We next
multiplied the reference irradiance spectrum by $(d_{\odot}/R_{\odot})^2$ to scale it to 
surface flux, then divided
by the smoothed continuum normalized trace.  We
identified regions of the ratio that corresponded to $>98$\,\% highpoints in the smoothed 
high resolution 
spectrum (in the interval 0.40--0.69~\micron)
and fitted a fifth-order polynomial to those points, with one pass of a 2\,$\sigma$ filter to suppress
outliers.  Because information
concerning the high resolution continuum level was encoded in the original residual flux spectrum
``100\,\% level,''
the resulting polynomial curve should correspond very well to the hypothetical pristine continuum
(in flux).  The final step was to divide the highpoint surface flux curve by the wavelength-dependent
${\cal F}_{\lambda}/I_{\lambda}$ (as taken from NL94 and
fitted by a fourth-order polynomial).  These steps are illustrated in Figure~7.  

Although the slight $\sim 1$\,\% downward adjustment by T04 of their
reference spectrum to match TSI values should yield an improved absolute accuracy, a more
recent TSI experiment, SORCE/TIM (Kopp et al.\ 2005), which utilizes a somewhat different
sensor technology than the earlier ACRIM-style devices, shows a systematic
5 W~m$^{-2}$ (0.4\,\%) lower TSI than contemporary ACRIM measurements over the same
time period.  The origin of the
difference is unknown, but it is comparable to systematic offsets seen in comparisons of individual
ACRIM instruments (Willson 2005).  The cited absolute accuracy of the
SORCE/TIM instrument is 0.01\,\%.  We therefore
further adjusted the derived continuum highpoints downward by 0.4\,\% to the SORCE/TIM
scale.  Table~1 lists representative line-free continuum 
central intensity highpoints for the range 0.45--0.70~\micron\ from
the polynomial model.
 
Incidentally, the solar luminosity that results from this exercise is
$L_{\odot}= 3.830\times 10^{33}$~ergs s$^{-1}$, and the effective temperature is
$(T_{\rm eff})_{\odot}= 5772$~K.  Asplund et al.\ (2000b) adopted $(T_{\rm eff})_{\odot}= 5767$~K for
their convective heat flux calculations, which corresponds to a (entirely negligible) third of a percent
lower surface flux.

We compared the new calibrated continuum intensity highpoints to those derived two decades ago by
Neckel \& Labs (1984, hereafter NL84), who applied an analogous approach with the best available measurements
of the time.  Over the 0.46--0.65~\micron\ range that we adopted to define the visible continuum,
our results for the 7 wavelengths in that interval reported by NL84
(their Table~VII) are on average $+0.9{\pm}1.7$\,\% higher; while for the interval between 0.4975--0.6200~\micron,
where the continuum potentially is less affected by line blanketing, our results are
$+0.1{\pm}0.8$\,\% higher (where the uncertainties are 1\,$\sigma$ dispersions about the
mean values).  The extent to which this remarkably good agreement is accidental cannot be
known, but we regard it as evidence that the new continuum highpoints 
do not suffer unrecognized systematic errors beyond the $\sim{\pm}1$\,\% level.

\subsubsection{Continuum Intensity Fluctuations}

Finally, we considered the additional information available from visible
and near-infrared continuum intensity fluctuations reported in several recent studies.  Such
``granular contrast''
measurements play a key role in validating the amplitude of thermal fluctuations in {\em ab initio}\/
convection simulations, and in estimating the $\Delta{T}$'s that one might impose
on a purely empirical model to capture the range of thermal inhomogeneities present at different levels
of the solar photosphere.

For example,
Leenaarts \& Wedemeyer-B\"{o}hm (2005) utilized the Dutch Open Telescope to obtain high resolution movies
simultaneously in the blue continuum (0.432~\micron) and the wing of the \ion{Ca}{2} H line
(0.397~\micron).  The latter forms in the mid- to high photosphere
and displays a ``reversed granulation'' signature like the strong CO fundamental lines
(Uitenbroek 2000a).  In their $I/\langle{I}\rangle$ histograms, the blue continuum
shows a roughly 25\,\% fluctuation amplitude at half maximum occurrence (corresponding to an rms of
about 20\,\% for a Gaussian distribution), while the H-line wing displays
about a 10\,\% amplitude, with reversed contrast (corresponding to $\Delta{T}_{\rm rms}\sim 50$~K).  

Puschmann et al.\ (2003) obtained a 50~minute time series of long slit spectrograms
at the German Vacuum Tower Telescope on Tenerife, and examined the behavior of several
0.65~\micron\ \ion{Fe}{1} lines with respect to the red continuum.  The latter displayed rms fluctuations in the 
neighborhood of 4\,\%.  The authors concluded that only relatively large structures
($\gtrsim 2\arcsec$) of the deep convective pattern print through to the upper photosphere, and
exhibit a reversed intensity contrast.  They estimated the reversal height to be about 140~km 
($m\sim 1.5$~gr cm$^{-2}$ in the FAL\,C model).  

S\'{a}nchez~Cuberes et al.\ (2003) 
described center-limb observations at 0.80~\micron\ and 
1.55~\micron---continuum opacity maximum and minimum,
respectively---recorded at
the Swedish Vacuum Solar Telescope (SVST) on La~Palma.  
They reported rms granulation contrasts of 6\,\% at 0.80~\micron,
and 3\,\% at 1.55~\micron, with decreasing contrasts at both wavelengths toward the limb.  
Frame selection 
ensured sharp images, although there was no explicit
correction for atmospheric degradation.  The authors estimated a convection
penetration height of $\sim 220$~km, from the presence of distinct granulation signatures out to the
extreme limb.  An earlier study by S\'{a}nchez~Cuberes et al.\ (2000) measured the center-limb behavior
of continuum intensity fluctuations at 0.67~\micron\ with the SVST during a partial eclipse, exploiting
the lunar limb to calibrate the image blur by telescope and atmosphere.  They found 
a disk center rms granular contrast at 0.67~\micron\ of 10\,\%, which translates
to about 13\,\% at 0.5~\micron, at the upper end of previously published values, but perhaps not unreasonable
given the accurate corrections enabled by the circumstance of the partial eclipse and the excellent observing
conditions.

These granular contrasts in the visible are entirely consistent with the predictions of the Asplund
convection model at 0.62~\micron, namely $\sim 15$\,\% from the full-resolution simulation and $\sim 10$\,\%
when smearing by telescope diffraction and seeing are taken into account (Asplund et al.\ 2000a).

\section{Atmospheric Modeling}

We carried out two general types of atmospheric modeling 
to constrain the photospheric thermal profile, at least 
from the special perspective
of carbon monoxide.  The first type involved simulations of the line-free continuum, in
the visible and infrared, and its center-limb behavior.  The second type involved 
synthesis of CO absorption lineshapes, and their center-limb behavior, to test thermal models 
(strong, saturated lines)
and derive self-consistent abundances (weak, unsaturated transitions).  

Within each class of simulation, we considered a range of single component models; as well as
multicomponent variations to mimic the thermally heterogeneous solar plasma (so-called ``1.5-D'' 
approximation in which temperature inhomogeneities are simulated by combining spectra
synthesized from a series of 1-D perturbed versions of the base stratification).  
We treated any specific model in hydrostatic equilibrium,
although we did allow for ``turbulent pressure'' support as described by Vernazza, Avrett, \& Loeser
(1973, hereafter VAL73).  
We adopted an anisotropic radial-tangential model
of the auxiliary turbulence parameter used in the Doppler width 
(and turbulent pressure) calculations, simplified
by assuming a depth independent velocity separately
for the radial and tangential lines of
sight.  We adopted $\mu= 1$ microturbulent velocities from other work, as 
described below (\S{3.3.1}), and
adjusted a macroturbulent parameter to achieve a satisfactory fit to the fully resolved
empirical CO profiles.  We used the same macroturbulence for the limb sightlines,
and adjusted the tangential microturbulent parameter to 
achieve a good match between synthesized and empirical limb lineshapes.  The
exact choice of these dynamical broadening parameters,
within the broad limits set by previous
studies, has only a very minor impact on deriving $\epsilon_{\rm O}$, primarily because the
selected abundance reference lines are weak and unstaturated.

We calculated the equation of state allowing for LTE formation of
H$_{2}$ (Auer, Heasley, \& Milkey 1973, hereafter AHM73), H$^{-}$ (VAL73), H$_{\rm 2}^{+}$ (AHM73), 
LTE ionization of the metals, a simple non-LTE Balmer-continuum photoionization
approximation for hydrogen ionization (Linsky 1968) as ``calibrated'' against
the actual electron densities of the FAL\,C reference model, 
and instantaneous chemical equilibrium for the important diatomic 
molecules composed of H, C, N, O, and Si (Kurucz 1985,
with the exception of the dissociation energy
of CO, as described later [\S{3.2}]).  Temperature dependent atomic 
partition functions were polynomial
fits to the data in Table~D2 of
Gray (1976).  We adopted the ``standard solar composition'' of GS98, to
maintain consistency with the Asplund model (see Asplund et al.\ 2000b), particularly 
with regard to the electron
donors which affect the crucial H$^{-}$ concentration.

Each model was represented by the run of temperature, $T$ (K),
with the logarithm of the column mass density, $\log{m}$ (gr cm$^{-2}$), on a scale from
$-4.50$ to $+0.86$ with a spacing of 0.02~dex.  The radiation transport was simulated using
Auer's (1976) Hermitian method implemented in a Feautrier-type ray solution, which has 
high accuracy especially with the finely spaced depth grids used here.   

\subsection{Background Opacities}

We adopted the H$^{-}$ b--f and f--f
opacities recommended by John (1988).  The absolute,
and relative, accuracies of the underlying cross sections are quoted as better than a few
percent.  H$^{-}$ b--f completely dominates the visible spectrum shortward of the photoionization
edge at 1.642~\micron, while the f--f opacity dominates out in the CO $\Delta{v}= 1$,\,2 portion
of the infrared (Vernazza, Avrett, \& Loeser 1981, hereafter VAL81).  Since our models
are ``calibrated'' against visible absolute intensities
and center-limb behavior, as described below (\S{3.3.1}),
it is important that the b--f and f--f opacities enjoy high relative accuracy, to
ensure an effective transfer of the visible temperature calibration to the infrared.

We also included the metal opacities of Peach (1970) for Mg, Si, and Al; hydrogen b--f according
to the Gaunt factors of Carbon \& Gingerich (1969), assuming LTE populations for the $n\ge 3$ levels 
that contribute
to the visible and IR opacity;  hydrogen f--f according
to the Gaunt factors of AHM73; Thomson scattering by free electrons 
and Rayleigh scattering by neutral hydrogen; and H$_{2}^{+}$ f--f from
Kurucz (1970).  These all are minor contributors at the wavelengths of
interest for the present work. 

We assumed LTE formation of
the continuous opacity sources and the CO $\Delta{v}=1$,\,2 bands.  This is
a good approximation for the collisionally controlled H$^{-}$ populations, and the 
H$^{-}$ f--f absorption process, in the deep
photosphere where the visible and infrared continua form; as well as for the CO bands
themselves (AW89).  Small
departures from LTE, and potentially large departures from instantaneous chemical
equilibrium, can affect the CO lines at high altitudes (Uitenbroek 2000b; 
Asensio Ramos et al.\ 2003; Wedemeyer-B\"ohm et al.\ 2005), but these
are less relevant for the present study. 

\subsection{Molecular Parameters}

We adopted the CO formation parameters of Kurucz (1985), except for
the dissociation energy of $^{12}$C$^{16}$O which we took from MN94, 
$D_{0}= 11.108\pm 0.002$~eV.  The dissociation energies for the isotopic variants of
CO typically are 0.002~eV higher.  The $D_{0}$ in the
Kurucz compilation was slightly lower (by 0.017~eV), but we prefer the more recent
Morton \& Noreau value.  The small difference in any event is inconsequential, and we consider the
dissociation energy to be very well determined.  

We adopted the excitation energies, oscillator strengths, and partition functions tabulated by
G94 for $^{12}$C$^{16}$O, and the isotopomers $^{13}$C$^{16}$O,
$^{12}$C$^{17}$O, and $^{12}$C$^{18}$O; based on experimentally determined
electric dipole moment functions (EDMFs).  The excitation energies are extremely well determined
through direct spectral measurements, but the line strengths 
are less certain (especially at high $vJ$: see, e.g.,
Chackerian et al.\ 1994), and potentially
are a major source of systematic error in an abundance analysis.  We therefore also considered
oscillator strengths derived from the dipole matrix elements of Hure \& Roueff (1996, hereafter HR96), based
on {\em ab initio}\/ EDMFs;
and the earlier pioneering study by Kirby-Docken \& Liu (1978, hereafter KL78).  Figure~8 compares these
two oscillator strength scales to the adopted G94.

The HR96 $gf$-values are systematically $\sim$3\,\% higher than G94 for the
$\Delta{v}=1$ bands; and up to 5\,\% lower for $\Delta{v}=2$ transitions, with a noticeable
trend with lower level excitation energy amounting to 2\,\% per $10^{4}$~cm$^{-1}$ interval
in $E_{\rm lower}$.  
The earlier KL78 values are lower than G94 by up to 20\,\%, and show
larger scatter than H96.  We consider the differences between the independent
G94 and HR96 scales to be a reasonable assessment
of the accuracy of the CO line strengths.  

That said, one
potential area of concern is that the $\Delta{v}=1$ abundance diagnostic lines
are mainly from high initial $vJ$ states: the lower lying
vibrational and rotational levels are heavily populated, and their usually saturated line transitions
mostly are useless as abundance tracers.  On the other hand, the $\Delta{v}=2$
transitions have oscillator strengths typically two orders of magnitude smaller
than the fundamental lines from the same $vJ$ initial
state.  Thus, the first overtone
abundance lines tend to be from low $vJ$ initial states
corresponding to peak populations of the rovibrational levels.  If there are any systematic
errors in the oscillator strength scales that depend, say, on $vJ$, our comparisons of
$\Delta{v}=1$ and $\Delta{v}=2$ abundances could be skewed,
and thermal profiling based on the stronger saturated $\Delta{v}=1$ lines could
be affected as well.  We therefore took special care to
select abundance diagnostic lines covering as wide a range of lower level excitation energies as
feasible for both the $\Delta{v}=1$ and 2 band systems.

As for the isotopic transitions, 
their oscillator strengths are essentially identical to those
of the parent lines, thus uncertainties in the $gf$-value scale effectively cancel
in the abundance ratio, say $^{16}$O/$^{18}$O.  In principle, then, it should be
possible to derive isotopic ratios more accurately than
$\epsilon_{\rm O}$ itself.

\subsection{Thermal Profiling}

The CO infrared absorption spectrum is highly temperature sensitive thanks to the 
diatomic molecular formation.  CO is more robust than other
diatoms owing to its high dissociation energy.  This allows
CO to persist over a wider
range of temperatures than any other molecule, extending even down into the relatively warm 
photospheric layers
where $T\gtrsim 5000$~K.  Thus CO can serve a thermal tracer over an
appreciable, important range of the middle photosphere.  Nevertheless, we must appeal
to visible continuum intensities, and visible and infrared center-limb behavior, to
constrain the thermal profile in the deepest molecule-free layers.

Despite some amount of standardization in opacities and equations
of state---thanks to historical codes
from major modeling groups, such as at Harvard and Uppsala, which have been adapted by
others and modified over the years---even small differences in cross sections, 
electron donor abundances,
or the level of sophistication in the equation of state is sufficient to render inaccurate one
group's calculations of, say, continuum intensities using another group's thermal model.  
One can minimize this ``model construction bias''
by beginning with the other group's $T(\tau_{0.5\,\micron})$ relation, if available, 
and decomposing it using ones
own opacities, abundances, equation of state, and so forth; but even so, small difficult to recognize
systematic differences might persist.

We worked around the bias by utilizing the absolute visible
continuum intensities not simply to test the accuracy of different models (L78), but rather
as a calibration which we force all model distributions to obey (A77).  
This ensures that the model temperatures
in the deep photosphere, where the visible continuum forms, are accurately slaved to the
calibrated empirical intensities, allowing us to compare
different models on as fair a basis as possible.  Then, the continuum center-limb behavior provides
an independent test of the adjusted models, specifically their temperature gradients
shallower than $\tau_{0.5\,\micron}\sim 1$.  A model scaled to match the visible absolute
intensities, and which reproduces the continuum center-limb behavior from 0.5~\micron\
out to the limit of the CO $\Delta{v}=1$ bands at $\sim 6$~\micron, should have the ``correct''
temperature distribution in the vicinity of the CO forming layers, which is a necessary
condition to ensure accurate abundance determinations from the molecular absorption
spectra. 

\subsubsection{Photospheric Models: Matching Absolute Continuum Intensities\\ and Center-limb Behavior}

Figure~9 compares three photospheric thermal profiles evaluated in the present
study: FAL\,C; the mean model
of Asplund et al.\ (2004), but including their rms temperature perturbations; and a
modified version of the Asplund model adjusted slightly in its deeper layers 
to better match the continuum center-limb behavior, and in its outer layers for CO center-limb behavior and
the off-limb CO emissions (for historical reasons, we call this model ``COmosphere''
[see, e.g., A02]). 

We caution that because the Asplund mean model was constructed by averaging the dynamic 3-D simulation
over surfaces of constant radial 0.5~\micron\ continuum optical depth, it not necessarily is the same
mean $T(m)$ stratification that would be obtained by a similar averaging at a very different
wavelength, say 5~\micron; or for that matter if the averaging were done along slanted rays, as for
limb viewing angles.  To the extent that 3-D effects are important, the conclusions we draw below
based on the 1-D Asplund mean model (or 1.5-D variants) might not be applicable to a proper treatment of
the multidimensional radiation transport in the full 3-D time dependent structure.  In a general
sense, the 3-D effects should be less important for disk center spectral modeling (from
which we derive abundances and isotopic ratios), but more
significant for center-limb behavior (which we exploit as a discriminator among the various
model stratifications).  We have adopted the 1.5-D approach here to allow a more detailed treatment
of the various model and parameter sensitivities of the CO and continuum formation
than would be possible with a full 3-D model.  This approach allows us to fairly
incorporate results based on previous 1-D semiempirical models (which still might have
future value for analyses of unresolved stars), and points toward additional
tests that could be applied to validate the full 3-D simulations (which---although
highly sophisticated in their present incarnations---after all still are only an
approximate representation of the complexity of the real photosphere of the Sun).

Figures~10a and 10b display absolute continuum intensities and
center-limb behavior predicted by the FAL\,C and Asplund mean model, 
relative to the measured values described
in \S{2.4}.  The mass scales of both models were adjusted slightly (typically a few percent)
to force agreement with the absolute continuum intensities, the fundamental model calibration
discussed earlier.  Although the Asplund mean model accurately reproduces the center-limb behavior
of 5~\micron\ continuum intensities, it does relatively poorly in the visible and near IR,
compared with the FAL\,C.  We also depict
the center-limb behavior in terms of
brightness temperatures, to remove the Planck function bias inherent in intensity comparisons
across large stretches of the spectrum (i.e., a $\Delta{I}$ of 1\,\% represents a much
larger temperature difference in the mid-IR than in the blue).  
The comparison indicates that the mean temperature profile of the Asplund model has too
steep a gradient in the visible continuum forming layers ($\tau_{0.5\,\micron}\sim 1$)
compared with the reference model, which can be seen clearly in
Fig.~9, as well.  The too steep predicted center-limb curves in the visible imply that
the deep photosphere temperature ``calibration'' will not be carried reliably out to the higher
layers in which the CO infrared bands form. 

Figure~10c depicts the same continuum comparison for the Asplund mean model, but now including
multicomponent effects in the schematic 1.5-D way by combining intensities synthesized with
a 5 component model.  The central profile is the
mean model $T_{\rm mean}(m)$, the outer temperature profiles
are $T_{\rm mean}\pm \sqrt{2}\times\Delta{T}_{\rm rms}$
(as given by Asplund et al.\ 2004), and the inner profiles are 
$T_{\rm mean}\pm \frac{\sqrt{2}}{2}\times\Delta{T}_{\rm rms}$.  The components were 
evenly weighted to
approximate a Gaussian distribution.  Predictions for the individual profiles
are depicted as thin curves, while the weighted averages are represented by heavier 
dashed curves.  Here, again, the underlying mean model was adjusted slightly in $m$,
with the indicated $\Delta{T}$ perturbations, to force the
average intensity to match the reference 
continuum highpoints.  (Accordingly, the scaling is slightly different
for the mean model by itself [Fig.~10b], versus the 5 component version [Fig.~10c].)   
Note that while the
range of intensity fluctuations for the individual 
components in the visible is substantial ($\sim 20$\,\%),
the center-limb behavior remains close to that of the single component model.  

The 1.5-D approach, of course, neglects the geometrical interaction along the same highly slanted
sightline of, say, the narrow dark subduction lanes
and the more puffed up, bright upwelling cells of the 3-D time dependent simulation.  
This effect is very important for understanding the
center-limb behavior of G-band bright points, for example,
which are very fine scale partially evacuated magnetic
structures found principally in the dark lanes of the granulation pattern (Carlsson et al.\ 2004).  However,
for the normal photospheric granulation pattern, which consists of much larger structures, the geometrical
interactions should be less important.  Furthermore, the empirical measurements of thermal
fluctuations in the 5~\micron\ continuum and weak CO lines---which at disk center bracket
the altitude range where
the visible continuum arises at the extreme limb---suggest that the 3-D model might exaggerate
the thermal contrasts in those layers, leading to even less of a sightline interaction effect in the
real Sun.  (This conclusion rests, however, on the assumption that the empirical measurements of
intensity fluctuations fully resolve the thermal inhomogeneities in the middle photosphere.
If the true structures are very fine in spatial scale, blurring by the telescope and atmosphere
will reduce the constrast.  This is particularly important in the thermal IR where diffraction
limits the resolution to barely subarcsecond.)

\subsubsection{Adjusted Thermal Structures and Model Sensitivities}

Although the scaled FAL\,C temperature profile in the deep photosphere reproduces
the continuum absolute intensities and center-limb behavior well, it has a warm chromospheric
temperature inversion 
at intermediate altitudes ($\sim 500$~km above $\tau_{0.5\,\micron}=1$), which is incompatible with
the center-limb behavior of strong CO $\Delta{v}=1$ lines and their off-limb emissions (e.g., A02).  The
Asplund mean model, as well, fails to reproduce the CO center-limb behavior
[in part because their $T(m)$ cuts off in the upper photosphere,
and had to be extended isothermally beyond the outermost point in their tabulation].  

We therefore crafted the COmosphere model by first
adjusting the scaled Asplund stratification in the deeper layers to 
match the continuum center-limb curves, while continuing to force a
fit to the absolute intensities.  The resulting deep photospheric temperature 
profile is nearly identical to that of the FAL\,C, and essentially
indistinguishable in terms of continuum center-limb behavior in the visible and near IR.  
We further spliced 
in a high altitude $T_{\rm min}$ (and temperature rise),
such as illustrated in Figure~3 of A02 (``Cool\,0'' model), and altered the mid-photospheric
profile to match CO $\Delta{v}=1$ center-limb behavior (as illustrated below [\S{3.3.4}]).  

Finally, like the
5 component stripped down version of the Asplund model, we introduced equally
weighted temperature perturbations around
the mean COmosphere $T(m)$ profile.  In order to match
the $\sim 10$\,\% rms granular contrast at 0.60~\micron, 
we retained the Asplund et al.\ $\Delta{T}_{\rm rms}$
in the deep photosphere ($m\gtrsim 3$~gr cm$^{-2}$), but imposed smaller
$\Delta{T}$'s in the higher layers based on the empirical behavior of strong and weak CO
lines and the 5~\micron\ continuum described in \S{2.3.2}.

We further added
a ``hot'' component identical to the FAL\,C in the deep photosphere, but 200~K warmer above about
$m= 1$~gr cm$^{-2}$.  This component was intended to mimic the influence of supergranulation
network elements, and was assigned a weight of 10\,\% (such bright, warm patches are seen
in large-area CO maps, but have small filling factors: AR96).  Since the ATMOS spectra
were accumulated over a substantial field of view at Sun center, it seemed prudent to incorporate at
least some contribution from the warm network elements that must have been captured.
As with the multicomponent version of the Asplund model, we found that the average
continuum center-limb behavior of the 6 component COmosphere model
was not appreciably different from that of the single component version
by itself (not illustrated).  Physical properties of the 6 component COmosphere model are
summarized in Table~2 (``$T_{\rm NW}$'' is the temperature profile of the hot network component).

\subsubsection{Initial Abundance Tests}

In this section, we describe our initial efforts to derive self-consistent C and O
abundances from the
reference models, as a prelude to several tests of the model sensitivities (varying such
parameters as the microturbulent velocity, mid-photosphere temperature profile,
$T_{\rm min}$ value, and so forth).  First, however, we depict in Figure~11 the
center-limb behavior and contribution functions for
representative CO $\Delta{v}= 1$ transitions synthesized with the 1-D version of the COmosphere
model and the G94 oscillator strengths.  The observations in the lower set of panels
are mainly from the McMath-Pierce FTS, although the ATMOS spectra are overlaid on the disk center 
($\mu= 1$) profiles (and are indistinguishable).  
The simulated profiles were based on
$\epsilon_{\rm O}= 846$~ppm (with C/O=\,0.5 and energy dependent corrections
to the oscillator strengths to be described later [\S{3.3.3}]).  
The optimum model matches the variation of CO core depths with increasing line
strength (right to left) very well at disk center, and the center-limb behavior
of the same features much better than the FAL\,C or Asplund models (shown later [\S{3.3.4}]).

In the upper two panels of Fig.~11,
normalized contribution functions are displayed for $\mu= 1$ (upper)
and $\mu= 0.152$ (lower).  The curves furthest to the right in each panel are 
the 4.66~\micron\ continuum ``intensity contribution functions,''
and the curves to the left are for the CO 
transitions, calculated as ``line depression contribution functions''
at the line center frequency, as described by Magain (1986).  The distinction between intensity
and line depression contribution functions is important for the weaker CO lines (e.g., 7--6~R68)
because the CO molecular concentration (top panel: diamonds) peaks higher in the photosphere than
the narrow deeper zone where the IR continuum arises, and the optically thin
line depression thus forms in
layers entirely separate from the background continuum.  However, for the stronger CO
lines, the distinction between the two types of contribution functions is less
important, because the
line opacity completely dominates the continuum, at least at line center.

Figure~12 displays oxygen abundances derived for single component
versions of the COmosphere and Asplund models (upper two, and middle two panels, respectively), 
and the two more recent sets of $gf$-values: G94 and
HR96.  In practice, we calculated CO line profiles for a grid of 
discrete oxygen abundances (10 covering the range of expected
individual $\epsilon_{\rm O}$'s with logarithmic spacing); fitted Gaussian profiles to the simulated
intensities to
determine theoretical equivalent widths $W_{\omega}$
as a function of $\epsilon_{\rm O}$ for that particular
transition; then used the empirical value of the equivalent width, obtained through
an identically motivated Gaussian fit to the observed lineshape, to 
solve for the oxygen abundance according to a low order polynomial match
to the theoretical $W_{\omega}(\epsilon_{\rm O})$ relation.

The resulting $\epsilon_{\rm O}$'s for the $\Delta{v}= 1$,\,2
samples are depicted in the right hand panels as a function of lower state excitation
energy, $E_{\rm lower}$.  The lighter dots generally to the left are the first overtone lines;
the darker dots generally to the right are the higher excitation fundamental transitions.  
The hatched areas indicate the means and
$\pm 1\,\sigma$ standard deviations of the two separate samples, while the green
line is a linear least squares fit to the two samples combined (utilizing 
a ``$2\,\sigma$'' filter to eliminate outliers).

The apparent slope of the 
derived abundances with excitation energy
could indicate either that the adopted thermal profile is too hot in
the CO layers (note that the cooler Asplund model displays
a shallower slope for the $\epsilon_{\rm O}$--$E_{\rm lower}$ relation), 
or that the oscillator strength scale is systematically skewed ($E_{\rm lower}$
generally is a proxy for $vJ$).  Since we have carefully
``calibrated'' the temperatures of the CO layers, we initially adopted the second viewpoint 
(although, we also will consider the first alternative later [\S{3.4.2}].)  We thus corrected
the oscillator strength scale according to the apparent slope of the 
$\epsilon_{\rm O}$--$E_{\rm lower}$
relation, taking the true abundance from the intercept of the fit, assuming that the lower energy
line strengths are better determined than those of the high-$vJ$ transitions.  (The
G94 line strengths were derived from
empirical EDMFs based on laboratory measurements of low excitation CO transitions.)
If the empirical abundances are represented as,
\begin{equation}
(\epsilon_{\rm O})_{i}= (\epsilon_{\rm O})_{0}\times(1\,+\,\Delta\times E_{i})\,\,, 
\end{equation}
then the corrected oscillator strengths would be,
\begin{equation}
(gf)^{\rm corr}_{i}= (gf)^{\rm orig}_{i}\,\times(1\,+\,\Delta\times E_{i})^2\,\,, 
\end{equation}
since the equivalent width in the unsaturated regime depends linearly on $gf$ but
quadratically on $\epsilon_{\rm O}$.
The effect of
correcting the $gf$-values is illustrated in the lower 
two panels [``($gf$)$\ast$''] for the alternative
oscillator-strength scales with the optimum COmosphere model.  To guard
against a systematic deviation between the $\Delta{v}= 1$ and 2 line strength scales (as 
seen for example
in Fig.~8 with HR96 versus G94) masquerading as a trend of $\epsilon_{\rm O}$ with $E_{\rm lower}$,
we also fitted the $\Delta{v}= 1$ and 2 samples separately, again employing
a 2\,$\sigma$ filter.  However, in these initial comparisons we relied mainly on the combined
fit.  In essence, we treated the
$\Delta{v}= 1$ and 2 samples as partially redundant realizations of the same phenomenon---a
dependence of $\epsilon_{\rm O}$ on $E_{\rm lower}$---and utilized the combined sample for
a collectively more refined view: $\Delta{v}= 1$ for the higher energies, $\Delta{v}= 2$ for
the lower.

Figure~13 is a montage of $\Delta{v}=1$,\,2 $^{12}$C$^{16}$O profiles and simulated lineshapes for
representative members of the abundance sample (40 fundamental and 40 first overtone lines in total:
20\,+\,20 are illustrated in the figure) and the 
single component COmosphere model with the best-fit $\epsilon_{\rm O}$ (846~ppm) and the 
empirically corrected $gf$-values.  
The weaker lines (with central residual intensities $\ge 85$\,\%)
were locally normalized to the continuum level according to bins (green points) on either side of
line center, away from the absorption depression itself.  The specific features were chosen
from a much larger sample on the basis of weakness and freedom from obvious blends:
other CO transitions are the most common spectral contaminant in
the thermal IR.  The 7--6~R68 transition provides a
connection to the groundbased center-limb line set. 
Table~3 lists the measured equivalent widths and FWHMs for
the abundance sample lines, derived from Gaussian fits to the
ATMOS/ATLAS-3 spectra.

\subsubsection{Model Sensitivities}

The
top panel of Figure~14 compares the effect of the different test temperature stratifications
on the center-limb behavior of strong and weak CO $\Delta{v}=1$ lines from the recent McMath-Pierce FTS
observations described earlier (\S{2.2}; the $\mu= 1$ traces include the ATMOS spectra for
comparison; the limb trace is for $\mu= 0.152$).  The darkening of the CO line cores with
increasing line strength (from right to left) at disk center, and the darkening of the individual
line cores from disk center to the extreme limb, carry a redundant imprint of the thermal
profile of the mid- and high photospheric layers. 

Here, the spectral simulations were based on $\epsilon_{\rm O}$'s specific to each thermal
model, as derived from the $\Delta{v}= 1$,\,2 abundance sample, and
with the empirical corrections to the oscillator strength scales.  
One notices that although all three models
fairly well reproduce the disk center line depths, the Asplund and FAL\,C models both are too
warm at high altitudes to match the center-limb behavior 
(i.e., yielding core intensities higher than observed), at least in the stronger $\Delta{v}= 1$ lines.
Furthermore, the Asplund continuum levels are slightly lower,
on the brightness temperature
scale, both at
disk center and the extreme limb.  This is a consequence of the too steep temperature
gradient of that model.  Multicomponent (1.5-D) versions of the Asplund
model, to mimic the thermal heterogeneity of the convection pattern, fail
to improve the center-limb comparison.

The remaining panels of Fig.~14 illustrate the sensitivity of the CO lines, and center-limb behavior, to
changes in a number of key parameters that affect the line formation, including
modifications of the thermal structure itself.  The panel labeled ``$\gamma$'' depicts lineshapes
synthesized for three values of the pressure broadening coefficient appearing in the dimensionless
Voigt $a$ parameter (which controls the Voigt profile shape adopted for the CO absorptions;
the radiation damping parameter $\gamma_{\rm rad}$ is completely negligible for these
lines owing to their small Einstein $A$ values):
\begin{equation}
a_{\rm V}= \frac{\gamma\, (n_{\rm H}/10^{16})\, (T/5000)^{\onehalf}}{\Delta\omega_{\rm D}}\,\,,
\end{equation}
where $n_{\rm H}$ is the neutral hydrogen density in cm$^{-3}$, $T$ is the temperature in
K, and $\Delta\omega_{\rm D}$ is the Doppler width in cm$^{-1}$ (including thermal and turbulent broadening; 
``macroturbulence'' was implemented by Gaussian smoothing the simulated profiles
after the fact).
The $\gamma$ coefficients used in the figure are 0, $2\times10^{-4}$, and $4\times10^{-4}$ cm$^{-1}$;
the middle one being our preferred value, after considering a large number of
very strong $\Delta{v}=1$ absorptions from the ATMOS traces.  The pressure broadening has a relatively
minor effect: exclusively in the line wings,
and mainly confined to the strong, saturated transitions, as can be seen from the diagram.
It usually is ignored in CO lineshape synthesis, but we include it here
for completeness.  

The panel labeled ``$\xi$'' illustrates the effect of changing the radial microturbulence parameter
over the range 0.25, 0.75, and 1.25 km s$^{-1}$.  Again, the middle
value is the preferred one (consistent with the depth-dependent velocity distributions
in the middle photosphere given by Asplund et al.\ [2004] and FAL\,C, and
nearly identical to that adopted by L78).  Since we adjusted the 
macrotubulence for
each specific value of the radial $\xi$ to reproduce the widths of (the fully
resolved) weak lines at disk
center, the effect of varying $\xi$ on the $\mu=1$ profiles of even
the stronger lines is relatively subtle.  However, changing $\xi$ also affects the 
turbulent pressure contribution to the total pressure, and thus modifies 
the density stratification to some extent,
requiring a small adjustment to the $m$ scale to maintain the continuum intensity calibration.  Still, the
overall effect on the disk center spectrum is relatively minor. 

The effect
of changing the tangential $\xi$ (over the corresponding range 1.5, 2.0, and 2.5 km s$^{-1}$)
is more pronounced.  The best fit value,
2~km s$^{-1}$, is substantially larger than the radial component.  The fact that 
7--6~R68---whose extreme limb contribution function overlaps the disk center peaks
of the stronger lines---exhibits enhanced broadening at the limb suggests
that the underlying $\xi$
distribution truly is anisotropic.  Like the pressure broadening, the turbulent
velocity plays mainly a minor role in the thermal profile and abundance analysis, but we optimize 
the match to minimize any (small) systematic influence.

The fourth panel down illustrates the effect of altering the oxygen abundance
over the range 
$\epsilon_{\rm O}/1.5$, $\epsilon_{\rm O}$, $\epsilon_{\rm O}\times 1.5$,
for the optimum value (846~ppm)
derived with the single component COmosphere model (and C/O=\,0.5).
Changing the oxygen abundance has the greatest influence on weak 7--6~R68 at disk
center, although the effect is suppressed at the limb.   The other stronger lines are much less 
affected at either viewing angle.  This is
because only 7--6~R68 at disk center is unsaturated ($\tau\sim 1$), and 
thus its equivalent width responds
strongly to abundance (as something like $\epsilon_{\rm O}^2$, given the imposed fixed
C/O ratio).  The stronger lines, on the other hand, are partially to fully saturated
($\tau\gg 1$),
particularly at the extreme limb.  They arise in the middle and upper photosphere where the
temperature profile is relatively flat compared with the continuum forming layers.  
The effect of altering the oxygen abundance
is to move the depth (and thus temperature) where the line core becomes opaque (and radiates
like a blackbody at the $\tau\sim 1$ temperature for these LTE lines).  
However, because the temperature gradient is shallow, 
the temperature at the new depth is not much different from the old one, and thus the influence
on the line core intensity is reduced.  

The next two panels illustrate the consequences of modifying the photospheric thermal structure of the
COmosphere test model.  The panel labeled ``$T_{\rm OUTER}$'' changes the mid-photospheric
temperature gradient.  The lower distribution closely matches that of
the FAL\,C model (which was derived largely from the behavior of the \ion{Ca}{2}
0.395~\micron\ H and K line wings, a powerful diagnostic for thermal structure
in the solar photosphere and stellar atmospheres in general
[cf., Ayres \& Linsky 1976, hereafter AL76]), while the upper two models
are 200~K and 400~K hotter, respectively, at $m= 0.01$~ gr cm$^{-2}$,
joining smoothly to the lower $T(m)$ at $m= 1$~ gr cm$^{-2}$.  Again, the middle model is our
preferred temperature distribution.  The main influence is 
concentrated on the strong lines at disk
center, and the two weaker lines at the limb; but the weak lines at disk center and the strong
lines at the limb are mostly unaffected.  The weak lines at disk center are minimally influenced because they 
form deeper
where the $T_{\rm OUTER}$ models converge in temperature.  As for the strong lines at disk center, they have
double-peaked contribution functions (Fig.~11), responding both to the $T_{\rm min}$ region as well
as the upper photosphere near $m= 0.01$.  It is the second bump that causes the larger changes in
the strong line cores at $\mu= 1$ for variations in $T_{\rm OUTER}$.  Similarly, the 
formation height for the weaker lines at the
extreme limb moves up from the $m\sim 1$ level, where 
the differences between the perturbed models are small, to the $m\sim 0.1$ level where the
temperature differences are larger.  Meanwhile, the stronger lines at the extreme limb form
almost exclusively in the $T_{\rm min}$ region, whose temperature is the same for 
the different $T_{\rm OUTER}$ models,
so the line cores are minimally affected.  

The bottom row of panels depict the influence of changing $T_{\rm min}$ over the range 3400, 3600,
3800~K, with again the middle value preferred.  Here, the main effect is on the cores of the
strongest lines at the extreme limb, since, as was mentioned above, their contribution functions peak 
at $T_{\rm min}$ for the highly slanted sight lines.
A smaller effect is seen in 6--5~R47 at the limb, because of the double-peaked nature of its
contribution function, but the influence on even weaker 7--6~R68 is minimal.  The stronger
lines at disk center show a similar level of response as 6--5~R47 at the extreme limb, again
because of twin peaks in their contribution functions.

The final COmosphere model balanced these several levels of sensitivity:
$T_{\rm min}$ altitude based on CO off-limb emission extensions, $T_{\rm min}$ temperature
based on cores of strong CO lines at the limb, mid-photospheric $T(m)$ from core-depth variation
with line strength at disk center, and deep photosphere thermal profile from continuum
absolute intensities and center-limb behavior; all
with a self-consistent $\epsilon_{\rm O}$ (and $\Delta$) derived from
the reference abundance sample.  The model is not perfect, because it does not fully
reproduce the center-limb behavior of 6--5~R47 and 7--6~R68.  However, models that do match
that observational constraint (like the cooler $T_{\rm OUTER}$ case) tend to fail other
criteria (in this case, the distribution of core depths with increasing line strength at disk
center).   
Nevertheless, the fact that a single $T(m)$ distribution does so well for a wide range of the criteria
gives us encouragement that we can exploit it as a basis to determine auxiliary properties,
like the oxygen abundance; estimating the
influence of, say, thermal inhomogeneities by treating them as 1.5-D perturbations of the reference
model.

\subsection{Abundance Simulations}

We conducted abundance simulations for a wide range of thermal models, to test the
sensitivities in a similar way as illustrated in Fig.~14.  The results are
summarized in Figures~15a (single- and multicomponent models) and 15b 
(parameter variations for 1-D COmosphere model),
and in Table~4.  The optimum single component COmosphere model yields a reference
oxygen abundance of 846~ppm (with a gradient in $\epsilon_{\rm O}$--$E_{\rm lower}$
of $\Delta= -5.5$\,\% per 10$^{4}$ cm$^{-1}$), considerably higher than the ``low'' $\epsilon_{\rm O}$'s
cited in \S{1} derived with the full 3-D Asplund
model from atomic oxygen and hydroxyl lines.  
The single component version of the Asplund model gives 
a lower 575~ppm, and better consistency between the fundamental and first overtone line
sets (i.e., a shallower gradient of $\Delta= -2.9$).
The 5 component Asplund model, again
forcing the continuum calibration on the weighted average intensity, yields an oxygen
abundance about 14\,\% lower than the single component model (496~ppm with a {\em positive}\/ $\Delta= +1.7$).
The 0.06~dex
drop in the oxygen abundance from the mean model to the 5 component version is entirely
consistent with the assessment by Asplund et al.\ (2004) of multicomponent effects; and the CO derived
oxygen abundance is similar to what those authors obtained from
\ion{O}{1} and OH using their full model.

However, the lower oxygen abundance from CO is a direct consequence of the too steep
temperature gradient of the Asplund model, which leads to cooler temperatures in the
CO forming layers, enhanced column densities of absorbers, and thus a lower 
inferred $\epsilon_{\rm O}$.  We do acknowledge that
the $\Delta{v}=1$/$\Delta{v}=2$ dichotomy (in principle temperature-sensitive because it
pits generally low excitation first overtone lines against mainly higher excitation fundamental
transitions) does seem to favor the cooler temperatures of the Asplund
model in those layers.  But, because even small systematic errors in the CO oscillator strength
scale could mimic that result, we prefer the more accurately known continuum properties
to constrain the temperatures.  (With the proviso that a full 3-D rendering of the 
continuum center-limb behavior might obtain a better match to the
observations than the 1.5-D version of the
Asplund model, thereby negating our fundamental test.)

Applying the milder IRIS derived thermal perturbations in the CO layers to the COmosphere profile, 
and adding in the hot network component (FAL\,C + 200~K at
high altitudes, covering
10\,\% of the area), yields 850~ppm ($\Delta= -5.3$),
virtually identical to the single component version.
The FAL\,C model predicts
820~ppm ($\Delta= -5.6$).  Changing the upper photosphere temperature gradient in the single component
setting (``$T_{\rm OUTER}$'' models) has
a $\sim \pm 10$\,\% influence, but is constrained by the
empirical distribution of CO core depths with increasing line strength.

In one of the cases, we altered
the H$^{-}$ f--f cross section relative to b--f by $\pm 10$\,\%, then carried out
the continuum calibration procedure and 
oxygen abundance synthesis.  Here, the photospheric temperature structure
remains essentially the same as before, but now the depth of formation
of the background continuum at 2.5~\micron\ and 5~\micron\ changes.
This apparently affects $\epsilon_{\rm O}$ by only $\sim \pm 2$\,\%.
In another case, we altered both the b--f and f--f
cross sections by the same $\pm 10$\,\%, then carried out the calibration/abundance procedure.
Here, the temperature structure is modified (by the continuum calibration) but the relative
depths of formation of the visible and IR continua remain the same.  The apparent effect
on $\epsilon_{\rm O}$ is somewhat larger, $\sim \pm 5$\,\%.

The first case tests possible systematic errors in the relative cross sections, and the second,
the absolute cross sections (assuming that the relative
behavior is well determined).  In both cases, the $\pm 10$\,\% perturbations in
the cross sections are much larger than the few percent accuracy cited in the original
studies.  We also altered the CO dissociation energy $D_{0}$ by $\pm 0.100$~eV
($\sim 1$\,\%) which affects the derived oxygen abundance by $\pm 10$\,\%.  We found
that the ``high'' oxygen abundance predicted by the
COmosphere model could be lowered to 450~ppm if $D_{0}$ were raised to (an implausibly large)
11.750~eV.  Changing the radial and tangential microturbulence parameters together
by $\pm 0.5$~km s$^{-1}$ alters the derived
oxygen abundance by an insignificant $<\pm 1$\,\%.

Our initial conclusion is that a solar oxygen abundance in the neighborhood of 850~ppm is more
compatible with the observed CO $\Delta{v}= 1$,\,2 spectra, 
than the $\sim 400$~ppm lower values obtained
from analyses of atomic oxygen lines in the visible, and hydroxyl bands in the infrared.  

\subsubsection{Uncertainties}

Random errors of measurement are essentially negligible owing to
the high S/N of the ATMOS spectra.   
Systematic errors must dominate the error budget, then, but are
more difficult to quantify.  One important source
is setting the continuum level, which becomes progressively
more important the weaker the absorption line.  The ATMOS traces are superior in this regard
to typical visible solar spectra, since line crowding is much less in the thermal IR and 
telluric absorption
is negligible.  Consequently, there usually is a high density of intervals where the 
intensity ostensibly is flat
and an accurate continuum level can be drawn in an unbiased way
using numerical filtering techniques (e.g., Bennett \& Ayres
1988).  We carefully selected the abundance sample lines by examining essentially
every possible CO transition in the 1600--2400~cm$^{-1}$ region, choosing only those which were 
minimally disturbed
by nearby blends whose wings might alter the surrounding continuum level.  Furthermore, 
we established a specific local continuum level for the weakest
transitions by averaging the intensities in two bands several FWHMs to
either side of line center.  

The typical 1\,$\sigma$ dispersion in derived abundances among the lines 
of the $\Delta{v}=1,\,2$ samples
and the COmosphere model is about $\pm 6$\,\%, before compensation for
the $\epsilon_{\rm O}$--$E_{\rm lower}$ effect, and $\pm 2$\,\% after.  The latter might
be viewed as a minimal estimate
of systematic errors due to 
the continuum normalization, and any higher order distortions of the G94 oscillator strength
scale beyond the linear correction.  Since we have
no idea how correlated these errors might be
over the line set, we cannot divide by $\sqrt{N}$ to 
characterize the accuracy in terms of the
standard error of the mean.  Thankfully, the size of the error is small in any event.  

Our tests indicated that the influence of uncertainties in continuum
opacity cross sections are likely to be small, about the same as the errors above.  Multicomponent
effects are larger, depending on the detailed amplitudes of the rms temperature fluctuations in
the line formation zone.  As mentioned previously, 
Asplund et al.\ estimated 3-D effects of the same order ($\sim 0.08$~dex [20\,\%]).

In order to exploit CO as an oxygen abundance indicator, we had to make an assumption concerning
the C/O ratio, which we fixed at 0.5, based on a variety of previous studies.  The range of
C/O ratios shown in Fig.~1 is about 0.45--0.60, or $\sim\pm 20$\,\% around the nominal value.
Meyer (1985) quotes a ``Local Galactic'' ratio of 0.56, and a solar energetic particle result
of 0.45, nicely bracketing our assumed value.  The oxygen abundance scales roughly
as $r^{-1/2}_{\rm C/O}$,
so the uncertainties in $r_{\rm C/O}$ translate to $\lesssim \pm 10$\,\%, perhaps as small
as $\pm 5$\,\% (taking Meyer's LG and SEP results as a gauge of the true range of uncertainty in
the C/O ratio).

The CO oscillator strengths also are an important potential source of systematic
error.  Intercomparison of the two recent compilations
suggest $<\pm 5$\,\% absolute (which translates
to a smaller error in $\epsilon_{\rm O}$ since $(W_{\omega})_{\rm obs}\sim gf\,\epsilon_{\rm O}^2\, 
r_{\rm C/O}$).
Uncertainties in the model thermal profile were meant to be
highly constrained by the continuum calibration
and the behavior of the strong lines of the CO spectrum, but any systematic errors here would
have a magnified effect owing to the temperature sensitivity of CO.  We assigned 
a conservative error to this source, $\pm 10$\,\% based on the $T_{\rm OUTER}$
perturbed models, which would dominate all the
other sources (if the others are uncorrelated, and can be combined in quadrature).

Given the uncertainties that are easily identified,
and quantified, we conclude that the solar oxygen abundance derived from CO is
significantly different---and larger---than that inferred in recent years
from other oxygen bearing species.

\subsubsection{An Alternative ``Optimum'' Thermal Profile}

In the preceding sections, we followed a chain of reasoning dictated by our
belief that the continuum temperature calibration should be given precedence over
the potential temperature proxy connected with
an apparent slope of the derived $\epsilon_{\rm O}$'s with $E_{\rm lower}$. 
After all, such a slope could result from a systematic skew in
the CO oscillator strength scales with $vJ$.  
However, it also is true that the two recent
studies on the issue, G94 and HR96, which took somewhat different tacks 
(see Chackerian et al.\ 1994), arrived at basically
the same conclusions.  While variations in the
$gf$-scales with $E_{\rm lower}$ of a few percent per $10^{4}$~cm$^{-1}$ might be plausible,
the effect on $\epsilon_{\rm O}$ should be only about half that; nevertheless we were
finding as much as 6\,\% per $10^{4}$~cm$^{-1}$ for the $\epsilon_{\rm O}$'s of our optimum
COmosphere model.  Furthermore, the optimum model was less successful in matching the
5~\micron\ continuum center-limb behavior than, say, the 1-D version of
the Asplund model (although much better in the visible), and
did not reproduce the 7--6~R68 center-limb behavior particularly well either.  

In the spirit of experimentation, we considered an alternative class of COmosphere
models with a small temperature dip in the middle photosphere.  If the
dip is narrow enough, it will have little effect on the visible absolute intensities and
center-limb behavior, and the disk center IR continuum; or on the stronger CO line cores, 
which form higher up.  
If the dip is deep
enough, it can steepen the IR continuum center-limb behavior, flatten
the $\epsilon_{\rm O}$--$E_{\rm lower}$ relation, and improve the
center-limb behavior
of the key 7--6~R68 line. 

With this in mind, we developed what we call the ``Double Dip'' model (the first, outer ``dip''
refers to the $T_{\rm min}$ region).  We experimented with
a variety of realizations of the model, changing the location $m$ of the inner dip, 
its width $\Delta{m}$ and
depth $\Delta{T}$ to optimize the match with the constraints discussed earlier.  A 250~K
temperature depression, at $m\sim 0.35$ gr~cm$^{-2}$ with a width of a few tenths in $m$,
and additionally lowering $T_{\rm min}$ by 100~K (to 3500~K),
was found to maximize the agreement with those constraints.

Figure~16 depicts continuum center-limb curves synthesized with a
6 component rendering of the Double Dip model, demonstrating the improved agreement for
the thermal IR wavelengths, while maintaining a good match to the visible behavior.  (Including
the 10\,\% network component caused us to further lower the 
$T_{\rm min}$ of the mean profile by an additional 100~K [to 3400~K].)  Figure~17a
illustrates abundance solutions for 1- and 6 component versions of the
model.  Here we find that
the fundamental displays an essentially flat slope, but the first overtone still is slightly
tilted and displaced upward.  (The
apparent separation between the fundamental and first overtone abundances would have 
worsened with the HR96 $gf$-values.)  Fig.~17b illustrates the CO center-limb
behavior calculated with the 1- and 6 component Double Dip models, 
using oxygen abundances derived exclusively from the
$\Delta{v}=1$ sample.  Again, there is improved agreement with the observed
center-limb behavior of 7--6~R68, while maintaining the core intensities of the stronger CO
lines at disk center. 
We consider the 6 component Double Dip model to be the
most self-consistent representation of the solar photosphere, of all the models we 
tested, at least in the empirical sense of matching
CO and continuum behavior both at disk center and the extreme limb.  Physical properties of the
6 component version of the Double Dip model are summarized in Table~5 (excluding $\Delta{T_{\rm rms}}$
and $T_{\rm NW}$, which would be the same as in Table~2).

The 1-component version of the
Double Dip model indicated $\epsilon_{\rm O}= 599$~ppm for
$\Delta{v}=1$, alone, with a gradient of $\Delta= -0.1$; and 654~ppm for $\Delta{v}=1$ and 2 combined (with $\Delta= -3.4$).
The 6 component model, thanks partly to the
hot ``network'' component, yielded a slightly higher 641~ppm ($\Delta= -1.1$) for $\Delta{v}=1$ alone; and 
685~ppm ($\Delta= -3.6$)
for $\Delta{v}=1$ and 2 combined.  These $\epsilon_{\rm O}$'s are very
similar to that recommended by GS98, prior to the more recent work of
Asplund and collaborators.

We advance the Double Dip model because it reproduces most of the imposed observational constraints
within the context of a 1.5-D thermal model (including
allowance for the empirical mild thermal fluctuations indicated
by the IRIS time series).  However, while the primary thermal dip 
(``$T_{\rm min}$'') above the photosphere
can be understood within the context of strong CO radiative cooling (Ayres 1981)
or dynamical effects connected with traveling acoustic waves (Carlsson \& Stein 1995), the presence of
a secondary depression in the average thermal profile of the middle photosphere would be
unexpected.  One
can imagine that these not-so-shallow layers still would be dominated by radiative heating and
cooling in the continuum, and even experience some convective overshooting (although empirically
we see little evidence of that, at least in CO).  It is worth noting, however, that the secondary dip
does occur in the zone where the $^{13}$C$^{16}$O $\Delta{v}=1$ bands
arise, as well as the $^{12}$C$^{16}$O first overtone, which potentially could provide
a small amount of localized radiative cooling, perhaps enough to create a modest temperature
depression.  Alternatively, the inner dip might simply mark the altitude above which nonradiative
heating imposes a few hundred
degree temperature ``hump,'' before the strong CO cooling can reverse the temperature profile in
higher layers.

Even though we must be skeptical concerning the reality of the Double Dip structure, we
retain it in the ensuing discussion as an example of the range of thermal profiles allowed by
the empirical constraints, and leave it for future work to explore its reality.

\subsubsection{The Solar Oxygen Abundance from CO}

A summary of the preceding sections is as follows.  An optimum smooth thermal 
profile---COmosphere---which includes mild thermal perturbations and a minority
hot ``network'' component, satisfies most of the imposed observational constraints and yields an
oxygen abundance of around 850~ppm.  However, the inferred dependence of derived
$\epsilon_{\rm O}$ on $E_{\rm lower}$ for the individual $\Delta{v}=1$,\,2
transitions could be viewed as too extreme to be a systematic
error in the $gf$-value scales (although that should be checked experimentally), and suggests that
the temperatures in the CO formation zone are too warm.  A stripped down
version of the Asplund 3-D convection simulation, 
rendered as a 5 component 1.5-D model centered on the reported mean thermal structure with flanking
smooth perturbations to capture the cited rms temperature fluctuations, yields an oxygen abundance of
around 500~ppm (now with a positive $\Delta\sim +2$), in
agreement with the original work of Asplund et al.; but rather 
poorly matches the pivotal continuum center-limb test, and the center-limb
behavior of strong CO $\Delta{v}=1$ lines.  
All models, or CO parameter modifications, crafted to reproduce the low O
abundance of 460~ppm, fail either the continuum or CO tests, or violate
cited laboratory or theoretical uncertainties.  A compromise---but perhaps
physically challenged---Double Dip model, including mild thermal fluctuations and a minority
network component, matches the observational constraints better than any of
the smooth thermal profiles, and yields an oxygen abundance of around 650~ppm.  We believe that
the truth lies somewhere between COmosphere and Double Dip, probably closer to the latter.  We
thus propose that the solar CO rovibrational spectrum is most consistent with an oxygen
abundance in the neighborhood of 700~ppm, but acknowledge that model dependent (and other)
systematic
errors could be large, perhaps as much as $\pm 100$~ppm.  (The corresponding carbon abundance
would be 350~ppm.)

\subsection{The Isotopes of Carbon and Oxygen}

Isotopic ratios of the abundant light elements provide insight into the history of 
galactic chemical evolution (Langer \& Penzias 1993), and
photochemical and other fractionation processes in primitive solar system material 
(Krot et al.\ 2005).  Theoretical predictions indicated that solar isotopic 
abundances would be only
a few tenths of a percent larger than terrestrial standard values 
(Wiens, Burnett, \& Huss 1997), providing the motivation
for NASA's recent
{\em Genesis}\/ Discovery mission 
to achieve extremely high accuracy through
direct capture of the light ions and their isotopes from the solar wind (Burnett et al.\ 2003).

The best recent spectroscopic study of the main isotopes of carbon and oxygen
in the Sun was that of Harris, Lambert, \& Goldman (1987), based on
CO fundamental lines observed with an FTS instrument in a
high altitude balloon flight.  They concluded that the
venerable Holweger--M\"{u}ller (1974, hereafter HM74) photospheric model was most consistent
with visible continuum center-limb behavior and properties of the CO fundamental
spectrum.  They obtained $^{12}$C/$^{13}$C=~$84{\pm}5$ and
$^{16}$O/$^{18}$O=~$440{\pm}50$, as well as $\epsilon_{\rm C}= 513{\pm}120$~ppm.  The $^{12}$C$^{17}$O
spectrum was too weak in their scans to measure.  The mid-photospheric temperature
profile of HM74 is very similar to
the FAL\,C (and the COmosphere class of models), since the FAL\,C was derived to fit
the photospheric damping wings of the \ion{Ca}{2} resonance lines, which the
HM74 model does very well (A77).

Figure~18, similar to Fig.~15, illustrates abundance ratio analyses applied
to samples of weak lines from the $\Delta{v}=1$ bands
of $^{13}$C$^{16}$O, $^{12}$C$^{17}$O, and $^{12}$C$^{18}$O.  ($\Delta{v}=2$ transitions
of $^{13}$C$^{16}$O are found in the ATMOS spectra, but are extremely weak and therefore
add little to the already extensive sample of stronger $\Delta{v}=1$ lines.)
The isotopic ratios were derived by a procedure similar to the oxygen abundances
described previously, although the observed
equivalent widths 
were determined by a more constrained Gaussian modeling utilizing the
predicted line position and a fixed FWHM (4.05~km s$^{-1}$, based on unconstrained fitting of the
highest S/N lines of the $^{13}$C$^{16}$O abundance sample).  The measured
values are listed in Table~6.
In the line synthesis, the 
(model dependent) empirical corrections to the oscillator strength scales,
derived from the $^{12}$C$^{16}$O sample, were applied to the isotopic transitions.   
Results from representative models are listed in Table~7.  Two sets of ratios are presented: one
based on $\epsilon_{\rm O}$ derived from the full sample of $\Delta{v}=1$,\,2 transitions; the second
based on $\epsilon_{\rm O}$ exclusively from $\Delta{v}=1$.   

One sees, in general, that
the isotopic ratios are much less sensitive to even rather
drastic changes in the model thermal structures than the absolute abundances.  
Furthermore, the large samples, especially
for $^{13}$C and $^{18}$O, imply small statistical uncertainties, particularly if
we are allowed to consider the standard error of the mean (assuming that the dispersion
of values arises from uncorrelated errors).
Figure~19 illustrates representative isotopic profiles synthesized with the single component 
COmosphere model.  

The isotopic abundance ratios derived from CO
should be more secure than $\epsilon_{\rm O}$, itself, since we obtained
consistent results nearly independently of the thermal model.  For the
final values, we considered the 6 component versions of both COmosphere and Double Dip.  These
empirically capture our
best estimates of the mid-photospheric temperature profile and thermal fluctuations about it,
including the presence of a small percentage of hot network elements.  The Double Dip model,
in particular, has the
flattest distribution of $\Delta{v}=1$ $\epsilon_{\rm O}$--$E_{\rm lower}$ of all the
models that fit the continuum calibration and the CO center-limb data.  Although we have argued that
the Double Dip model might not be as physically plausible as some of the others, this criticism probably
is less germane in a differential comparison, such as performed here for the 
isotopic abundance ratios.  Flatness
of the $\epsilon_{\rm O}$--$E_{\rm lower}$ relation is important because we are comparing
generally low excitation isotopomers to generally higher excitation parent molecules, thus any
temperature dependent biases can skew the derived isotopic ratios.  In this regard, the Double Dip
model does display an apparent separation between the $\Delta{v}=1$ and 2 abundances,
particularly conspicuous in the single component version.  We therefore elected to consider only
the $\epsilon_{\rm O}$--$E_{\rm lower}$ relation for $\Delta{v}=1$ to derive the isotopic abundances,
whereas for the other classes of models---COmosphere in particular---we considered the  $\Delta{v}=1$,\,2 results together;
imagining that the fundamental and overtone bands were suffering the same systematic error
in their $gf$-values, and exploiting the $\Delta{v}=2$ transitions to provide additional information on
the abundance-excitation gradient, and abundance intercept, where the $\Delta{v}=1$ coverage is sparse.

We point out that the 1-D mean version of the Asplund model yields isotopic ratios in
good agreement with those from COmosphere and Double Dip, but the 5 component variant predicts larger
ratios (actually in quite good agreement with terrestrial values, unlike those from
the other models: see below).  The increase (corresponding to smaller isotopic abundances)
undoubtedly arises from the same general effect that lowers the $\epsilon_{\rm O}$ abundance itself
going from the single component thermal profile to the large temperature contrasts of the
multicomponent model, accentuated by the fact that the isotopic transitions, as mentioned above, are
mostly low excitation whereas the abundance sensitive $\Delta{v}=1$ lines are mainly high excitation.
However, we have dismissed the large temperature excursions of the Asplund model in the CO formation zone
owing to the empirical lack of significant temperature fluctuations in either the cores of
weak CO lines like 7--6~R68 or in the deeper seated 5~\micron\ continuum. 

Our recommended values of the isotopic ratios, taken
as an average of COmosphere*6 ($\epsilon_{\rm O}[\Delta{v}=1,\,2]$) and Double Dip*6
($\epsilon_{\rm O}[\Delta{v}=1]$) are: $^{12}$C/$^{13}$C=\,$80{\pm}1$, 
$^{16}$O/$^{17}$O=\,$1700{\pm}220$, and 
$^{16}$O/$^{18}$O=\,$440{\pm}6$, where the cited uncertainties are standard errors of the mean.  
Ironically,
these values possibly will be more accurate than those obtained {\em in situ}\/
from the solar wind by NASA's ill-fated {\em Genesis}\/ mission, depending on
how successfully extraterrestrial
material can be recovered from the damaged return capsule (McNamara et al.\ 2005). 

The specific values of the isotopic ratios deserve comment.  In the
solar system context, isotopic
abundances conventionally are expressed as differences with respect to 
reference terrestrial values in 
parts per thousand (per~mil: $^{\circ}\!\!/\!_{\circ\circ}$)
according to, for example, $\delta{^{13}\mbox{C}}\equiv [\epsilon(^{13}{\rm C})_{\odot}/
\epsilon(^{13}{\rm C})_{\rm standard})\,-\,1]\times10^{3}$.  In previous solar
studies, the uncertainties associated with the isotopic ratios were large enough
that the observed solar values generally were taken to be consistent with the terrestrial ratios
(based on the Vienna Standard Mean Ocean Water [V-SMOW] mixture for the oxygen isotopes
and Vienna Peedee Belemnite [V-PDB] for carbon: Gonfiantini, Stichler, \& Rozanski 1995;
$^{12}$C/$^{13}$C=\,$89.2{\pm}0.2$ [ibid., Table~3], 
$^{16}$O/$^{17}$O=\,$2632{\pm}5$ [ibid., Table~1], and 
$^{16}$O/$^{18}$O=\,$498.7{\pm}0.1$ [ibid., Table~1]).  
Lunar and Martian isotopic ratios are very close to Earth's; while
those inferred for asteroids from
meteoritic analyses cover a wider range, but still deviate from terrestrial
by less than 20\,$^{\circ}\!\!/\!_{\circ\circ}$ (Burnett et al.\ 2003).  Some
extraterrestrial materials show large isotopic
deficiencies, reaching $-$60\,$^{\circ}\!\!/\!_{\circ\circ}$ for the oxygen isotopes in 
calcium-aluminum inclusions of
chondritic meteorites in extreme cases (Wiens et al.\ 2004).  

Theoretical predictions of solar isotopic ratios---thought to be essentially identical to those
of the original solar nebula---based on fractionation processes in
the gas-dust chemistry were only a few~$^{\circ}\!\!/\!_{\circ\circ}$ larger than the terrestrial values,
necessitating the very high precision measurements intended for {\em Genesis,} as mentioned above. 
Now, with the much smaller
uncertainties of the present work, we see that the solar ratios deviate 
from the terrestrial values by much larger factors than
anticipated by the previous theoretical models:
$\delta{^{13}\mbox{C}}\sim +115{\pm}14$\,$^{\circ}\!\!/\!_{\circ\circ}$, 
$\delta{^{17}\mbox{O}}\sim +550{\pm}200$\,$^{\circ}\!\!/\!_{\circ\circ}$,
and $\delta{^{18}\mbox{O}}\sim +133{\pm}16$\,$^{\circ}\!\!/\!_{\circ\circ}$.  It seems that 
isotopic fractionation
theories for primitive solar system material incorporated into the inner planets
are in need of some revision.

The carbon isotope $^{13}$C is a secondary product of nuclear processing, through the CNO cycle, requiring a seed 
abundance of $^{12}$C from previous generations of primary production (helium burning); thus 
$^{12}$C/$^{13}$C is expected generally to decrease as the Galaxy ages, and 
display a distinct gradient with Galactocentric distance (Wilson \& Rood 1994, hereafter WR94).
The solar $^{12}$C/$^{13}$C ratio, however, now is closer to, although
still larger than, those seen in nearby interstellar clouds 
($62{\pm}4$: Langer \& Penzias 1993; $77{\pm}7$: WR94; $59{\pm}2$: Lucas \& Liszt 1998) 
and star-forming regions like Orion (40--70: Savage et al.\ [2002]).  This might modify the notion that
substantial enrichment of $^{13}$C has occurred in the Galaxy over the past 4.6~Gyr since the
formation of the solar system (Savage et al.\ 2002).

\section{Discussion}

Through a multifaceted analysis of CO infrared 
rovibrational bands, we have investigated the thermal structure of the
solar photosphere, and as a byproduct derived abundances of C, O, and their main
isotopes.  The $\epsilon_{\rm O}$, in particular, is
similar to  
values recommended a decade ago, compatible with helioseismological
constraints, but is significantly larger than reported in recent studies of OH and \ion{O}{1}
exploiting sophisticated
simulations of solar surface convection.  
We now can understand at least the molecular side of
the dichotomy straightforwardly: as was demonstrated
in \S{3.3.1}, the temperature gradient of the mean Asplund model is too steep in the visible
continuum forming layers.  Consequently,
the model is somewhat too cool in the higher layers
where the infrared OH lines arise.  This
leads to the inference of a smaller oxygen abundance from the hydroxyl spectrum (and from CO, as well).  

On the other hand, the behavior of the \ion{O}{1} 0.630~\micron\ forbidden 
line---main oxygen diagnostic
in the previous work---cannot be explained similarly.  
The transition arises from the ground state of a majority species, and thus
should be relatively immune to thermal effects (see L78).  
Perhaps the 0.630~\micron\ $gf$-value is not as securely established
as believed, or the \ion{Ni}{1} blend is less significant than thought.  One certainly
should be wary, in any case, when an important abundance is tied so strongly to
essentially a singular spectral feature.

At the same time, one should ask what is the proper role of {\em ab initio}\/
models such as that of Asplund and colleagues, versus semiempirical thermal
stratifications such as derived by E.\ H.\
Avrett and collaborators?  Clearly, an {\em ab initio}\/ model is intended to provide direct insight
into the fundamental physical processes that govern, say, the structure and dynamics of the solar
photosphere.  The Asplund model meets this
requirement by accurately reproducing the details of velocity induced lineshape distortions (encoded in the
so-called C-shaped line bisectors), and general characteristics of 
the time dependent brightness pattern
of granular convection cells.  Yet, the model is not a perfect representation of the solar
atmosphere because it does not predict other key properties, such
as the chromospheric temperature rise (or, the temperature {\em fall}\/ of the COmosphere), or
magnetic-related phenomena such as G-band bright points and \ion{Ca}{2} network elements.  This,
of course, is not the point of the Asplund model, and it should not be faulted for
failing tests outside of its intended purview.  Nevertheless, because the model is in some
respects not a flawless representation of the solar atmosphere, it should be used with caution
in roles outside of its defined domain.  Deriving solar abundances would be a good example of
an area where it would be prudent to exercise such caution.

Semiempirical models, on the other hand, provide little in the way of direct physical 
insight, in and of themselves; but can be used to calculate ancillary
quantities (such as chromospheric radiative losses) which
are of interpretive value (VAL81).  A semiempirical model, nevertheless, to the extent that it
successfully captures mean thermal properties of the atmosphere, could be considered more
satisfactory for abundance work.  Deficiencies of 1-D models relative
to, say, a fully time dependent spatially resolved convection simulation probably are not
as important as accurately mimicking the mean thermal profile, as far as most abundance
studies are concerned; and the influence of thermal heterogeneity can be investigated to some
extent by imposing 1.5-D thermal perturbations around the basis model, as
in the present work.  
Again, the Asplund et al.\ study, itself, demonstrated that
the thermal inhomogeneities of the deep photosphere have a relatively minor influence on the
oxygen abundance, at least as derived from the \ion{O}{1} spectrum.

Nevertheless, it would be foolish for us to claim a definitive solution to the oxygen crisis.
Deriving $\epsilon_{\rm O}$ from CO clearly is highly model dependent, and there are enough 
inconsistencies for the 
optimum COmosphere model---with regard to the
center-limb behavior of weak CO lines like 7--6~R68, and the small but significant 
discrepancies between
abundances derived from $\Delta{v}=1$ versus $\Delta{v}=2$ lines---that
we view our contrary result mainly as an indication that a low solar oxygen abundance is not
fully consistent with all the available diagnostics, and certainly merits 
further consideration.  One important issue
is the C/O ratio.  Increasing it above the assumed 0.5 would allow a decrease of the
inferred oxygen abundance, although to achieve the low proposed $\epsilon_{\rm O}$ would require
C/O\,$\gtrsim 1$.  Elevating the Sun to the status of a ``carbon star'' certainly
would be more controversial than the low O problem itself.

With regard to the apparent variations of the theoretical rovibrational
oscillator strengths with
lower state excitation energy, which we have treated as a systematic error and
corrected:  the viability of our proposal could be
pursued in refined calculations of the dipole matrix elements, and new laboratory measurements
of high temperature CO.  However, it also is possible that the consistent behavior of the predicted CO
equivalent widths with excitation energy for the Asplund class of models, which---coincidentally or 
not---yields 
a low O abundance of $\sim$500~ppm, is not accidental, but rather is conveying a fundamental
message concerning our ability to constrain the thermal properties of the middle
photosphere empirically.  The Double Dip model is one answer to that possibility, but its physical
reality remains to be tested by independent observations, such as of the
\ion{Ca}{2} wings (AL76), or other potential mid-photosphere thermal tracers.  

Finally, while atmospheric thermal profiling and abundance studies might seem less fashionable than
searches for exosolar planets or the elusive dark matter (or dark energy, for that
matter), the 
fundamental importance of such 
basic quantities across diverse areas of
astrophysics should not be questioned. 

\acknowledgments 

This work was supported by grant AST-9987414 from the National 
Science Foundation and NAG5-13058 from the 
National Aeronautics and Space Administration.  
We thank T.\ Woods for providing an electronic version of
the Thuillier et al.\ ATLAS-3 solar irradiance reference spectrum; R.~L.~Kurucz
for providing an advance copy of his high resolution
solar irradiance atlas; and R.\ L.\ Kurucz and J.\ L.\ Linsky
for helpful comments.  More detailed
parameters for the thermal models described in this study are available from
the first author by request.

%%%%%%%%%%%%%%%%%%%%%%%%%%%%%%%%%%%%%%%%%%%%%%%%%%%%%%%%%%%%%%%%%%%%%%%%%%%%%%%%
%\clearpage

\clearpage
\begin{deluxetable}{ccc}
%%%%\rotate
%%%%\tabletypesize{\small}
\tablenum{1}
\tablecaption{Absolute Continuum Intensities}
\tablewidth{0pt}
\tablecolumns{3}
\tablehead{
\colhead{$\lambda$} &  \colhead{~~~~${\cal F}_{\lambda}/{\pi}\,I_{\lambda}$~~~~} & \colhead{$I_{\lambda}$} \\[3pt]                
\colhead{(\micron)} & \colhead{} & \colhead{($10^{6}$ ergs cm$^{-2}$ s$^{-1}$ sr$^{-1}$ \AA$^{-1}$)}  
}                
\startdata   
%===============================================================================
   0.400  &      0.728  &      4.501 \\[3pt]
   0.420  &      0.743  &      4.710 \\[3pt]
   0.440  &      0.756  &      4.687 \\[3pt]
   0.460  &      0.767  &      4.536 \\[3pt]
   0.480  &      0.777  &      4.330 \\[3pt]
   0.500  &      0.785  &      4.110 \\[3pt]
   0.520  &      0.793  &      3.903 \\[3pt]
   0.540  &      0.800  &      3.718 \\[3pt]
   0.560  &      0.806  &      3.555 \\[3pt]
   0.580  &      0.812  &      3.406 \\[3pt]
   0.600  &      0.818  &      3.263 \\[3pt]
   0.620  &      0.824  &      3.118 \\[3pt]
   0.640  &      0.829  &      2.967 \\[3pt]
   0.660  &      0.834  &      2.817 \\[3pt]
   0.680  &      0.839  &      2.682 \\[3pt]
   0.700  &      0.843  &      2.597 \\[3pt]
%-------------------------------------------------------------------%
\enddata
\tablecomments{
The 1\,$\sigma$ standard deviation on the polynomial fit
to the calibrated highpoints was $\pm 0.9$\,\%.  This represents
an estimate of the limiting precision due to
uncorrelated systematic (and random) errors.  The absolute accuracy of the intensity
scale, as normalized to the TSI, is thought to be better than 0.1\,\%.  }
\end{deluxetable}

\clearpage
\begin{deluxetable}{ccrcrc}
%%%%\rotate
%%%%\tabletypesize{\small}
\tablenum{2}
\tablecaption{The COmosphere Model}
\tablewidth{0pt}
\tablecolumns{6}
\tablehead{
\colhead{$\log{m}$} & \colhead{$\log{\tau_{0.5}}$} & \colhead{$z$} & \colhead{~~~~$T$~~~~} & 
 \colhead{~~$\Delta{T}_{\rm rms}$} & \colhead{~~$T_{\rm NW}$} \\[3pt]                
\colhead{(gr cm$^{-2}$)} & \colhead{} & \colhead{(km)} & \colhead{(K)} & \colhead{(K)} & 
\colhead{(K)}  
}                
\startdata   
%===============================================================================
  $-4.00$   &   $-5.07$   &   $-1280$   &   $6625$   &   $100$   &   $6853$  \\[3pt]
  $-3.80$   &   $-4.88$   &   $-1204$   &   $6475$   &   $100$   &   $6702$  \\[3pt]
  $-3.60$   &   $-4.75$   &   $-1130$   &   $6332$   &   $100$   &   $6557$  \\[3pt]
  $-3.40$   &   $-4.64$   &   $-1058$   &   $6107$   &   $100$   &   $6426$  \\[3pt]
  $-3.20$   &   $-4.60$   &   $-995$   &   $4857$   &   $100$   &   $6299$  \\[3pt]
  $-3.00$   &   $-4.59$   &   $-946$   &   $3844$   &   $100$   &   $6179$  \\[3pt]
  $-2.80$   &   $-4.58$   &   $-903$   &   $3543$   &   $99$   &   $6058$  \\[3pt]
  $-2.60$   &   $-4.55$   &   $-862$   &   $3500$   &   $96$   &   $5926$  \\[3pt]
  $-2.40$   &   $-4.51$   &   $-821$   &   $3610$   &   $92$   &   $5760$  \\[3pt]
  $-2.20$   &   $-4.43$   &   $-778$   &   $4019$   &   $86$   &   $5516$  \\[3pt]
  $-2.00$   &   $-4.32$   &   $-728$   &   $4399$   &   $80$   &   $5228$  \\[3pt]
  $-1.80$   &   $-4.17$   &   $-677$   &   $4450$   &   $72$   &   $4949$  \\[3pt]
  $-1.60$   &   $-3.95$   &   $-626$   &   $4473$   &   $64$   &   $4744$  \\[3pt]
  $-1.40$   &   $-3.68$   &   $-575$   &   $4497$   &   $55$   &   $4613$  \\[3pt]
  $-1.20$   &   $-3.36$   &   $-523$   &   $4520$   &   $47$   &   $4592$  \\[3pt]
  $-1.00$   &   $-3.03$   &   $-471$   &   $4548$   &   $40$   &   $4627$  \\[3pt]
  $-0.90$   &   $-2.86$   &   $-444$   &   $4580$   &   $37$   &   $4662$  \\[3pt]
  $-0.80$   &   $-2.69$   &   $-418$   &   $4621$   &   $34$   &   $4705$  \\[3pt]
  $-0.70$   &   $-2.51$   &   $-391$   &   $4660$   &   $32$   &   $4744$  \\[3pt]
  $-0.60$   &   $-2.34$   &   $-364$   &   $4700$   &   $29$   &   $4785$  \\[3pt]
  $-0.50$   &   $-2.16$   &   $-337$   &   $4744$   &   $28$   &   $4831$  \\[3pt]
  $-0.40$   &   $-1.99$   &   $-310$   &   $4791$   &   $26$   &   $4879$  \\[3pt]
  $-0.30$   &   $-1.81$   &   $-282$   &   $4838$   &   $25$   &   $4926$  \\[3pt]
  $-0.20$   &   $-1.64$   &   $-255$   &   $4887$   &   $24$   &   $4975$  \\[3pt]
  $-0.10$   &   $-1.46$   &   $-226$   &   $4935$   &   $24$   &   $5021$  \\[3pt]
  $+0.00$   &   $-1.28$   &   $-198$   &   $4977$   &   $24$   &   $5077$  \\[3pt]
  $+0.05$   &   $-1.19$   &   $-184$   &   $5002$   &   $24$   &   $5111$  \\[3pt]
  $+0.10$   &   $-1.10$   &   $-169$   &   $5050$   &   $24$   &   $5150$  \\[3pt]
  $+0.15$   &   $-1.02$   &   $-155$   &   $5112$   &   $24$   &   $5199$  \\[3pt]
  $+0.20$   &   $-0.93$   &   $-140$   &   $5180$   &   $24$   &   $5258$  \\[3pt]
  $+0.25$   &   $-0.84$   &   $-125$   &   $5260$   &   $24$   &   $5327$  \\[3pt]
  $+0.30$   &   $-0.75$   &   $-110$   &   $5350$   &   $23$   &   $5402$  \\[3pt]
  $+0.35$   &   $-0.65$   &   $-95$   &   $5454$   &   $24$   &   $5490$  \\[3pt]
  $+0.40$   &   $-0.56$   &   $-79$   &   $5572$   &   $26$   &   $5591$  \\[3pt]
  $+0.45$   &   $-0.46$   &   $-63$   &   $5705$   &   $37$   &   $5702$  \\[3pt]
  $+0.50$   &   $-0.35$   &   $-46$   &   $5841$   &   $55$   &   $5859$  \\[3pt]
  $+0.55$   &   $-0.24$   &   $-30$   &   $6006$   &   $108$   &   $6070$  \\[3pt]
  $+0.60$   &   $-0.10$   &   $-12$   &   $6224$   &   $215$   &   $6319$  \\[3pt]
  $+0.65$   &   $+0.06$   &   $+6$   &   $6566$   &   $387$   &   $6650$  \\[3pt]
  $+0.70$   &   $+0.29$   &   $+25$   &   $7185$   &   $634$   &   $7146$  \\[3pt]
  $+0.75$   &   $+0.62$   &   $+47$   &   $7970$   &   $790$   &   $7798$  \\[3pt]
  $+0.80$   &   $+1.00$   &   $+71$   &   $8606$   &   $772$   &   $8530$  \\[3pt]
  $+0.85$   &   $+1.33$   &   $+96$   &   $8955$   &   $646$   &   $9279$  \\[3pt]
%-------------------------------------------------------------------%
\enddata
%%%%%\tablecomments{}
\end{deluxetable}

\clearpage
\begin{deluxetable}{rcrcc}
%%%%\rotate
%%%%\tabletypesize{\small}
\tablenum{3}
\tablecaption{$^{12}$C$^{16}$O FWHMs and Equivalent Widths for $\Delta{v}=1$,\,2 Bands}
\tablewidth{0pt}
\tablecolumns{5}
\tablehead{
\colhead{Transition} &  \colhead{$\omega_0$} & \colhead{$E_{\rm lower}$} & 
 \colhead{FWHM} & \colhead{$W_{\omega}$} \\[3pt]                
\colhead{} & \colhead{(cm$^{-1}$)} & \colhead{(cm$^{-1}$)} & \colhead{(km~s$^{-1}$)} & \colhead{($10^{-3}$ cm$^{-1}$)}  
}                
\startdata   
%===============================================================================
  1--0~P~96 &  1635.859 &   17376 & $4.35{\pm}0.03$ & $2.060{\pm}0.014$ \\[3pt]
  1--0~P~90 &  1674.640 &   15337 & $4.51{\pm}0.01$ & $3.203{\pm}0.009$ \\[3pt]
  1--0~R106 &  2322.725 &   21025 & $4.19{\pm}0.02$ & $2.353{\pm}0.013$ \\[3pt]
  1--0~R109 &  2320.247 &   22180 & $4.20{\pm}0.02$ & $1.815{\pm}0.007$ \\[3pt]
  2--1~R~~0 &  2120.566 &    2143 & $4.73{\pm}0.01$ & $6.905{\pm}0.010$ \\[3pt]
  2--1~R112 &  2286.814 &   25283 & $4.29{\pm}0.02$ & $1.440{\pm}0.008$ \\[3pt]
  2--1~R118 &  2279.315 &   27701 & $3.96{\pm}0.06$ & $0.723{\pm}0.012$ \\[3pt]
  3--2~P~88 &  1640.640 &   18670 & $4.53{\pm}0.02$ & $3.062{\pm}0.014$ \\[3pt]
  3--2~R108 &  2260.499 &   25640 & $4.23{\pm}0.02$ & $1.735{\pm}0.008$ \\[3pt]
  4--3~P~80 &  1666.616 &   18212 & $4.64{\pm}0.01$ & $3.800{\pm}0.012$ \\[3pt]
  4--3~P~~2 &  2056.515 &    6361 & $4.76{\pm}0.01$ & $6.468{\pm}0.015$ \\[3pt]
  4--3~R118 &  2218.067 &   31415 & $4.19{\pm}0.06$ & $0.444{\pm}0.007$ \\[3pt]
  5--4~P~75 &  1672.976 &   18775 & $4.65{\pm}0.01$ & $3.748{\pm}0.010$ \\[3pt]
  5--4~P~~3 &  2026.533 &    8436 & $4.69{\pm}0.02$ & $5.979{\pm}0.027$ \\[3pt]
  5--4~R~~0 &  2041.423 &    8414 & $4.30{\pm}0.01$ & $3.120{\pm}0.011$ \\[3pt]
  5--4~R105 &  2202.920 &   28281 & $4.10{\pm}0.02$ & $1.150{\pm}0.006$ \\[3pt]
  5--4~R110 &  2198.043 &   30129 & $4.18{\pm}0.04$ & $0.718{\pm}0.007$ \\[3pt]
  6--5~P~72 &  1666.903 &   19928 & $4.56{\pm}0.01$ & $3.314{\pm}0.009$ \\[3pt]
  6--5~R107 &  2171.091 &   30847 & $4.10{\pm}0.09$ & $0.637{\pm}0.015$ \\[3pt]
  7--6~P~~5 &  1966.891 &   12518 & $4.51{\pm}0.01$ & $4.310{\pm}0.007$ \\[3pt]
  7--6~P~~4 &  1970.664 &   12500 & $4.40{\pm}0.01$ & $3.788{\pm}0.011$ \\[3pt]
  7--6~P~~1 &  1981.778 &   12467 & $4.12{\pm}0.03$ & $1.346{\pm}0.011$ \\[3pt]
  7--6~R~68 &  2143.690 &   20857 & $4.60{\pm}0.01$ & $5.237{\pm}0.010$ \\[3pt]
  7--6~R101 &  2146.363 &   30543 & $3.94{\pm}0.07$ & $0.653{\pm}0.012$ \\[3pt]
  8--7~P~71 &  1625.476 &   23491 & $4.40{\pm}0.03$ & $1.828{\pm}0.011$ \\[3pt]
  8--7~P~66 &  1653.678 &   22289 & $4.41{\pm}0.02$ & $2.321{\pm}0.010$ \\[3pt]
  8--7~P~15 &  1901.762 &   14880 & $4.78{\pm}0.01$ & $5.302{\pm}0.011$ \\[3pt]
  8--7~P~~8 &  1929.586 &   14578 & $4.48{\pm}0.01$ & $3.991{\pm}0.013$ \\[3pt]
  8--7~R~~5 &  1980.214 &   14503 & $4.39{\pm}0.01$ & $3.600{\pm}0.013$ \\[3pt]
  8--7~R~~6 &  1983.566 &   14524 & $4.44{\pm}0.01$ & $3.992{\pm}0.013$ \\[3pt]
  8--7~R~61 &  2108.392 &   21169 & $4.59{\pm}0.01$ & $4.971{\pm}0.007$ \\[3pt]
  9--8~P~22 &  1847.119 &   17308 & $4.59{\pm}0.01$ & $4.192{\pm}0.010$ \\[3pt]
  9--8~R~~9 &  1967.073 &   16568 & $4.40{\pm}0.02$ & $3.451{\pm}0.015$ \\[3pt]
  9--8~R~23 &  2008.100 &   17390 & $4.72{\pm}0.01$ & $5.382{\pm}0.011$ \\[3pt]
  9--8~R~66 &  2085.165 &   24171 & $4.37{\pm}0.01$ & $3.005{\pm}0.009$ \\[3pt]
  9--8~R~75 &  2091.185 &   26370 & $4.19{\pm}0.03$ & $1.933{\pm}0.016$ \\[3pt]
  10--9~P~66 &  1606.423 &   26027 & $4.25{\pm}0.05$ & $1.035{\pm}0.012$ \\[3pt]
  10--9~P~30 &  1787.007 &   19978 & $4.50{\pm}0.01$ & $3.085{\pm}0.008$ \\[3pt]
  10--9~P~3 &  1896.769 &   18363 & $4.11{\pm}0.04$ & $0.834{\pm}0.008$ \\[3pt]
  10--9~R~38 &  2016.404 &   20944 & $4.45{\pm}0.02$ & $3.909{\pm}0.014$ \\[3pt]
  2--0~P~41 &  4046.692 &    3292 & $4.14{\pm}0.02$ & $3.006{\pm}0.012$ \\[3pt]
  2--0~P~25 &  4143.315 &    1247 & $4.46{\pm}0.01$ & $4.323{\pm}0.007$ \\[3pt]
  2--0~P~~9 &  4222.954 &     172 & $4.10{\pm}0.03$ & $2.297{\pm}0.020$ \\[3pt]
  2--0~P~~1 &  4256.217 &       3 & $4.00{\pm}0.13$ & $0.306{\pm}0.010$ \\[3pt]
  2--0~R~~2 &  4271.177 &      11 & $3.99{\pm}0.16$ & $0.845{\pm}0.036$ \\[3pt]
  2--0~R~20 &  4324.410 &     806 & $4.27{\pm}0.01$ & $5.224{\pm}0.014$ \\[3pt]
  2--0~R~81 &  4323.688 &   12501 & $4.52{\pm}0.06$ & $0.919{\pm}0.013$ \\[3pt]
  3--1~P~60 &  3859.911 &    9033 & $4.09{\pm}0.03$ & $1.959{\pm}0.015$ \\[3pt]
  3--1~P~38 &  4014.514 &    4953 & $4.23{\pm}0.01$ & $4.821{\pm}0.009$ \\[3pt]
  3--1~P~19 &  4122.974 &    2866 & $4.33{\pm}0.01$ & $6.858{\pm}0.010$ \\[3pt]
  3--1~P~~2 &  4199.478 &    2154 & $4.32{\pm}0.05$ & $1.048{\pm}0.012$ \\[3pt]
  3--1~R~20 &  4270.782 &    2942 & $4.27{\pm}0.01$ & $7.278{\pm}0.016$ \\[3pt]
  3--1~R~59 &  4302.391 &    8810 & $4.17{\pm}0.02$ & $4.609{\pm}0.020$ \\[3pt]
  3--1~R~65 &  4296.835 &   10203 & $4.28{\pm}0.01$ & $3.857{\pm}0.009$ \\[3pt]
  3--1~R~80 &  4270.324 &   14232 & $3.87{\pm}0.05$ & $1.322{\pm}0.016$ \\[3pt]
  3--1~R~93 &  4232.443 &   18332 & $4.78{\pm}0.08$ & $0.619{\pm}0.011$ \\[3pt]
  4--2~P~76 &  3678.606 &   15097 & $4.70{\pm}0.08$ & $0.828{\pm}0.015$ \\[3pt]
  4--2~P~58 &  3824.488 &   10647 & $4.17{\pm}0.03$ & $2.313{\pm}0.015$ \\[3pt]
  4--2~P~13 &  4099.925 &    4603 & $4.20{\pm}0.01$ & $4.397{\pm}0.012$ \\[3pt]
  4--2~P~~1 &  4150.632 &    4263 & $3.66{\pm}0.11$ & $0.403{\pm}0.013$ \\[3pt]
  4--2~R~36 &  4243.636 &    6763 & $4.36{\pm}0.00$ & $8.403{\pm}0.007$ \\[3pt]
  4--2~R~61 &  4245.925 &   11311 & $4.26{\pm}0.02$ & $4.578{\pm}0.021$ \\[3pt]
  4--2~R~63 &  4244.002 &   11771 & $4.27{\pm}0.01$ & $4.154{\pm}0.014$ \\[3pt]
  4--2~R~72 &  4231.403 &   14012 & $4.20{\pm}0.02$ & $2.584{\pm}0.010$ \\[3pt]
  4--2~R~83 &  4207.105 &   17124 & $4.15{\pm}0.05$ & $1.274{\pm}0.017$ \\[3pt]
  4--2~R~95 &  4169.200 &   20969 & $4.12{\pm}0.12$ & $0.495{\pm}0.016$ \\[3pt]
  5--3~P~73 &  3654.208 &   16274 & $4.23{\pm}0.06$ & $0.844{\pm}0.013$ \\[3pt]
  5--3~P~~1 &  4098.043 &    6354 & $4.19{\pm}0.12$ & $0.413{\pm}0.012$ \\[3pt]
  5--3~R~~3 &  4116.042 &    6372 & $4.30{\pm}0.03$ & $1.664{\pm}0.011$ \\[3pt]
  5--3~R~~5 &  4122.749 &    6406 & $4.17{\pm}0.03$ & $2.318{\pm}0.018$ \\[3pt]
  5--3~R~~6 &  4125.995 &    6428 & $4.19{\pm}0.01$ & $2.741{\pm}0.009$ \\[3pt]
  5--3~R~68 &  4182.763 &   14990 & $4.25{\pm}0.01$ & $2.920{\pm}0.010$ \\[3pt]
  5--3~R~74 &  4172.463 &   16541 & $4.30{\pm}0.04$ & $2.092{\pm}0.018$ \\[3pt]
  6--4~R~~3 &  4063.418 &    8436 & $4.05{\pm}0.06$ & $1.193{\pm}0.019$ \\[3pt]
  6--4~R~~4 &  4066.772 &    8451 & $4.31{\pm}0.04$ & $1.595{\pm}0.015$ \\[3pt]
  6--4~R~~7 &  4076.408 &    8518 & $4.32{\pm}0.02$ & $2.639{\pm}0.011$ \\[3pt]
  6--4~R~75 &  4115.314 &   18775 & $4.23{\pm}0.03$ & $1.605{\pm}0.012$ \\[3pt]
  7--5~R~20 &  4057.639 &   11221 & $4.25{\pm}0.01$ & $4.052{\pm}0.013$ \\[3pt]
  7--5~R~76 &  4058.158 &   20982 & $4.19{\pm}0.08$ & $1.116{\pm}0.024$ \\[3pt]
  8--6~R~70 &  4015.183 &   21346 & $4.19{\pm}0.03$ & $1.221{\pm}0.008$ \\[3pt]
%-------------------------------------------------------------------%
\enddata
\tablecomments{~Line center frequencies, $\omega_{0}$, and lower level energies, $E_{\rm lower}$,
are from Goorvitch (1994).  Uncertainties on the Gaussian fit parameters (full width at half maximum
absorption, FWHM, and the equivalent width, $W_{\omega}$) 
are $\pm$1\,$\sigma$, based on Lenz \& Ayres (1992); 
the photometric noise was estimated by the
dispersion of intensities about the empirically fitted Gaussian lineshape.}
\end{deluxetable}

\clearpage
\begin{deluxetable}{lccccccccc}
\vskip 5mm
\rotate
\tabletypesize{\footnotesize}
\tablenum{4}
\tablecaption{Model- and Parameter-Dependent Oxygen Abundances}
\tablewidth{0pt}
\tablecolumns{10}
\tablehead{
\colhead{Model} &  \colhead{Components} & \colhead{Cont} &\colhead{CO} &\colhead{CO} & 
 \multicolumn{3}{c}{$(\epsilon_{\rm O})_{0}~{\pm}~\sigma~~(\Delta)$} & \multicolumn{2}{c}{$\langle\epsilon_{\rm O}\rangle~{\pm}~\sigma$}  \\[3pt]                
\colhead{} &          \colhead{} &        \colhead{c--l}         &\colhead{$\mu=1$}          &\colhead{c--l} &
 \colhead{$\Delta{v}=1$,\,2} &  \colhead{$\Delta{v}=1$} &  \colhead{$\Delta{v}=2$} & 
\colhead{$\Delta{v}=1$} &  \colhead{$\Delta{v}=2$}  \\[3pt]                
\colhead{(1)} & \colhead{(2)} & \colhead{(3)} & \colhead{(4)} & \colhead{(5)} & \colhead{(6)} & 
\colhead{(7)} & \colhead{(8)} & \colhead{(9)} & \colhead{(10)}
 } 
\startdata   
%===============================================================================
FAL\,C & 1 & 4 & 3 & 0 & $820{\pm}16$~($-5.6$)  &  $809{\pm}15$~($-4.9$)  &  
$824{\pm}14$~($-6.2$)  &  $731{\pm}37$  & $780{\pm}44$  \\[3pt] 
Asplund & 1 & 1 & 5 & 0 & $575{\pm}13$~($-2.9$)  &  $566{\pm}14$~($-2.3$)  &  
$579{\pm}11$~($-3.1$)  &  $540{\pm}18$  & $565{\pm}24$  \\[3pt] 
Asplund/HR96~$gf$ & 1 & 1 & 5 & 0 & $585{\pm}16$~($-4.2$)  &  
$558{\pm}14$~($-2.4$)  &  $594{\pm}10$~($-4.0$)  &  
$531{\pm}18$  & $575{\pm}26$  \\[3pt] 
Asplund*5 & 5\,$\Delta{T}$ & 1 & 5 & 0 & $496{\pm}12$~($+1.7$)  &  
$530{\pm}9$~($-1.1$)  &  $490{\pm}11$~($+1.5$)  &  
$520{\pm}18$  & $499{\pm}19$  \\[3pt] 
COmosphere & 1 & 4 & 5 & 4 & $846{\pm}18$~($-5.5$)  &  $871{\pm}14$~($-6.4$)  &  
$848{\pm}15$~($-6.6$)  &  $762{\pm}43$  & $798{\pm}47$  \\[3pt] 
COmosphere/HR96~$gf$ & 1 & 4 & 5 & 4 & $864{\pm}14$~($-6.8$)  &  
$858{\pm}14$~($-6.4$)  &  $870{\pm}15$~($-7.5$)  &  
$749{\pm}42$  & $812{\pm}52$  \\[3pt] 
COmosphere*6 & 5\,$\Delta{T}$+1\,NW & 4 & 5 & 4 & $851{\pm}20$~($-5.1$)  &  
$901{\pm}13$~($-6.9$)  &  $853{\pm}13$~($-6.8$)  &  
$777{\pm}47$  & $803{\pm}48$  \\[3pt] 
COmosphere $D_0-0.10$ & 1 & 4 & 5 & 4 & $939{\pm}19$~($-5.4$)  &  
$966{\pm}15$~($-6.2$)  &  $939{\pm}15$~($-6.5$)  &  
$846{\pm}47$  & $887{\pm}51$  \\[3pt] 
COmosphere $D_0+0.10$ & 1 & 4 & 5 & 4 & $764{\pm}16$~($-5.5$)  &  
$787{\pm}13$~($-6.5$)  &  $764{\pm}12$~($-6.7$)  &  
$686{\pm}39$  & $720{\pm}43$  \\[3pt] 
COmosphere $D_0+0.64$ & 1 & 4 & 5 & 4 & $451{\pm}10$~($-5.9$)  &  
$465{\pm}8$~($-6.8$)  &  $453{\pm}8$~($-7.4$)  &  
$403{\pm}26$  & $423{\pm}28$  \\[3pt] 
COmosphere H$^{-}~f-f\,\div{1.1}$ & 1 & 4 & 5 & 4 & $831{\pm}15$~($-6.1$)  &  
$837{\pm}15$~($-6.2$)  &  $833{\pm}15$~($-6.7$)  &  
$734{\pm}40$  & $784{\pm}46$  \\[3pt] 
COmosphere H$^{-}~f-f\,\times{1.1}$ & 1 & 4 & 5 & 4 & $860{\pm}21$~($-4.7$)  &  
$913{\pm}13$~($-6.7$)  &  $862{\pm}13$~($-6.5$)  &  
$791{\pm}47$  & $813{\pm}47$  \\[3pt] 
COmosphere H$^{-}\,\div{1.1}$ & 1 & 4 & 5 & 4 & $802{\pm}17$~($-5.5$)  &  
$825{\pm}13$~($-6.3$)  &  $804{\pm}14$~($-6.6$)  &  
$722{\pm}40$  & $757{\pm}44$  \\[3pt] 
COmosphere H$^{-}\,\times{1.1}$ & 1 & 4  & 5  & 4  & $893{\pm}19$~($-5.5$)  &  
$920{\pm}14$~($-6.4$)  &  $892{\pm}14$~($-6.6$)  &  
$804{\pm}45$  & $842{\pm}49$  \\[3pt] 
COmosphere $\xi-0.5$ & 1 & 4 & 5 & 4 & $848{\pm}20$~($-5.1$)  &  
$903{\pm}13$~($-7.1$)  &  $848{\pm}13$~($-6.7$)  &  
$777{\pm}51$  & $799{\pm}47$  \\[3pt] 
COmosphere $\xi+0.5$ & 1 & 4 & 5 & 4 & $837{\pm}19$~($-5.7$)  &  
$816{\pm}19$~($-4.7$)  &  $845{\pm}15$~($-6.4$)  &  
$741{\pm}35$  & $797{\pm}46$  \\[3pt] 
COmosphere $T_{\rm min}-200$ & 1 & 4 & 4 & 2 & $840{\pm}16$~($-5.1$)  &  
$858{\pm}12$~($-5.8$)  &  $840{\pm}13$~($-6.2$)  &  
$759{\pm}39$  & $795{\pm}45$  \\[3pt] 
COmosphere $T_{\rm min}+200$ & 1 & 4 & 4 & 2 & $852{\pm}18$~($-5.6$)  &  
$878{\pm}14$~($-6.5$)  &  $855{\pm}15$~($-7.0$)  &  
$766{\pm}47$  & $802{\pm}49$  \\[3pt] 
COmosphere $T_{\rm outer}-200$ & 1 & 4 & 3 & 4 & $742{\pm}19$~($-4.1$)  &  
$709{\pm}17$~($-2.4$)  &  $751{\pm}12$~($-4.1$)  &  
$676{\pm}23$  & $727{\pm}33$  \\[3pt] 
COmosphere $T_{\rm outer}+200$ & 1 & 4 & 3 & 4 & $928{\pm}31$~($-6.0$)  &  
$1036{\pm}20$~($-9.4$)  &  $928{\pm}16$~($-8.3$)  &  
$845{\pm}70$  & $860{\pm}59$  \\[3pt] 
COmosphere--200 & 1 & 1 & 3 & 4 & $472{\pm}14$~($-0.7$)  &  $451{\pm}13$~($+1.1$)  &  
$479{\pm}9$~($-0.4$)  &  $460{\pm}15$  & $479{\pm}17$  \\[3pt] 
Double Dip & 1 & 5 & 5 & 5 & $654{\pm}26$~($-3.4$)  &  $599{\pm}22$~($-0.1$)  &  
$663{\pm}12$~($-2.1$)  &  $598{\pm}22$  & $653{\pm}26$  \\[3pt] 
Double Dip*6 & 5\,$\Delta{T}$+1\,NW & 5 & 5 & 5 & $685{\pm}24$~($-3.6$)  &  
$641{\pm}21$~($-1.1$)  &  $693{\pm}13$~($-2.5$)  &  
$627{\pm}21$  & $680{\pm}27$  \\[3pt] 
%-------------------------------------------------------------------%
\enddata
\tablecomments{Columns 3
(continuum center-limb behavior), 4 (CO $\Delta{v}=1$ core depths at disk center),
and 5 (CO $\Delta{v}=1$ center-limb behavior)
grade the success of the model in reproducing the particular observational
constraint, on a scale of 0 (poor) to 5 (very good).  In columns 6--8 the cited uncertainties are standard
deviations about the linear fit to
the $\epsilon_{\rm O}$--$E_{\rm lower}$ relation, with mild filtering to discard
outliers.  In columns 9 and 10, the uncertainties are the 1\,$\sigma$ dispersions in each
40-member abundance sample, with no filtering and not accounting for a possible correlation
between $\epsilon_{\rm O}$ and $E_{\rm lower}$.
}
\end{deluxetable}

\clearpage
\begin{deluxetable}{ccrc}
%%%%\rotate
%%%%\tabletypesize{\small}
\tablenum{5}
\tablecaption{The Double Dip Model}
\tablewidth{0pt}
\tablecolumns{4}
\tablehead{
\colhead{$\log{m}$} & \colhead{$\log{\tau_{0.5}}$} & \colhead{$z$} & \colhead{~~~~$T$~~~~} \\[3pt]                
\colhead{(gr cm$^{-2}$)} & \colhead{} & \colhead{(km)} & \colhead{(K)}  
}                
\startdata   
%===============================================================================
  $-4.00$   &   $-5.14$   &   $-1264$   &   $6527$    \\[3pt]
  $-3.80$   &   $-4.94$   &   $-1189$   &   $6377$    \\[3pt]
  $-3.60$   &   $-4.81$   &   $-1116$   &   $6233$    \\[3pt]
  $-3.40$   &   $-4.71$   &   $-1046$   &   $6015$    \\[3pt]
  $-3.20$   &   $-4.67$   &   $-983$   &   $4776$    \\[3pt]
  $-3.00$   &   $-4.66$   &   $-935$   &   $3750$    \\[3pt]
  $-2.80$   &   $-4.64$   &   $-894$   &   $3445$    \\[3pt]
  $-2.60$   &   $-4.61$   &   $-854$   &   $3400$    \\[3pt]
  $-2.40$   &   $-4.56$   &   $-814$   &   $3531$    \\[3pt]
  $-2.20$   &   $-4.47$   &   $-771$   &   $3967$    \\[3pt]
  $-2.00$   &   $-4.36$   &   $-722$   &   $4385$    \\[3pt]
  $-1.80$   &   $-4.19$   &   $-671$   &   $4450$    \\[3pt]
  $-1.60$   &   $-3.97$   &   $-620$   &   $4473$    \\[3pt]
  $-1.40$   &   $-3.68$   &   $-568$   &   $4496$    \\[3pt]
  $-1.20$   &   $-3.37$   &   $-517$   &   $4520$    \\[3pt]
  $-1.00$   &   $-3.03$   &   $-464$   &   $4546$    \\[3pt]
  $-0.90$   &   $-2.86$   &   $-438$   &   $4571$    \\[3pt]
  $-0.80$   &   $-2.69$   &   $-412$   &   $4584$    \\[3pt]
  $-0.70$   &   $-2.51$   &   $-386$   &   $4560$    \\[3pt]
  $-0.60$   &   $-2.33$   &   $-360$   &   $4514$    \\[3pt]
  $-0.50$   &   $-2.15$   &   $-334$   &   $4499$    \\[3pt]
  $-0.40$   &   $-1.98$   &   $-308$   &   $4557$    \\[3pt]
  $-0.30$   &   $-1.80$   &   $-281$   &   $4678$    \\[3pt]
  $-0.20$   &   $-1.63$   &   $-254$   &   $4810$    \\[3pt]
  $-0.10$   &   $-1.45$   &   $-226$   &   $4909$    \\[3pt]
  $0.00$   &   $-1.28$   &   $-198$   &   $4984$    \\[3pt]
  $+0.05$   &   $-1.19$   &   $-184$   &   $5024$    \\[3pt]
  $+0.10$   &   $-1.10$   &   $-169$   &   $5074$    \\[3pt]
  $+0.15$   &   $-1.01$   &   $-155$   &   $5135$    \\[3pt]
  $+0.20$   &   $-0.92$   &   $-140$   &   $5203$    \\[3pt]
  $+0.25$   &   $-0.83$   &   $-125$   &   $5282$    \\[3pt]
  $+0.30$   &   $-0.74$   &   $-110$   &   $5372$    \\[3pt]
  $+0.35$   &   $-0.65$   &   $-94$   &   $5477$    \\[3pt]
  $+0.40$   &   $-0.56$   &   $-78$   &   $5595$    \\[3pt]
  $+0.45$   &   $-0.46$   &   $-62$   &   $5728$    \\[3pt]
  $+0.50$   &   $-0.35$   &   $-46$   &   $5864$    \\[3pt]
  $+0.55$   &   $-0.23$   &   $-29$   &   $6029$    \\[3pt]
  $+0.60$   &   $-0.10$   &   $-11$   &   $6242$    \\[3pt]
  $+0.65$   &   $+0.06$   &   $+7$   &   $6580$    \\[3pt]
  $+0.70$   &   $+0.29$   &   $+26$   &   $7186$    \\[3pt]
  $+0.75$   &   $+0.62$   &   $+48$   &   $7973$    \\[3pt]
  $+0.80$   &   $+1.00$   &   $+72$   &   $8633$    \\[3pt]
  $+0.85$   &   $+1.33$   &   $+97$   &   $8991$    \\[3pt]

%-------------------------------------------------------------------%
\enddata
%%%\tablecomments{}
\end{deluxetable}

\clearpage
\begin{deluxetable}{rccrcc}
%%%%\rotate
%%%%\tabletypesize{\small}
\tablenum{6}
\tablecaption{CO Isotopomer Equivalent Widths}
\tablewidth{0pt}
\tablecolumns{6}
\tablehead{
\colhead{Transition} & \colhead{Isotopomer} & \colhead{$\omega_0$} & \colhead{$E_{\rm lower}$} & 
 \colhead{FWHM} & \colhead{$W_{\omega}$} \\[3pt]                
\colhead{} & \colhead{} & \colhead{(cm$^{-1}$)} & \colhead{(cm$^{-1}$)} & \colhead{(km~s$^{-1}$)} & \colhead{($10^{-3}$ cm$^{-1}$)}  
}                
\startdata   
%===============================================================================
  1--0~P65 & $^{13}$C$^{16}$O &  1795.191 &    7782 & $4.05$ & $0.435{\pm}0.011$ \\[3pt]
  1--0~P62 &   &  1811.598 &    7094 & $4.05$ & $0.540{\pm}0.013$ \\[3pt]
  1--0~P10 &   &  2057.857 &     202 & $4.05$ & $1.165{\pm}0.004$ \\[3pt]
  1--0~R14 &   &  2147.205 &     385 & $4.05$ & $1.809{\pm}0.011$ \\[3pt]
  1--0~R42 &   &  2221.409 &    3301 & $4.05$ & $2.057{\pm}0.012$ \\[3pt]
  1--0~R64 &   &  2258.716 &    7549 & $4.05$ & $0.979{\pm}0.011$ \\[3pt]
  1--0~R76 &   &  2270.724 &   10565 & $4.05$ & $0.518{\pm}0.016$ \\[3pt]
  1--0~R81 &   &  2273.917 &   11962 & $4.05$ & $0.365{\pm}0.012$ \\[3pt]
  2--1~P52 &   &  1841.059 &    7074 & $4.05$ & $0.960{\pm}0.009$ \\[3pt]
  2--1~P30 &   &  1947.828 &    3785 & $4.05$ & $1.669{\pm}0.014$ \\[3pt]
  2--1~P12 &   &  2024.913 &    2380 & $4.05$ & $1.212{\pm}0.016$ \\[3pt]
  2--1~R13 &   &  2118.261 &    2427 & $4.05$ & $1.632{\pm}0.014$ \\[3pt]
  2--1~R18 &   &  2133.604 &    2718 & $4.05$ & $2.047{\pm}0.010$ \\[3pt]
  2--1~R82 &   &  2246.386 &   14236 & $4.05$ & $0.314{\pm}0.007$ \\[3pt]
  2--1~R83 &   &  2246.817 &   14526 & $4.05$ & $0.331{\pm}0.007$ \\[3pt]
  3--2~P67 &   &  1737.948 &   12275 & $4.05$ & $0.346{\pm}0.013$ \\[3pt]
  3--2~R~5 &   &  2066.470 &    4220 & $4.05$ & $0.604{\pm}0.011$ \\[3pt]
  3--2~R21 &   &  2116.416 &    4999 & $4.05$ & $1.653{\pm}0.010$ \\[3pt]
  3--2~R36 &   &  2154.960 &    6561 & $4.05$ & $1.710{\pm}0.011$ \\[3pt]
  3--2~R37 &   &  2157.233 &    6693 & $4.05$ & $1.764{\pm}0.009$ \\[3pt]
  3--2~R81 &   &  2217.967 &   15911 & $4.05$ & $0.278{\pm}0.013$ \\[3pt]
  4--3~P17 &   &  1955.148 &    6759 & $4.05$ & $0.815{\pm}0.009$ \\[3pt]
  4--3~P~7 &   &  1994.585 &    6312 & $4.05$ & $0.429{\pm}0.010$ \\[3pt]
  4--3~P~6 &   &  1998.356 &    6287 & $4.05$ & $0.384{\pm}0.011$ \\[3pt]
  4--3~R36 &   &  2128.561 &    8585 & $4.05$ & $1.226{\pm}0.007$ \\[3pt]
  4--3~R37 &   &  2130.802 &    8716 & $4.05$ & $1.244{\pm}0.006$ \\[3pt]
  4--3~R43 &   &  2143.447 &    9576 & $4.05$ & $1.105{\pm}0.016$ \\[3pt]
  5--4~P57 &   &  1745.031 &   14031 & $4.05$ & $0.272{\pm}0.012$ \\[3pt]
  6--5~P45 &   &  1781.720 &   13839 & $4.05$ & $0.301{\pm}0.010$ \\[3pt]
  6--5~P38 &   &  1814.887 &   12818 & $4.05$ & $0.325{\pm}0.016$ \\[3pt]
  6--5~P31 &   &  1846.698 &   11964 & $4.05$ & $0.372{\pm}0.011$ \\[3pt]
  6--5~R11 &   &  2009.683 &   10459 & $4.05$ & $0.339{\pm}0.008$ \\[3pt]
  6--5~R27 &   &  2054.702 &   11552 & $4.05$ & $0.530{\pm}0.007$ \\[3pt]
  6--5~R35 &   &  2073.746 &   12431 & $4.05$ & $0.517{\pm}0.009$ \\[3pt]
  7--6~R20 & $^{13}$C$^{16}$O   &  2010.445 &   12927 & $4.05$ & $0.292{\pm}0.007$ \\[3pt]
\hline\\
  1--0~P13 & $^{12}$C$^{17}$O  &  2064.996 &     340 & $4.05$ & $0.111{\pm}0.012$ \\[3pt]
  1--0~R22 &   &  2192.917 &     946 & $4.05$ & $0.174{\pm}0.010$ \\[3pt]
  1--0~R33 &   &  2222.770 &    2095 & $4.05$ & $0.090{\pm}0.017$ \\[3pt]
  2--1~P34 &   &  1946.217 &    4318 & $4.05$ & $0.089{\pm}0.008$ \\[3pt]
  2--1~R37 &   &  2205.393 &    4715 & $4.05$ & $0.078{\pm}0.010$ \\[3pt]
  3--2~R39 &   &  2182.856 &    7063 & $4.05$ & $0.081{\pm}0.012$ \\[3pt]
  3--2~R45 & $^{12}$C$^{17}$O   &  2195.376 &    7991 & $4.05$ & $0.102{\pm}0.006$ \\[3pt]
\hline\\
  1--0~P57 & $^{12}$C$^{18}$O  &  1835.568 &    5992 & $4.05$ & $0.191{\pm}0.008$ \\[3pt]
  1--0~P43 &   &  1907.045 &    3444 & $4.05$ & $0.282{\pm}0.009$ \\[3pt]
  1--0~P35 &   &  1945.548 &    2298 & $4.05$ & $0.330{\pm}0.008$ \\[3pt]
  1--0~P27 &   &  1982.267 &    1381 & $4.05$ & $0.372{\pm}0.007$ \\[3pt]
  1--0~P18 &   &  2021.359 &     625 & $4.05$ & $0.303{\pm}0.013$ \\[3pt]
  1--0~P11 &   &  2050.081 &     241 & $4.05$ & $0.222{\pm}0.009$ \\[3pt]
  1--0~R11 &   &  2133.490 &     241 & $4.05$ & $0.299{\pm}0.006$ \\[3pt]
  1--0~R26 &   &  2178.262 &    1282 & $4.05$ & $0.500{\pm}0.010$ \\[3pt]
  1--0~R48 &   &  2229.106 &    4275 & $4.05$ & $0.345{\pm}0.010$ \\[3pt]
  1--0~R49 &   &  2230.979 &    4452 & $4.05$ & $0.345{\pm}0.009$ \\[3pt]
  1--0~R50 &   &  2232.813 &    4632 & $4.05$ & $0.322{\pm}0.015$ \\[3pt]
  1--0~R63 &   &  2253.037 &    7292 & $4.05$ & $0.196{\pm}0.008$ \\[3pt]
  2--1~P43 &   &  1883.226 &    5505 & $4.05$ & $0.244{\pm}0.014$ \\[3pt]
  2--1~P31 &   &  1939.926 &    3886 & $4.05$ & $0.348{\pm}0.010$ \\[3pt]
  2--1~P25 &   &  1966.756 &    3269 & $4.05$ & $0.341{\pm}0.009$ \\[3pt]
  2--1~P17 &   &  2000.888 &    2646 & $4.05$ & $0.322{\pm}0.007$ \\[3pt]
  3--2~R~7 &   &  2069.339 &    4259 & $4.05$ & $0.126{\pm}0.006$ \\[3pt]
  3--2~R18 &   &  2103.755 &    4773 & $4.05$ & $0.279{\pm}0.009$ \\[3pt]
  3--2~R29 &   &  2133.930 &    5719 & $4.05$ & $0.317{\pm}0.010$ \\[3pt]
  3--2~R37 &   &  2153.111 &    6676 & $4.05$ & $0.356{\pm}0.008$ \\[3pt]
  4--3~P42 &   &  1840.585 &    9401 & $4.05$ & $0.140{\pm}0.008$ \\[3pt]
  4--3~P23 &   &  1926.712 &    7182 & $4.05$ & $0.184{\pm}0.009$ \\[3pt]
  4--3~R10 &   &  2053.683 &    6396 & $4.05$ & $0.131{\pm}0.008$ \\[3pt]
  4--3~R26 &   &  2100.159 &    7449 & $4.05$ & $0.212{\pm}0.010$ \\[3pt]
  5--4~P31 &   &  1867.672 &    9963 & $4.05$ & $0.132{\pm}0.009$ \\[3pt]
  5--4~R16 & $^{12}$C$^{18}$O   &  2046.578 &    8697 & $4.05$ & $0.099{\pm}0.006$ \\[3pt]
%-------------------------------------------------------------------%
\enddata
\tablecomments{~Line center frequencies, $\omega_{0}$, and lower level energies, $E_{\rm lower}$,
are from Goorvitch (1994).  Uncertainties on the Gaussian fit parameters
(for the equivalent width $W_{\omega}$, only; $\omega$ and FWHM were fixed) 
are $\pm$1\,$\sigma$, based on Lenz \& Ayres (1992), with empirical estimates of the photometric noise.}
\end{deluxetable}

\clearpage
\begin{deluxetable}{lcccccc}
\rotate
\tabletypesize{\small}
\tablenum{7}
\tablecaption{Model-Dependent Carbon and Oxygen Isotopic Ratios}
\tablewidth{0pt}
\tablecolumns{7}
\tablehead{
\colhead{Model} &  \multicolumn{3}{c}{$R~{\pm}~\sigma$~(s.e.)~[$\epsilon_{\rm O}(\Delta{v}=1$,\,2)]} & 
 \multicolumn{3}{c}{$R~{\pm}~\sigma$~(s.e.)~[$\epsilon_{\rm O}(\Delta{v}=1$)]}\\[3pt] 
\colhead{} &     \colhead{$^{12}$C/$^{13}$C}  &  \colhead{$^{16}$O/$^{17}$O}  &   
\colhead{$^{16}$O/$^{18}$O}   &  \colhead{$^{12}$C/$^{13}$C}  &  \colhead{$^{16}$O/$^{17}$O}  &   
\colhead{$^{16}$O/$^{18}$O} }                
\startdata   
%===============================================================================
FAL\,C &  $81.4{\pm}3.9$~(0.7)    & $1806{\pm}620$~(234)  & $451{\pm}32$~(7)  & $80.7{\pm}3.8$~(0.6)    & $1775{\pm}609$~(230)  & $440{\pm}28$~(6)  \\[3pt] 
Asplund  &  $81.8{\pm}3.9$~(0.7)    & $1810{\pm}621$~(235)  & $453{\pm}31$~(7)  & $80.7{\pm}3.8$~(0.6)    & $1771{\pm}606$~(229)  & $443{\pm}31$~(7)  \\[3pt] 
Asplund*5 &  $89.0{\pm}4.2$~(0.7)    & $2022{\pm}684$~(258)  & $506{\pm}35$~(7)  & $94.7{\pm}3.3$~(0.6)    & $2182{\pm}728$~(275)  & $546{\pm}36$~(7)  \\[3pt] 
COmosphere  &  $80.8{\pm}3.8$~(0.7)    & $1790{\pm}613$~(232)  & $448{\pm}31$~(6)  & $83.9{\pm}3.7$~(0.6)    & $1898{\pm}649$~(245)  & $467{\pm}32$~(7)  \\[3pt] 
COmosphere*6 &  [{$80.4{\pm}3.6$~(0.6)}]    & [{$1726{\pm}592$~(224)}]  & [{$441{\pm}28$~(6)}]  & $86.3{\pm}3.4$~(0.6)    & $1923{\pm}651$~(246)  & $482{\pm}33$~(7)  \\[3pt] 
COmosphere--200  &  $83.3{\pm}3.3$~(0.6)    & $1777{\pm}605$~(229)  & $458{\pm}31$~(6)  & $78.8{\pm}4.0$~(0.7)    & $1680{\pm}569$~(215)  & $429{\pm}27$~(6)  \\[3pt] 
Double Dip   &  $85.4{\pm}3.0$~(0.5)    & $1901{\pm}645$~(244)  & $473{\pm}30$~(6)  & $77.0{\pm}4.3$~(0.7)    & $1667{\pm}568$~(215)  & $421{\pm}29$~(6)  \\[3pt] 
Double Dip*6 & $85.5{\pm}2.9$~(0.5)    & $1868{\pm}626$~(236)  & $475{\pm}30$~(6)  & [{$78.9{\pm}4.2$~(0.7)}]    & [{$1682{\pm}566$~(214)}]  & [{$436{\pm}29$~(6)}]  \\[3pt] 
%-------------------------------------------------------------------%
\enddata
\tablecomments{
Favored isotopic ratios are enclosed in square brackets.  Values in parentheses are standard errors of the mean.}
\end{deluxetable}

\clearpage
\begin{figure}
\figurenum{1}
\epsscale{1.0}
\plotone{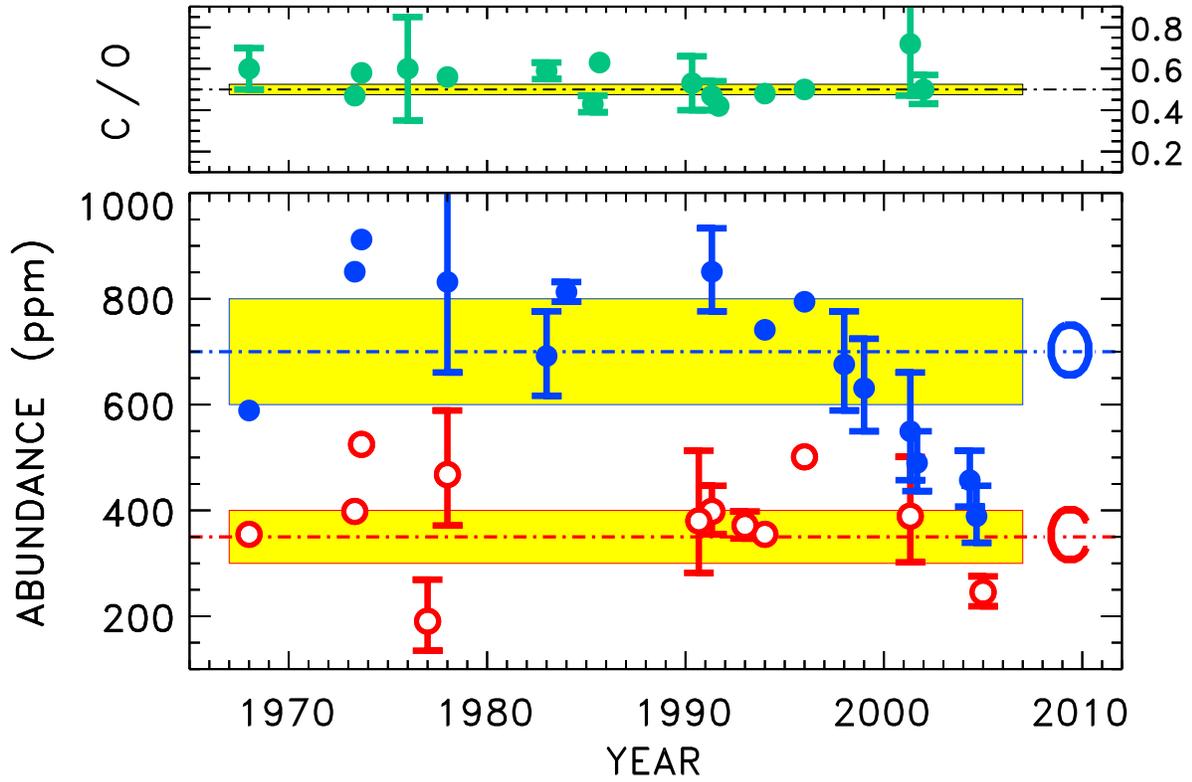}
\caption[]{Solar abundances of oxygen (solid blue dots, lower panel) and carbon 
(red circles), and C/O ratio
(green dots, upper panel), over time.  Within past few
years, reported solar $\epsilon_{\rm O}$ has fallen precipitously to historical
level of the carbon abundance.  Although recent measurements of solar carbon are fewer,
it too has kept pace with drop in solar oxygen, maintaining a ratio of about
0.5\,.  At recent rate of decline, Sun will run out of oxygen in around 2015.
(Shaded bands indicate values recommended in present work.)}
\end{figure}

\clearpage
\begin{figure}
\figurenum{2}
\epsscale{1.0}
\plotone{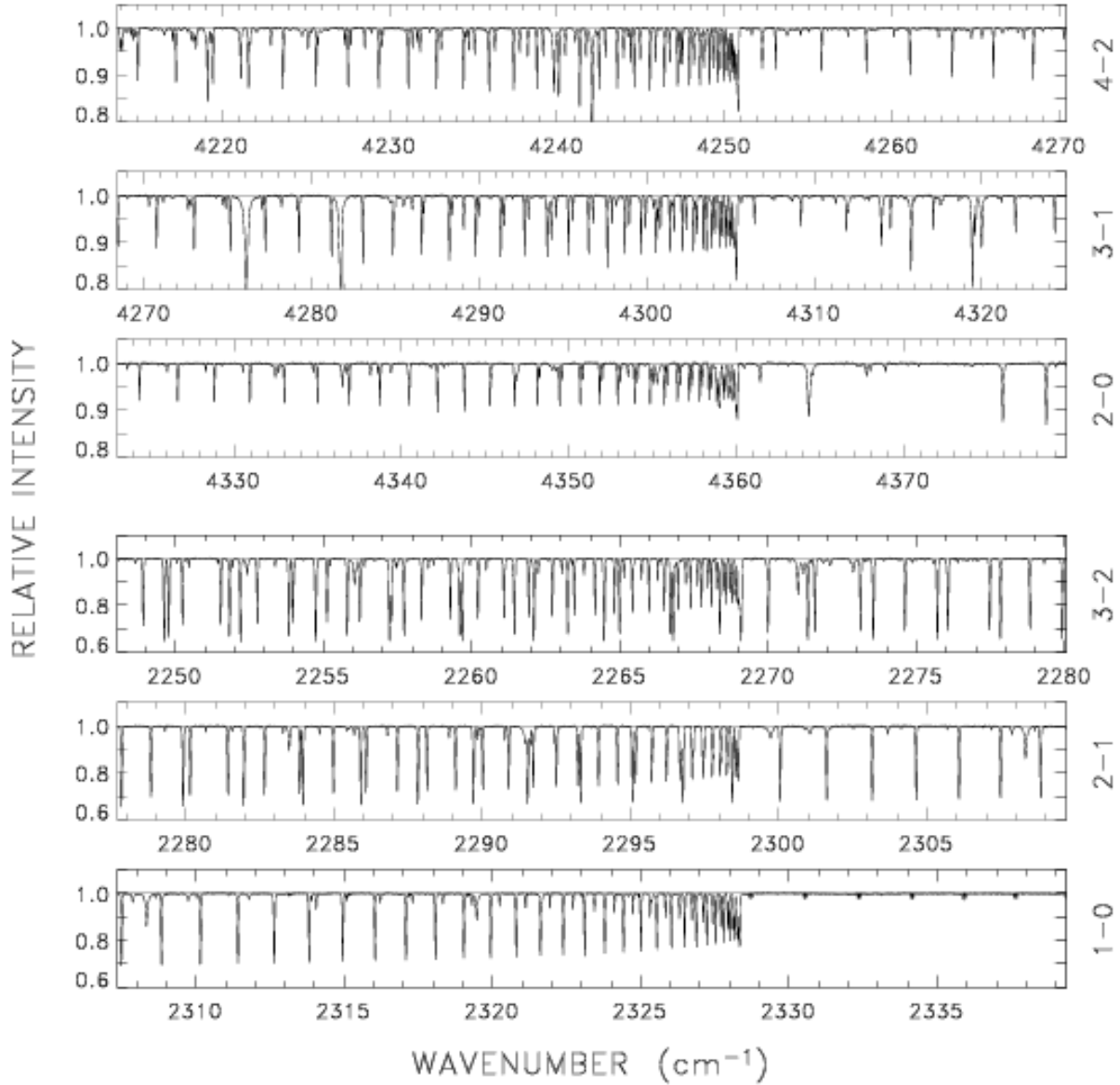}
\caption[]{Representative intervals from ATMOS/ATLAS-3 solar infrared spectrum.  Upper
three panels depict heads of first overtone ($\Delta{v}= 2$)
2--0, 3--1, and 4--2 bands;
lower three panels depict heads of fundamental ($\Delta{v}= 1$) 
1--0, 2--1, and 3--2 bands.  Most absorption features in these
tracings are solar CO.  (However, faint evenly spaced dips longward of
1--0 bandhead are due to residual water vapor in the instrument shroud, which
was unable to fully outgas owing to a protective window.) }
\end{figure}

\clearpage
\begin{figure}
\figurenum{3}
\caption[]{Representative transitions from 1--0 R-branch 
($\Delta{v}=+1$, $\Delta{J}=+1$) of CO fundamental 
from the ATMOS spectra (dots).  Simulated line profiles (red curves) were synthesized with
an ``optimum'' thermal profile, to be described later.  Note graduated relative intensity ordinates:  
values are scaled linearly between major tick marks.  First interval above, and below, 
unit level is first
$\pm$1\,\%, emphasizing continuum and (photometric) scatter
of points around it.  Full range extends to 40\,\% below unit line to show details of
stronger features.  Red ticks indicate line center ($\omega_{0}$, noted below each panel),
and major ticks on abscissa are 0.02~cm$^{-1}$ apart.  Thick blue vertical ticks indicate
$^{12}$C$^{16}$O blends; orange for blends with
CO isotopomers, which---aside from a few moderate-strength
$^{13}$C$^{16}$O transitions---usually
are quite weak.  Lowest excitation 1--0 lines ($J\le 10$) are strongly
saturated in Earth's atmosphere, and must be observed from space. }
\end{figure}

\clearpage
\begin{figure}
%\figurenum{3}
\hskip -7mm
%\vskip -10mm
\includegraphics[scale=0.90]{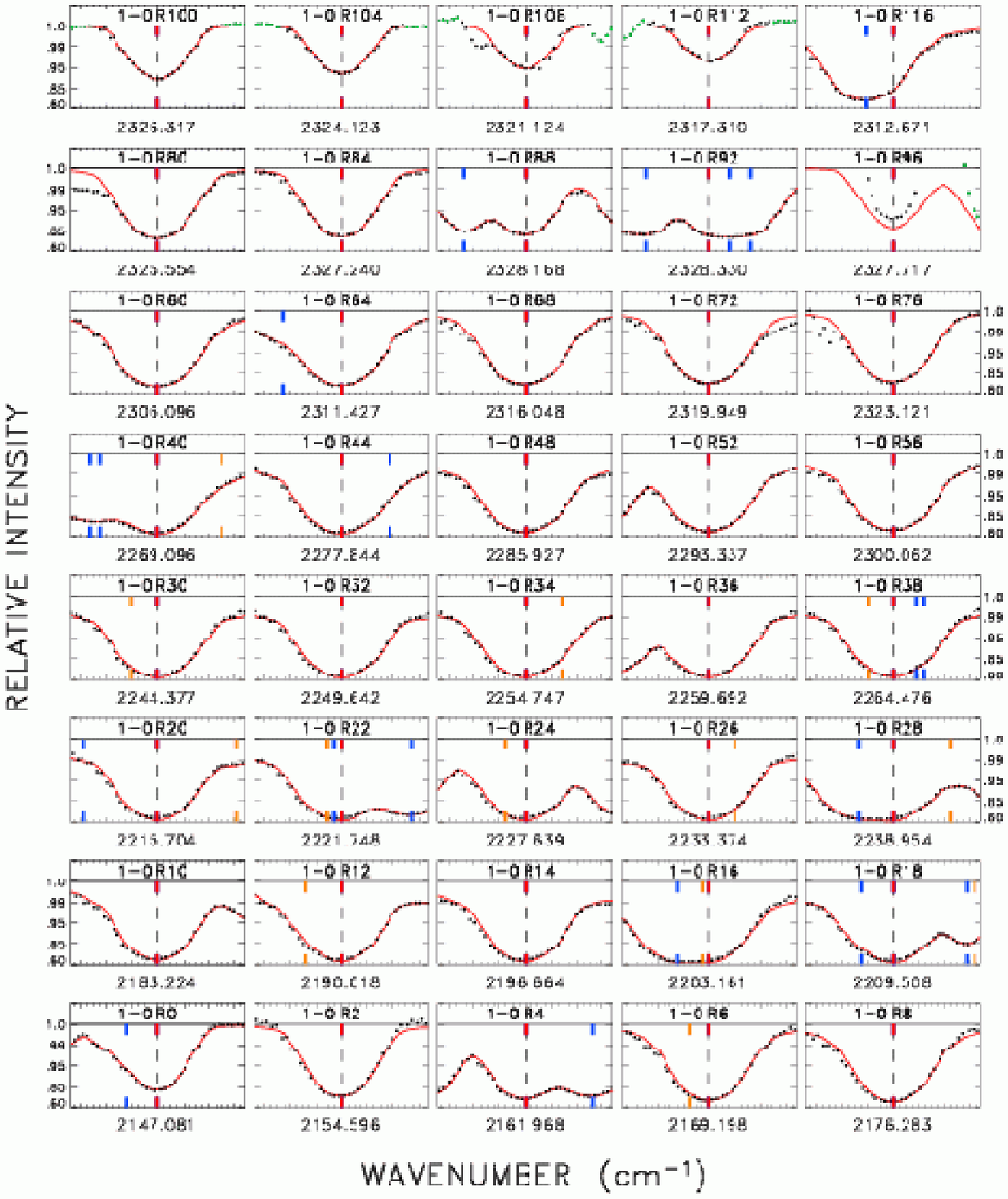} % fig_2.eps}
\centerline{Fig.~3}
%\caption[]{}
\end{figure}

\clearpage
\begin{figure}
\figurenum{4}
%%%\epsscale{0.5}
\hskip 5mm
\includegraphics[scale=0.675,angle=90]{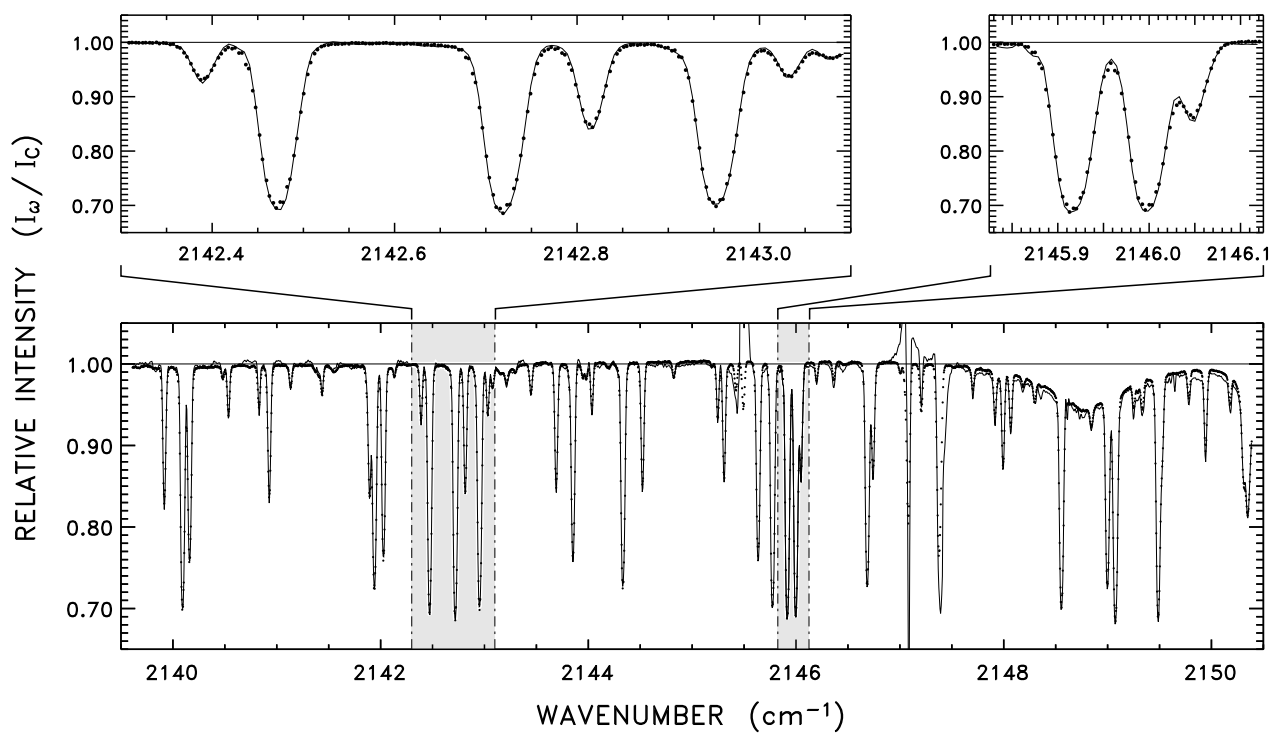}
\vskip 5mm
\caption[]{Comparison of disk center McMath-Pierce FTS scan (solid curve,
corrected for terrestrial absorptions) and
ATMOS/ATLAS-3 traces (dots) for 2145~cm$^{-1}$ (4.66~\micron) region favored for
groundbased observations of solar CO, owing to low contamination
by telluric absorptions.  Broad shallow dip at 2149~cm$^{-1}$
is \ion{H}{1} Pf$\beta$, but otherwise all other narrow
absorptions are solar CO.  In a few places (e.g., 2147~cm$^{-1}$), the groundbased FTS
scan was affected by a large telluric absorption correction and is unreliable.  ATMOS
spectrum was obtained above Earth's atmosphere and is largely
free of such contamination.  
Upper panels are expanded views of two regions.  Careful examination finds that 
Shuttle-borne FTS scans are very slightly lower in resolution than groundbased counterparts,
although ATMOS resolution still is extremely high ($R\sim 2\times10^5$), fully resolving
the narrow CO absorptions. }
\end{figure}

\clearpage
\begin{figure}
\vskip -5mm
\figurenum{5}
\epsscale{0.875}
\plotone{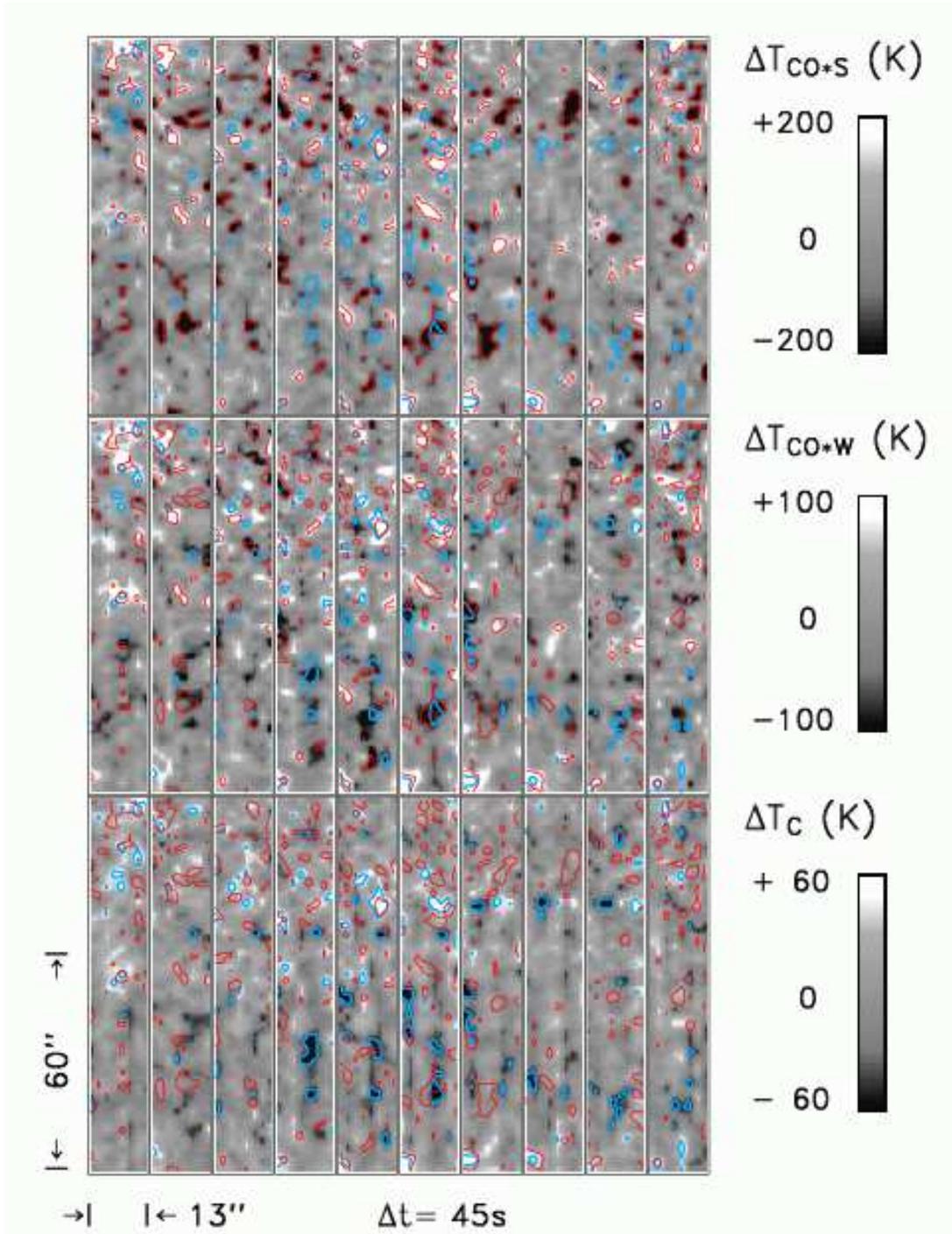}
\caption[]{Time sequences of thermal maps
recorded at disk center: sum of three strong CO lines (upper), sum of two weak CO
lines (middle), and 2145~cm$^{-1}$ (4.66~\micron) continuum (lower).  Cadence was
45~s per 13\arcsec$\times$97\arcsec\ field of view.  Color bars on 
right indicate range of temperature fluctuations displayed
in specific row.  Contours
outline patches of highest and lowest temperature fluctuations for the strong CO lines (red)
and continuum (blue), and are repeated in each panel.}
\end{figure}

\clearpage
\begin{figure}
\figurenum{6}
%%%\epsscale{0.6}
\hskip 5mm
\includegraphics[scale=1.00,angle=90]{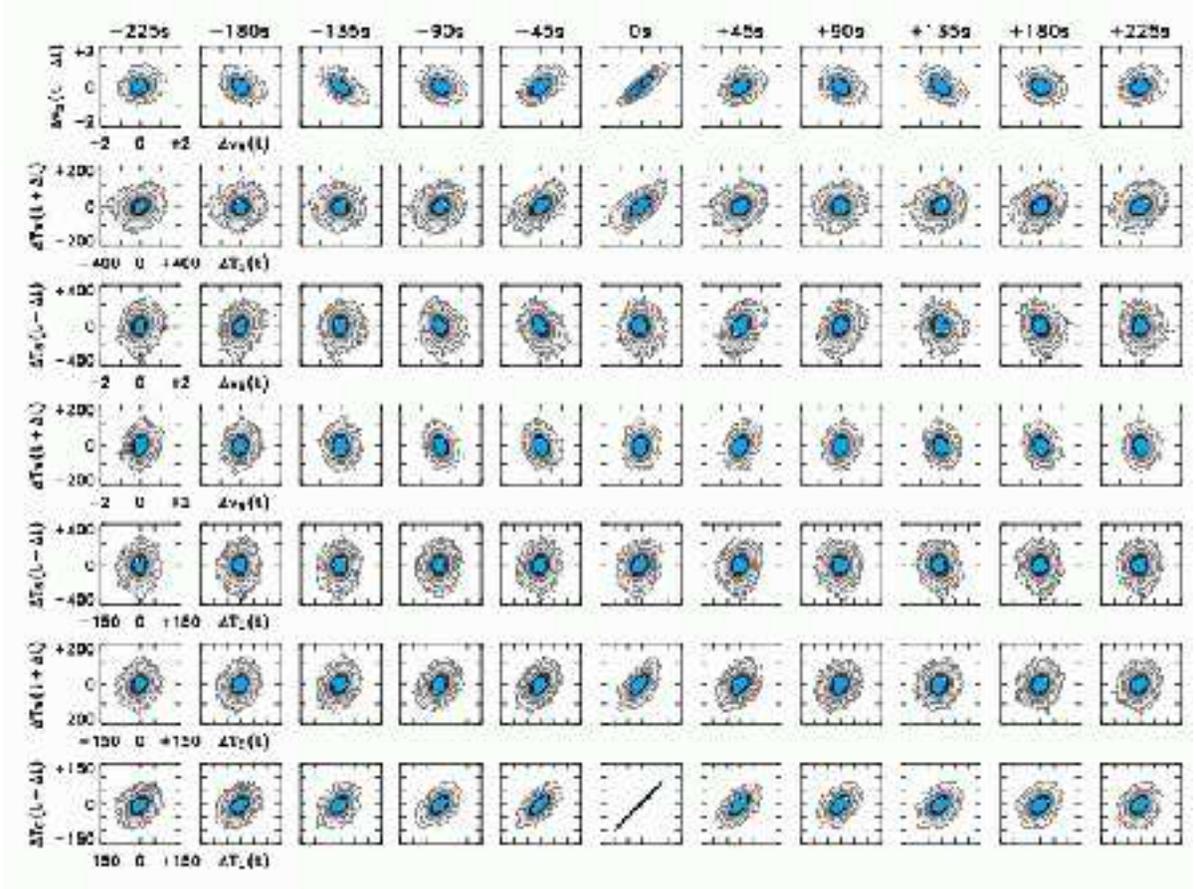}  
\caption[]{Cross-correlation diagrams comparing CO and 2145~cm$^{-1}$ continuum parameters
for range of temporal lags.  Subscript ``S'' indicates sum of three strong
CO lines; ``W'' is for sum of two weaker features.  Correlations
are broadly consistent with domination of temperature and velocity fields
by photospheric $p$-modes.  Temperature fluctuation rms values are: 22~K (continuum);
42~K (weak CO line set), 83~K (strong CO line set).  Doppler rms fluctuations are
270~m~s$^{-1}$ (weak CO) and 314~m~s$^{-1}$ (strong CO).}
\end{figure}

\clearpage
\begin{figure}
\figurenum{7}
%%%\epsscale{1.0}
%%%\plotone{f7.eps}  
\caption[]{Schematic illustration of steps to derive calibrated disk center
continuum intensities.  Upper panel depicts conversion factor between
disk average and disk center intensity (former is surface flux density ${\cal F}_{\lambda}$
divided by $\pi$).  Smaller green dots in lower portion of panel are deviations of
individual measurements from polynomial fit: horizontal dot-dashed lines are
$\pm$1\,\%. 
In middle panel, lighter curve (orange: left hand scale) is high dispersion residual flux spectrum
smoothed to resolution of Thuillier et al.\ (2004) irradiances, while darker curve is
Thuillier tracing adjusted from irradiance to surface flux (red: right hand scale).  
Larger dots are ratio of Thuillier to smoothed residual
flux spectrum in places where smoothed trace exceeds 98\,\% of running maximum.  
Curve through these points is a polynomial fit,
utilizing a 2\,$\sigma$ filter to eliminate outliers (dark points), representing
100\,\% continuum flux level.  Bottom panel shows result of dividing continuum
flux level by ${\cal F}_{\lambda}/I_{\lambda}$ ($\pi$ times upper panel) to obtain
absolute specific intensities (thin solid curve), then multiplying by original
residual flux spectrum (last step is purely for illustration).}
\end{figure}

\clearpage
%\begin{figure}
%\figurenum{7}
\epsscale{1.00}
\plotone{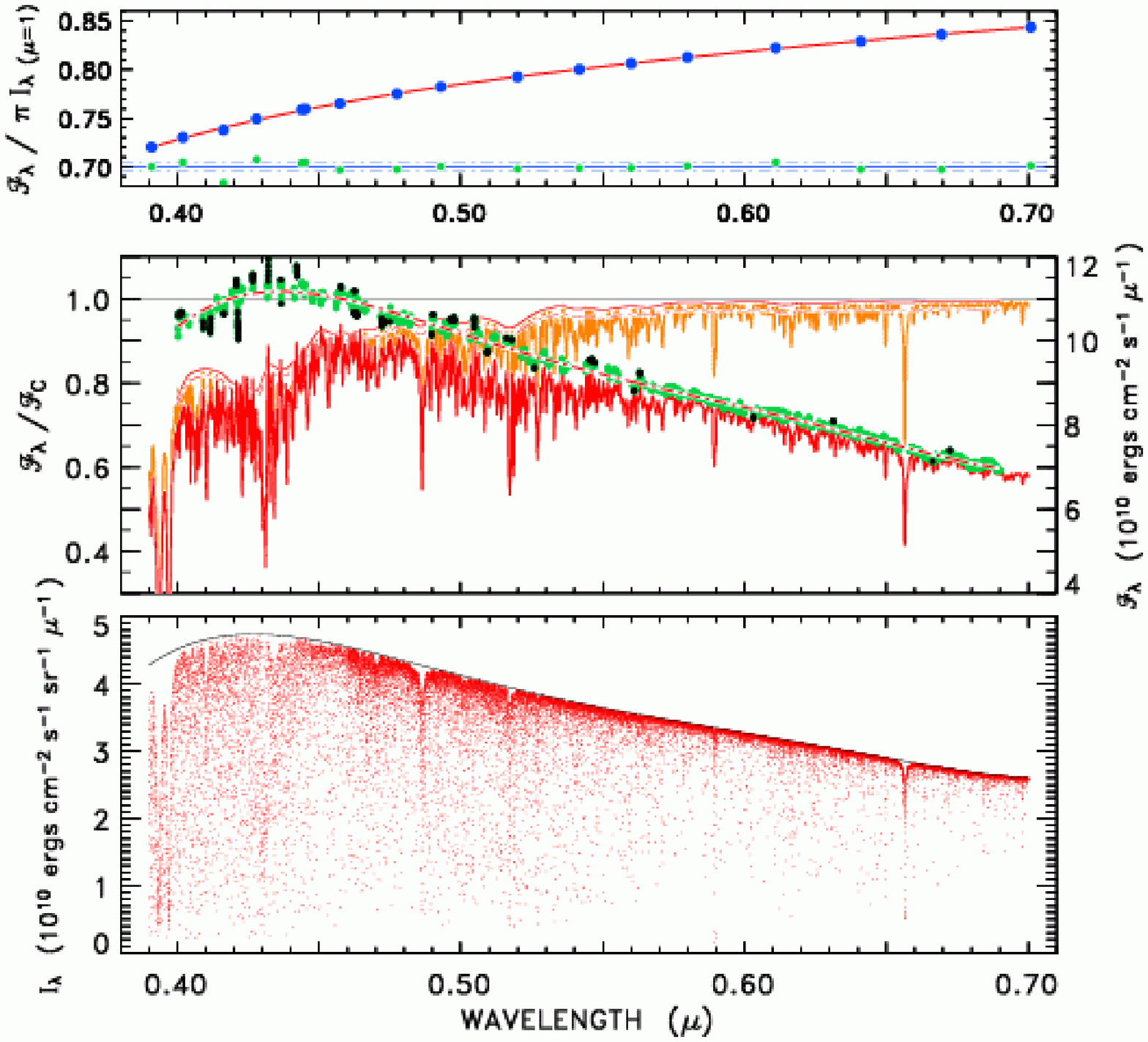}  
\centerline{Fig.~7}
%\caption[]{}
%\end{figure}

\clearpage
\begin{figure}
\figurenum{8}
\epsscale{1.00}
\plotone{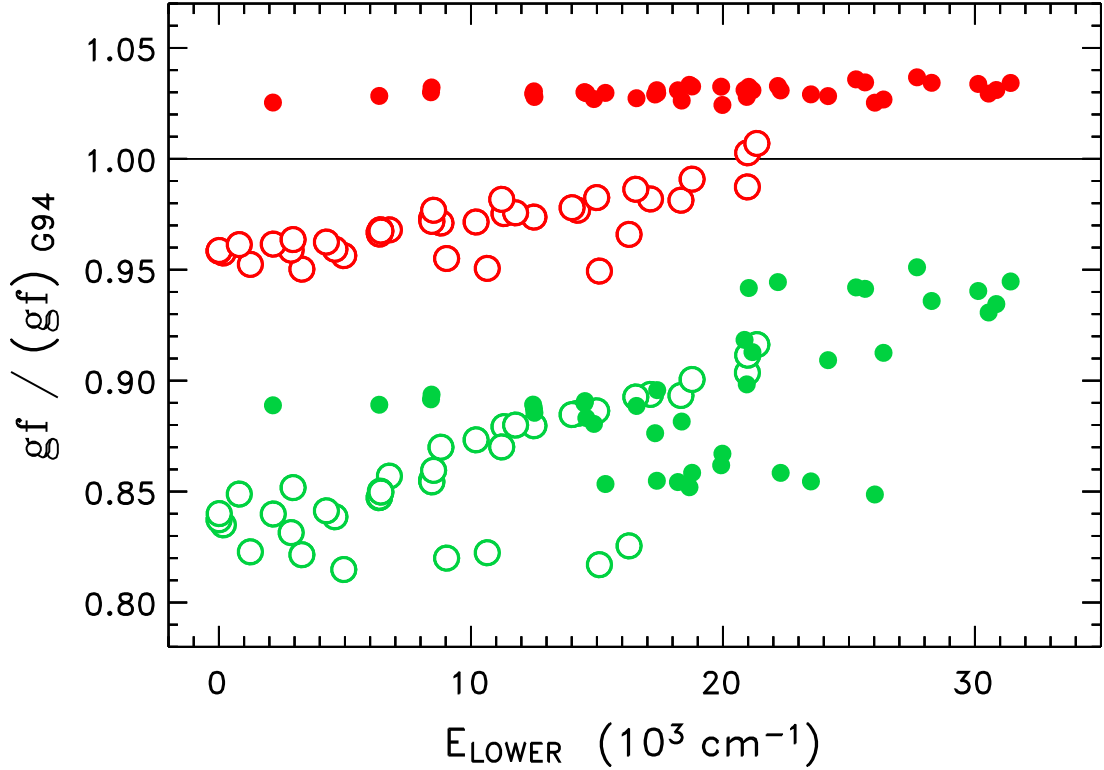}  
\caption[]{Comparison of CO $\Delta{V}=1$,\,2 oscillator strengths
versus lower state excitation energy.  
Goorvitch (1994) was used as reference, and 
relative behavior of Hure \& Roueff (1996) $gf$-values are depicted as
darker (red) dots ($\Delta{v}=1$) and open circles ($\Delta{v}=2$) in 
upper part of diagram.  Specific points refer to CO abundance sample
described in text: 40 lines from fundamental, 40 from first overtone.  
Lower, lighter (green) points are
from earlier CO line strength compilation by Kirby-Docken \& Liu (1978), again
displayed relative to Goorvitch scale.  }
\end{figure}

\clearpage
\begin{figure}
\vskip -15mm
\figurenum{9}
\epsscale{0.75}
\plotone{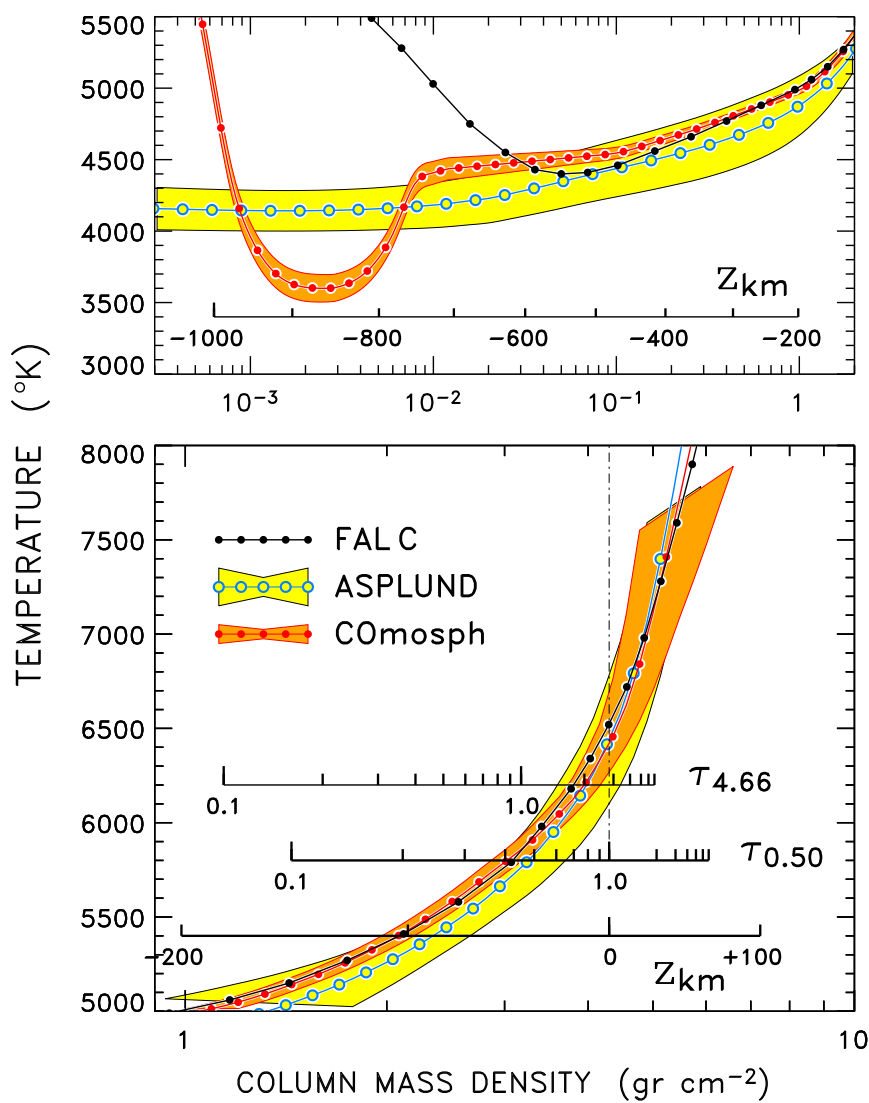}  
\caption[]{Photospheric thermal profiles considered in present study.  Black
dots refer to FAL\,C semi-empirical model of Fontenla, Avrett, \& Loeser (1993); 
larger, lighter (blue) dots to mean temperature distribution of Asplund et al.\ (2004) 3-D
convection model;
and medium (red) dots to our optimum ``COmosphere.''  Shading on latter
two profiles indicate rms thermal excursions to capture heterogeneous nature of solar
atmosphere at different levels.  Asplund range is as reported by those authors for their
{\em ab initio}\/ model,
and deeper part well matches continuum granulation contrasts
measured at high spatial resolution in visible.
Smaller COmosphere rms range at mid- and high altitudes
was based on empirical measurements of thermal fluctuations observed
in CO $\Delta{v}=1$ lines and 2145~cm$^{-1}$ (4.66~\micron) continuum.  
COmosphere model was specifically adjusted to match visible continuum
center-limb behavior, and empirical properties of CO $\Delta{v}=1$ spectrum.  Optical depth
and physical depth scales are for that model.
}
\end{figure}

\clearpage
\begin{figure}
\figurenum{10}
%%%\epsscale{1.0}
%%%\plotone{f10a.eps}  
\caption[]{(a) Continuum properties of FAL\,C model.  Larger left hand
panel depicts $T(m)$ profile and contribution functions for continuum
intensities at representative wavelengths
(from right to left, 0.4--5~\micron).  Inset panel illustrates (enforced)
fit to continuum highpoints: ordinate scale is in $10^{10}$~ergs cm$^{-2}$ s$^{-1}$
ster$^{-1}$ \micron$^{-1}$, and shaded bands indicate $\pm10$\,\% (inner)
and $\pm20$\,\% (outer).  Upper right hand panel compares observed (dots) and
predicted relative intensity center-limb curves
for representative wavelengths, from top to bottom (in \micron): 5.3, 2.5, 0.8, 0.5, 0.4\,.
Lower right hand panel also depicts continuum center-limb behavior, but now expressed
in brightness temperatures (to suppress Planck function bias) and calculated for larger
sample of wavelengths: generally, thermal IR wavelengths are at top (in red), near red wavelengths
toward bottom (orange), and visible in middle (green/blue).  
FAL\,C reproduces empirical center-limb behavior very well in
visible, although less well in thermal IR.

(b) Asplund mean temperature model.  
While somewhat hotter in 
deep photosphere than FAL\,C, Asplund $T(m)$ profile is noticeably cooler 
above $m= 3$~gr cm$^{-2}$,
which has a negative impact on center-limb behavior in visible, although thermal IR curves are
well matched.  However, the too steep predicted center-limb curves in visible indicate that
the deep-photosphere temperature ``calibration'' will not be carried reliably out to higher
layers where CO rovibrational bands form.

(c) 5 component (``1.5-D'') version of Asplund model, to
investigate influence of thermally heterogeneous photosphere on
continuum intensities and center-limb behavior.  Thin curves depict predictions for
individual thermal profiles, while heavier dot-dashed curves show
averaged intensities.  Here, rms intensity excursions in visible
continuum are comparable to observed $\sim 13$\,\% at 0.5~\micron.

(d) 6 component version of ``optimum'' COmosphere model.}
\end{figure}

\clearpage
%\begin{figure}
%\figurenum{10a}
\epsscale{1.0}
\plotone{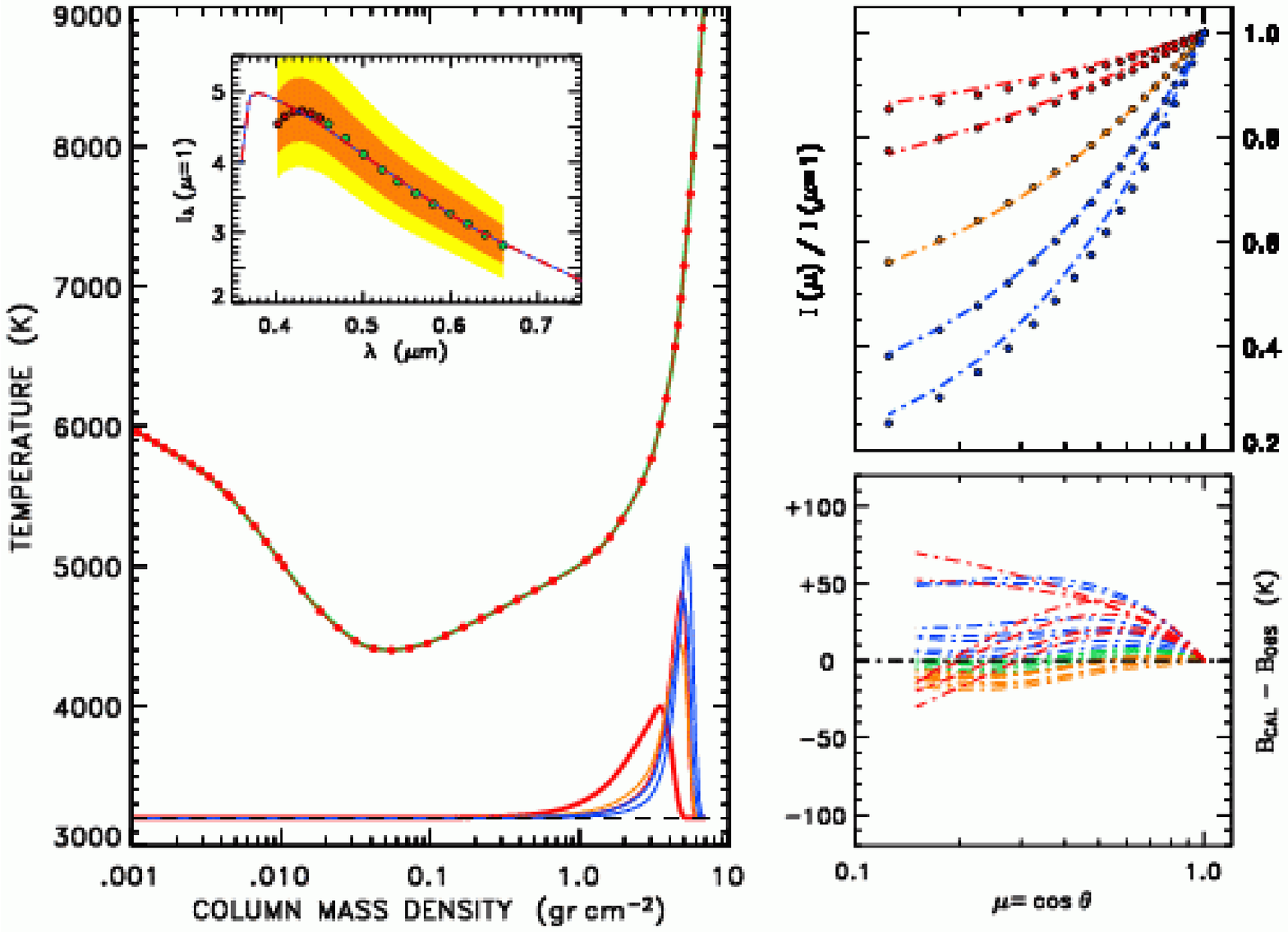}  
\centerline{Fig.~10a}
%\caption[]{}
%\end{figure}

\clearpage
%\begin{figure}
%\figurenum{10b}
%\epsscale{1.0}
\plotone{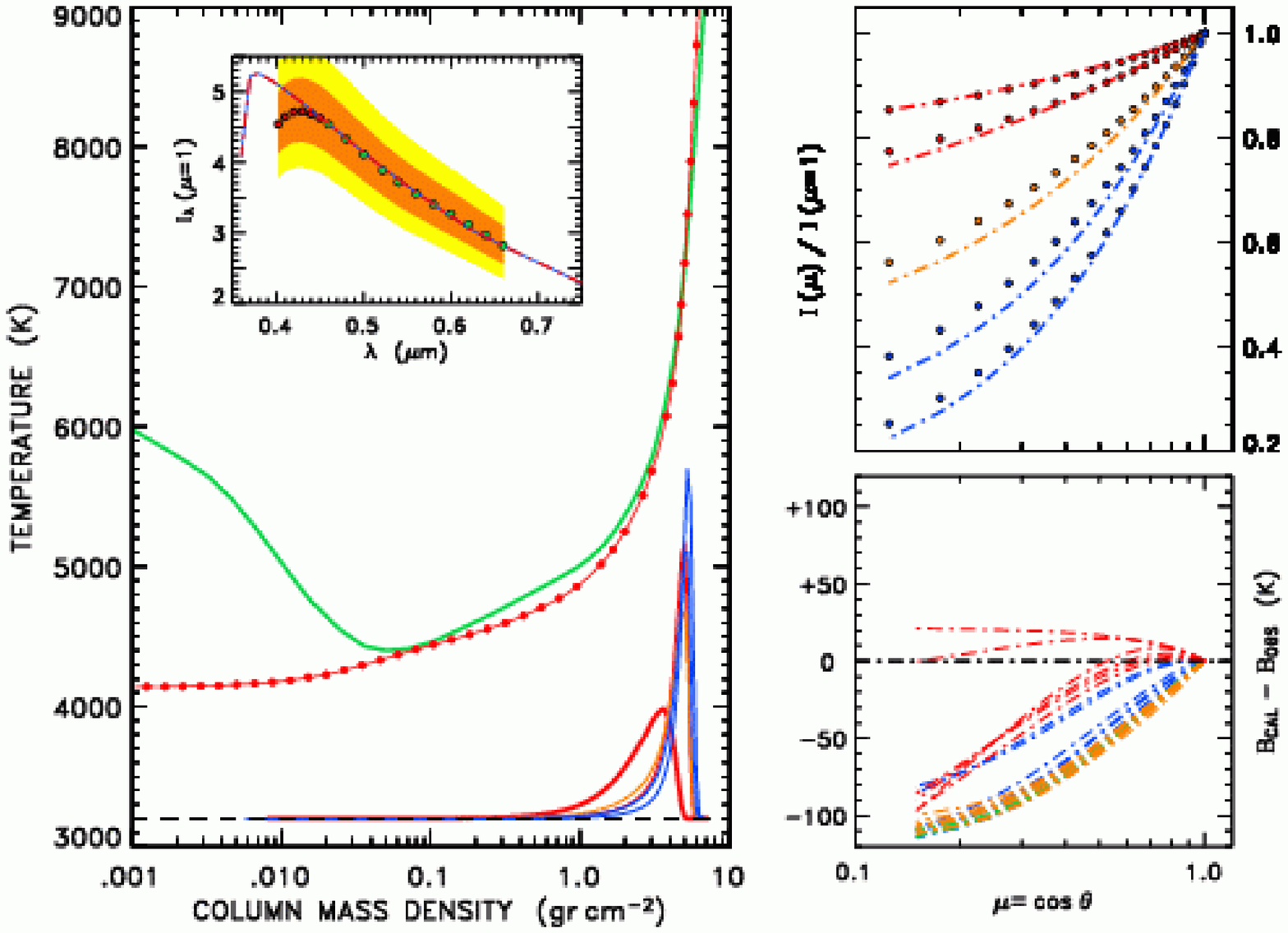}  
\centerline{Fig.~10b}
%\caption[]{}
%\end{figure}

\clearpage
%\begin{figure}
%\figurenum{10c}
%\epsscale{1.0}
\plotone{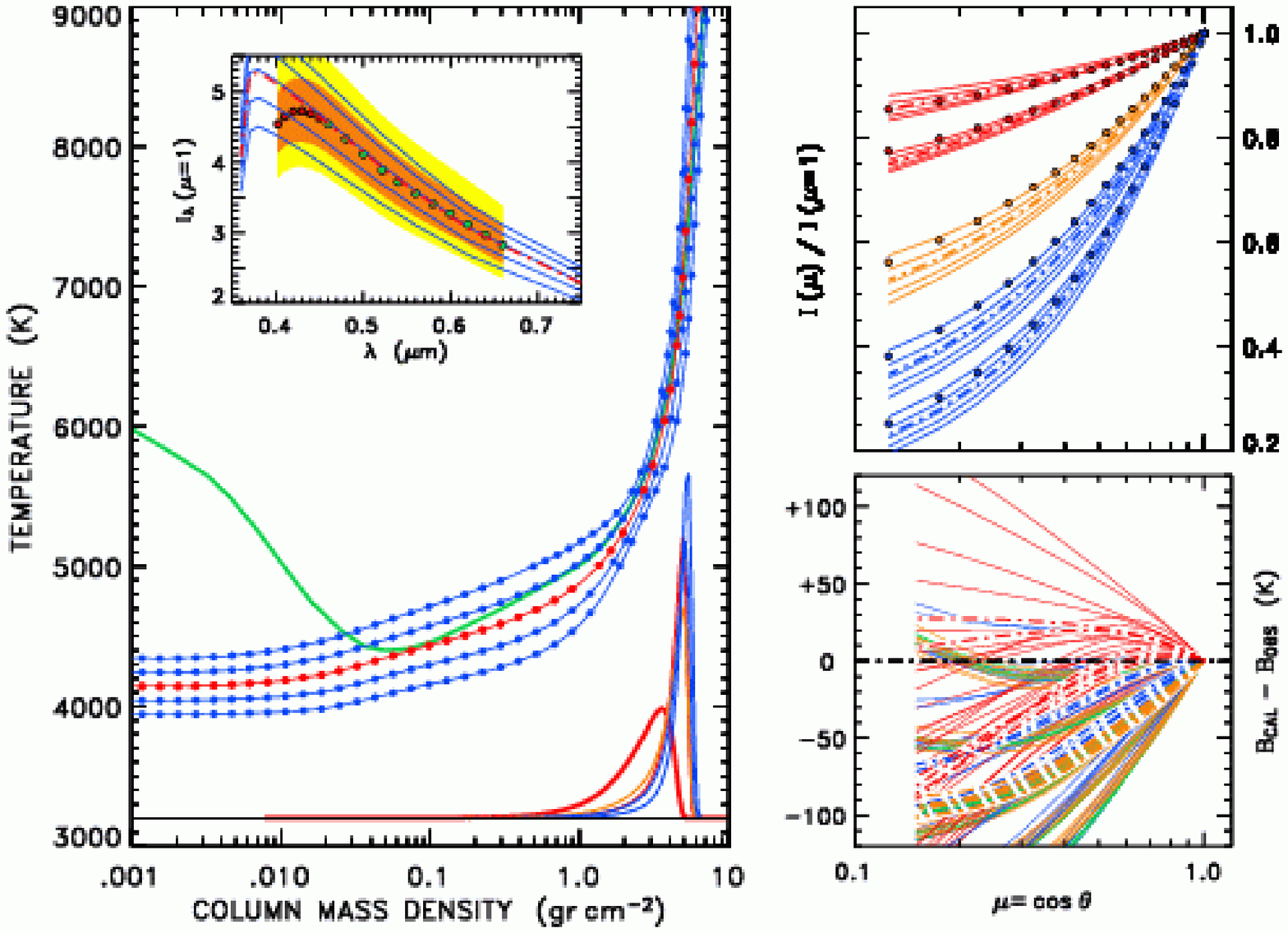}  
\centerline{Fig.~10c}
%\caption[]{   
%}
%\end{figure}

\clearpage
%\begin{figure}
%\figurenum{10d}
%\epsscale{1.0}
\plotone{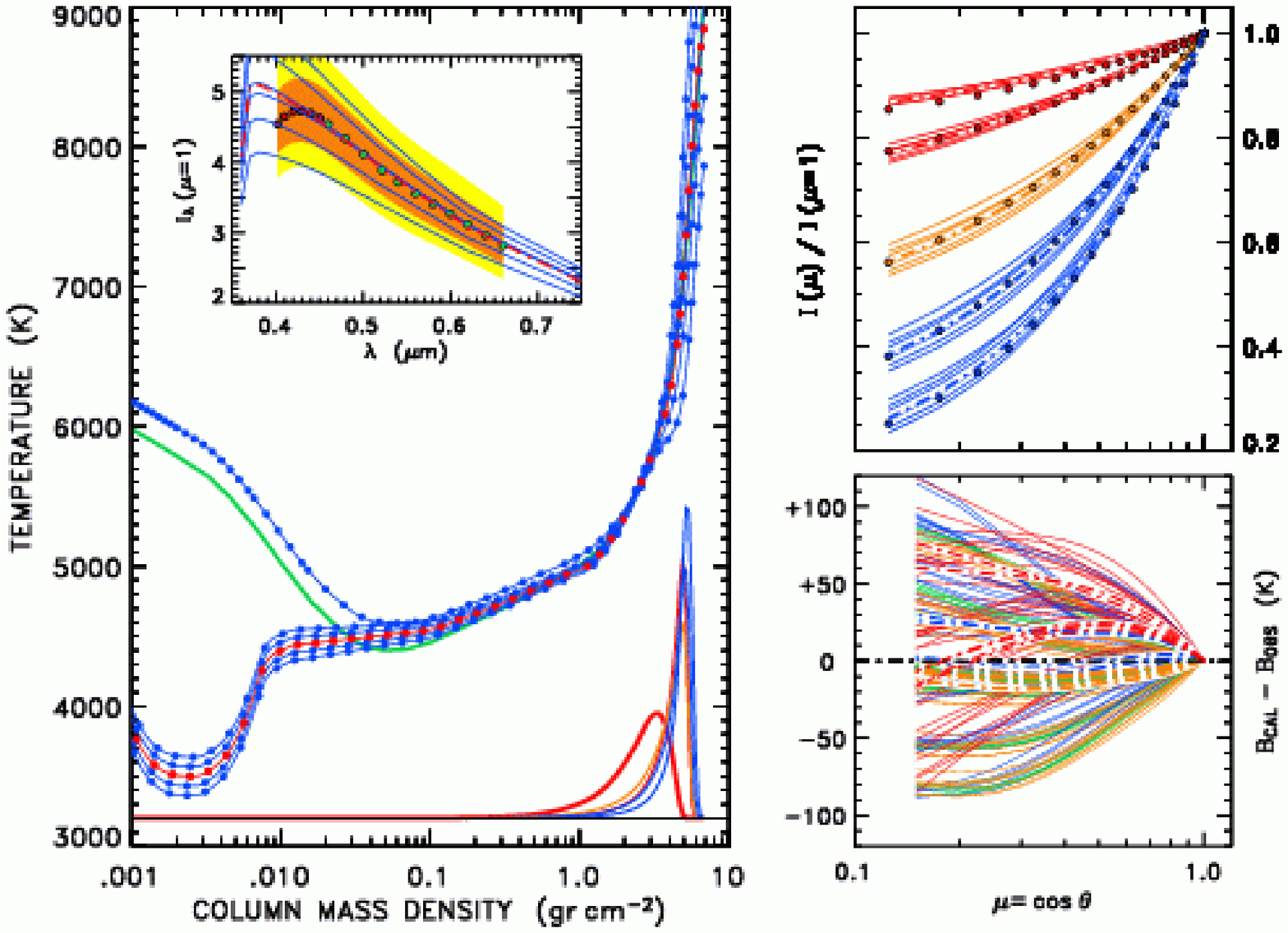}  
\centerline{Fig.~10d}
%\caption[]{ 
%}
%\end{figure}

\clearpage
\begin{figure}
\hskip -10mm
\figurenum{11}
\epsscale{1.0}
\plotone{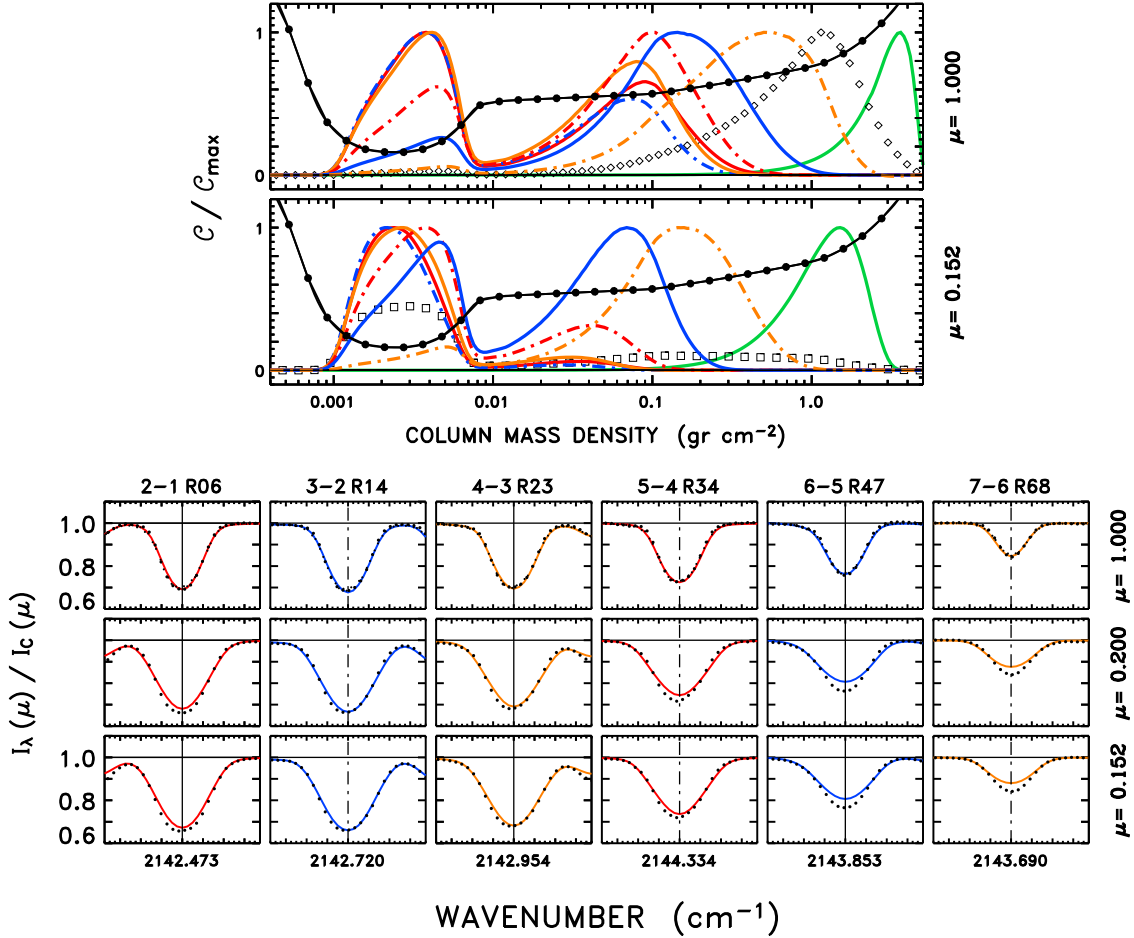}  
\caption[]{Center-limb behavior and contribution functions for representative CO
$\Delta{v}=1$ lines synthesized with 1-D version of 
COmosphere model and derived $\epsilon_{\rm O}$. 
Major ticks on relative wavenumber scales (abscissa) are 0.020~cm$^{-1}$ (2.8~km s$^{-1}$).
In upper two panels,
normalized contribution functions are displayed for $\mu= 1$ (upper)
and $\mu= 0.152$ (lower).  Green curves furthest to right in each panel are 
intensity contribution functions
for 2145~cm$^{-1}$ (4.66~\micron) continuum, while curves to left are 
line depression contribution functions
for CO 
transitions (note double peaks).  CO curves are color coded, and solid or
dot-dashed, to correspond with spectral panels (vertical line at $\omega_{0}$
indicates linestyle).  In uppermost panel, small diamonds depict relative
concentration of CO: $n_{\rm CO}/(n_{\rm CO})_{\rm max}$.  In second panel, small squares
depict fraction of oxygen bound in CO: $n_{\rm CO}/(\epsilon_{\rm O}\,n_{\rm H})$.
In $T_{\rm min}$ region, ratio is nearly 0.5 reflecting that virtually all 
available carbon (C/O= 0.5) has been captured into CO.}
\end{figure}

\clearpage
\begin{figure}
\figurenum{12}
%%%\epsscale{0.75}
%%%\plotone{f12.eps}  
\caption[]{Oxygen abundances determined from single component
versions of COmosphere and Asplund models (upper two, and middle two panels, respectively), 
and two sets of $gf$-values: Goorvitch (1994) and
Hure \& Roueff (1996). 
Panels to left are ``cartoons'' of particular models used in the simulations; FAL\,C
temperature distribution is provided (as a thin curve) in each panel for
comparison.  Long ticks at right and left sides of panel indicate
the radial-tangential (blue and red, respectively)
velocity model used in the simulation (usually [0.75,\,2.00]~km s$^{-1}$); horizontal
line marks velocity zero (and $T=3500$~K level); major ordinate ticks are spaced
1~km s$^{-1}$ for $\xi$ and 500~K for $T(m)$.  
Resulting $\epsilon_{\rm O}$'s for $\Delta{v}= 1$,\,2
samples are depicted in right hand panels as a function of lower state excitation
energy.  Lighter (red) dots generally to left are first overtone lines, while
black dots generally to right are fundamental transitions.  
Hatched areas indicate means and
${\pm}1$ standard deviations for the two separate samples, while green
line is linear least squares fit to the two samples combined (with mild filtering to eliminate
outliers).  Effect of
correcting $gf$-values---according to apparent slope of $\epsilon_{\rm O}$--$E_{\rm ex}$
relation, taking true abundance from intercept of fit---is 
illustrated in lower two panels [``($gf$)$\ast$''] for the alternative
oscillator-strength scales with optimum COmosphere model.}
\end{figure}

\clearpage
%\begin{figure}
%\figurenum{12}
\epsscale{.90}
\plotone{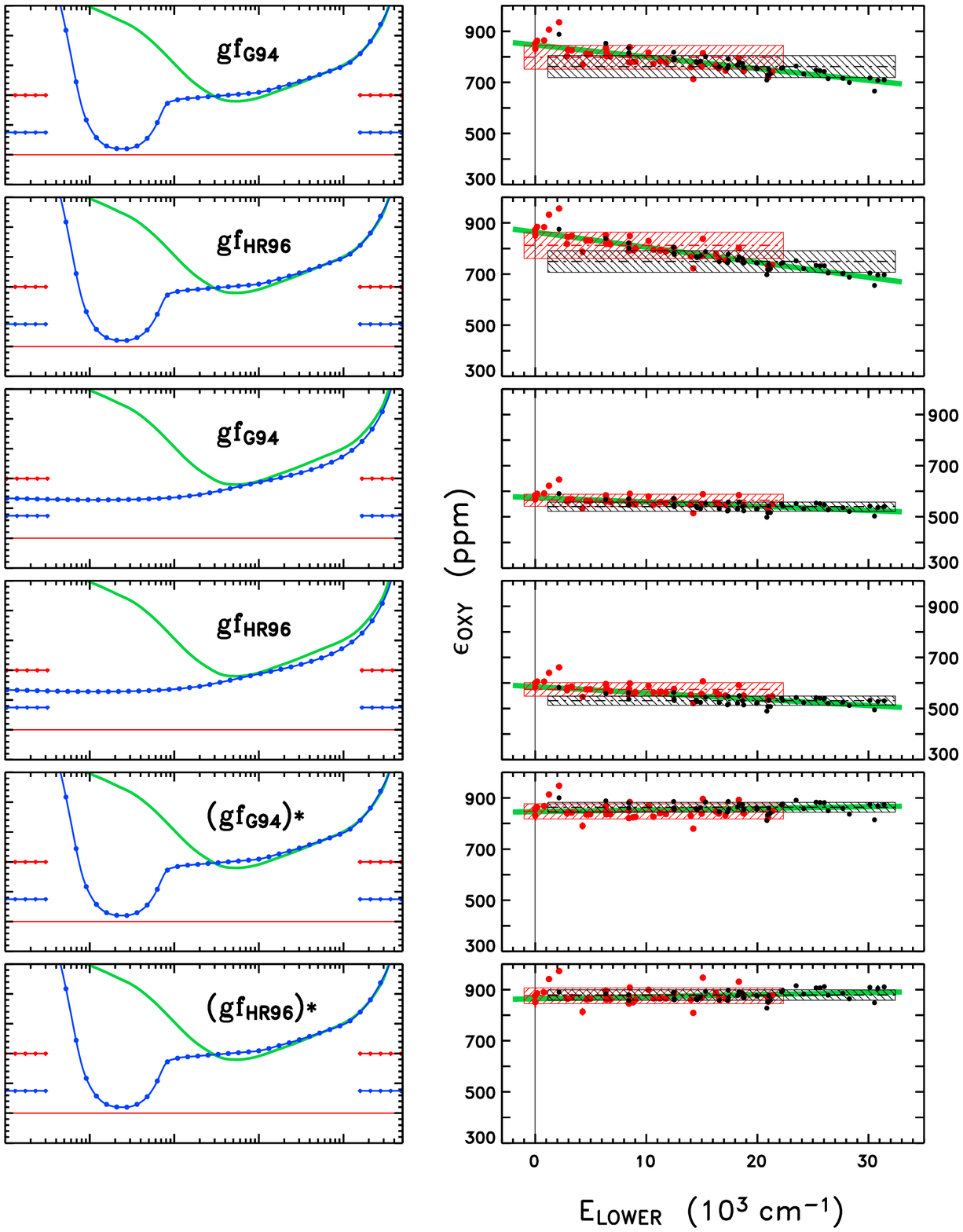}  
\centerline{Fig.~12}
%\caption[]{}
%\end{figure}

\clearpage
\begin{figure}
\figurenum{13}
%%%%\epsscale{1.0}
\vskip -5mm
\caption[]{Montage of 
$^{12}$C$^{16}$O $\Delta{v}=1$,\,2 rovibrational profiles, for
representative members of abundance sample, and synthesized lineshapes for
single component COmosphere model with best-fit $\epsilon_{\rm O}= 846$~ppm (and 
empirical corrections to $gf$-values).  Like Fig.~3, major ticks on 
abscissa are 0.02~cm$^{-1}$ apart, and relative intensity scales are graduated.  First interval around
unit level is first
$\pm$1\,\%, emphasizing continuum and (photometric) scatter
around it.  Full range extends to 20\,\% below continuum level to show details of
deeper features (although still weak in comparison to strongest
$\Delta{v}= 1$ absorptions of Fig.~3).  
Shallower lines (with central residual intensities $\ge 85$\,\%)
were locally normalized to continuum level according to bins (green points) on either side of
line center, away from absorption depression itself.  For stronger features, original
``long range'' continuum fit was retained.  
The 7--6~R68 transition
connects to center-limb line set.}
\end{figure}

\clearpage
\begin{figure}
%\figurenum{13}
\hskip -5mm
%\vskip -10mm
\includegraphics[scale=0.90]{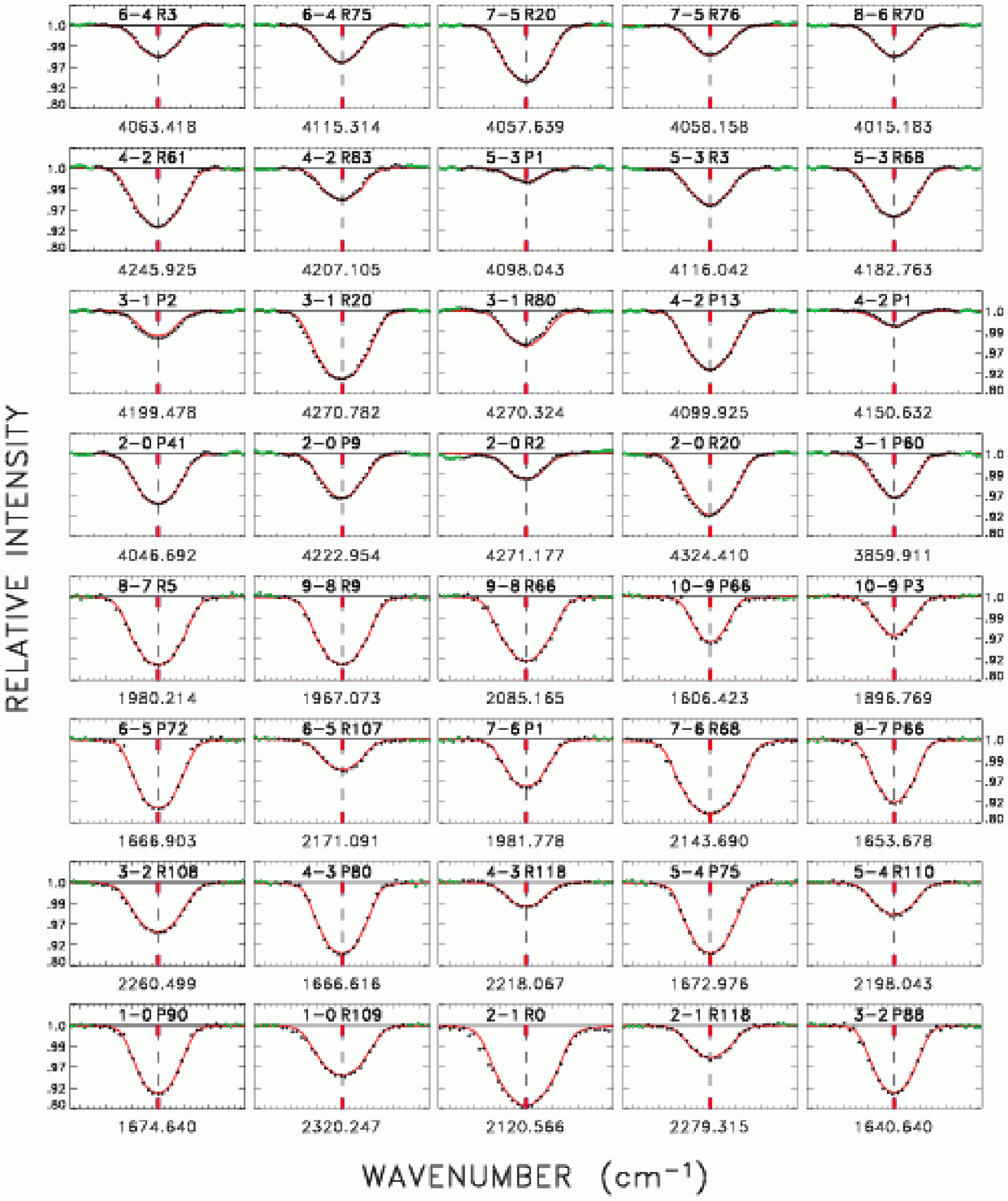}  
\centerline{Fig.~13}
%\caption[]{}
\end{figure}

\clearpage
\begin{figure}
\figurenum{14}
%%%%\epsscale{0.5}
\includegraphics[scale=0.75,angle=90]{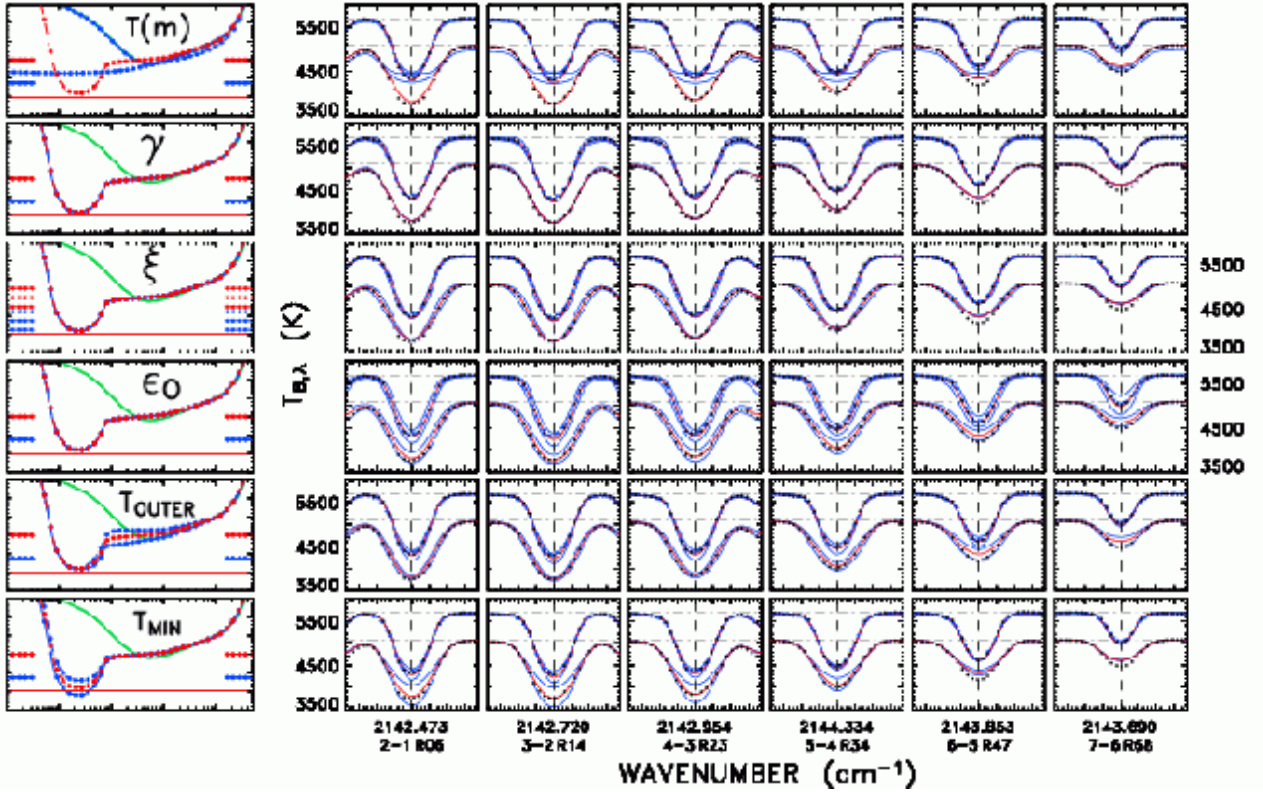}  
\caption[]{Sensitivity of representative $\Delta{v}=1$ transitions at $\mu=1$,
and their center-limb behavior, to changes in models and various key parameters (see text).   
As in Fig.~12, model temperature structures are illustrated schematically
in left hand panels.  Observations
are same as in Fig.~11, although only 
disk center (upper, in each panel) and extreme limb
($\mu= 0.152$: lower) are shown here.  
Absolute continuum levels were set according to $I_{\rm C, \lambda}(\mu)$ 
calculated with
optimum COmosphere model, and profiles now are displayed on brightness temperature
scale.  Major ticks on abscissa are 0.02~cm$^{-1}$.
General decrease in core brightness temperature with increasing
line strength (from right to left), and darkening of individual lines toward extreme
limb, carry redundant information concerning the thermal profile of the outer photosphere. }
\end{figure}

\clearpage
\begin{figure}
\figurenum{15}
\caption[]{(a) Oxygen abundances determined for representative single- and multicomponent
models.  As in Fig.~12, model temperature profiles are illustrated schematically
in left hand panels, and resulting $\epsilon_{\rm O}$'s for $\Delta{v}= 1$,\,2
samples in right hand panels as function of $E_{\rm lower}$.  In general, multicomponent
versions of the base models have relatively little effect on derived oxygen abundance.
Main influence is average mid-photospheric temperature where CO concentration
peaks.  Asplund mean model is cooler in this region than optimum COmosphere, and consequently
predicts a lower oxygen abundance.  Curiously, agreement between $\Delta{v}= 1$ and
2 abundances is improved: this contrast is temperature sensitive in principle because one is
comparing generally low excitation $\Delta{v}= 2$ lines to mainly high excitation 
$\Delta{v}= 1$ transitions, although one also must be
wary of systematic errors in line absorption strengths which potentially could mimic
a temperature effect, as seen in Fig.~8 for Hure \& Roueff (1996) $\Delta{v}= 2$ $gf$-values.  
However, continuum center-limb test indicates Asplund model
is too cool in middle photosphere.  Dropping 
entire COmosphere temperature stratification systematically by 200~K (lowest panel) reproduces
low O abundance ($\sim$450~ppm) obtained by Asplund et al.\ (2004), and removes 
excitation gradient between fundamental and first overtone samples, but
low-$T$ model completely fails continuum center-limb test, like 
Asplund model itself.

(b) For additional model or parameter perturbations described in
text.  Here we see, for example, that changing $T_{\rm min}$ temperature or
microturbulent velocity $\xi$ has a negligible effect on $\epsilon_{\rm O}$, whereas
altering thermal profile of outer photosphere ($T_{\rm OUTER}$) or
scaling total H$^{-}$ cross-section (b--f and f--f together) have more noticeable influences.  However,
none of the changes, for base COmosphere model, removes discrepancy between
$\Delta{v}=1$ and $\Delta{v}=2$ abundances, although
cooler $T_{\rm OUTER}$ profile is improvement in that direction.  Nevertheless, that model
performs more poorly in simulating run of $\Delta{v}=1$ 
core brightness temperatures
with increasing line strength at $\mu= 1$ (Fig.~14), although it
does reproduce center-limb
behavior of weaker 6--5~R47 and 7--6~R68 (which base model does less well).}

\end{figure}

\clearpage
%\begin{figure}
%\figurenum{15a}
\epsscale{.90}
\plotone{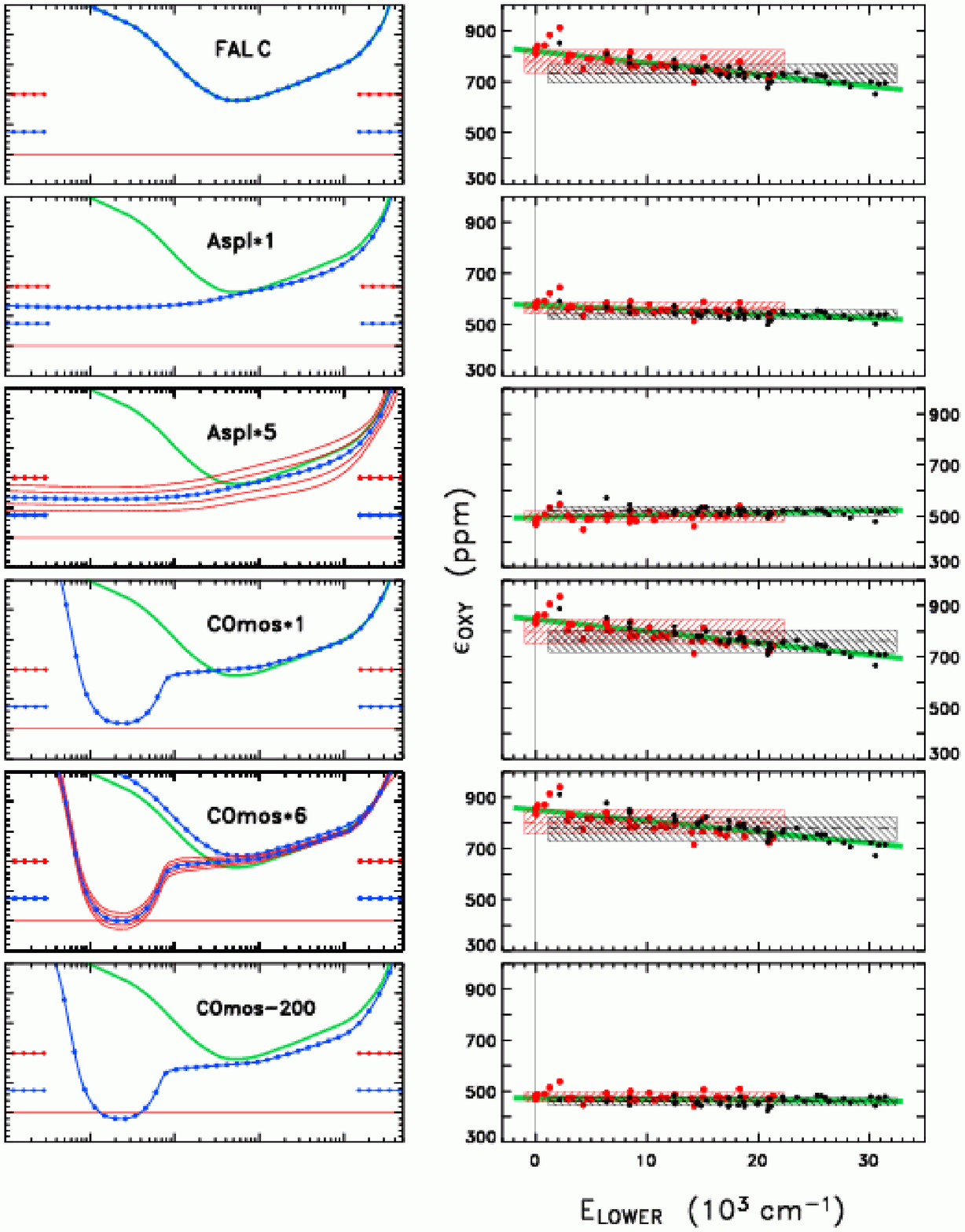}  
\centerline{Fig.~15a}
%\caption[]{}
%\end{figure}

\clearpage
%\begin{figure}
%\figurenum{15b}
%\epsscale{1.00}
\plotone{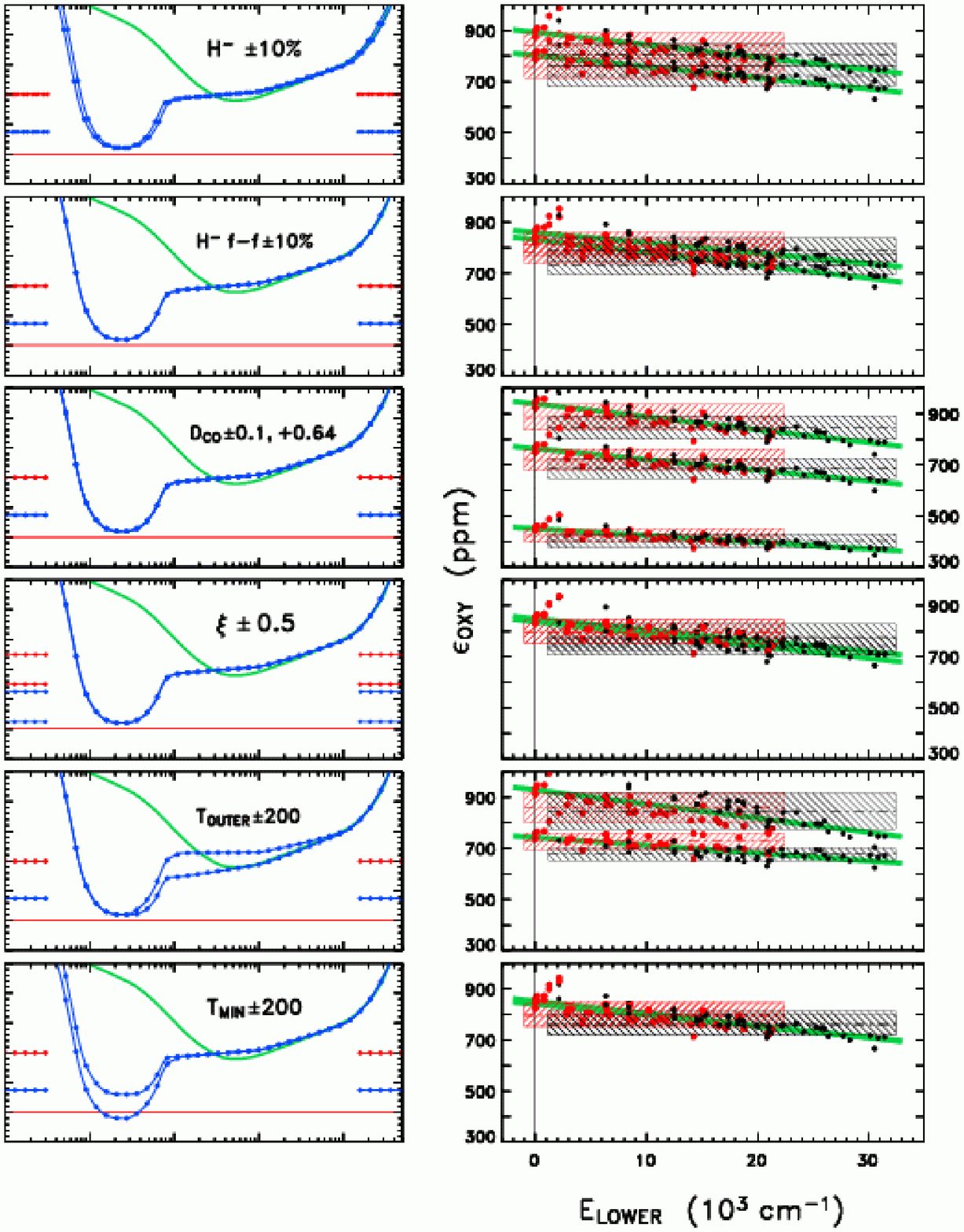}  
\centerline{Fig.~15b}
%\caption[]{
%}
%\end{figure}

\clearpage
\begin{figure}
\figurenum{16}
\epsscale{1.0}
\plotone{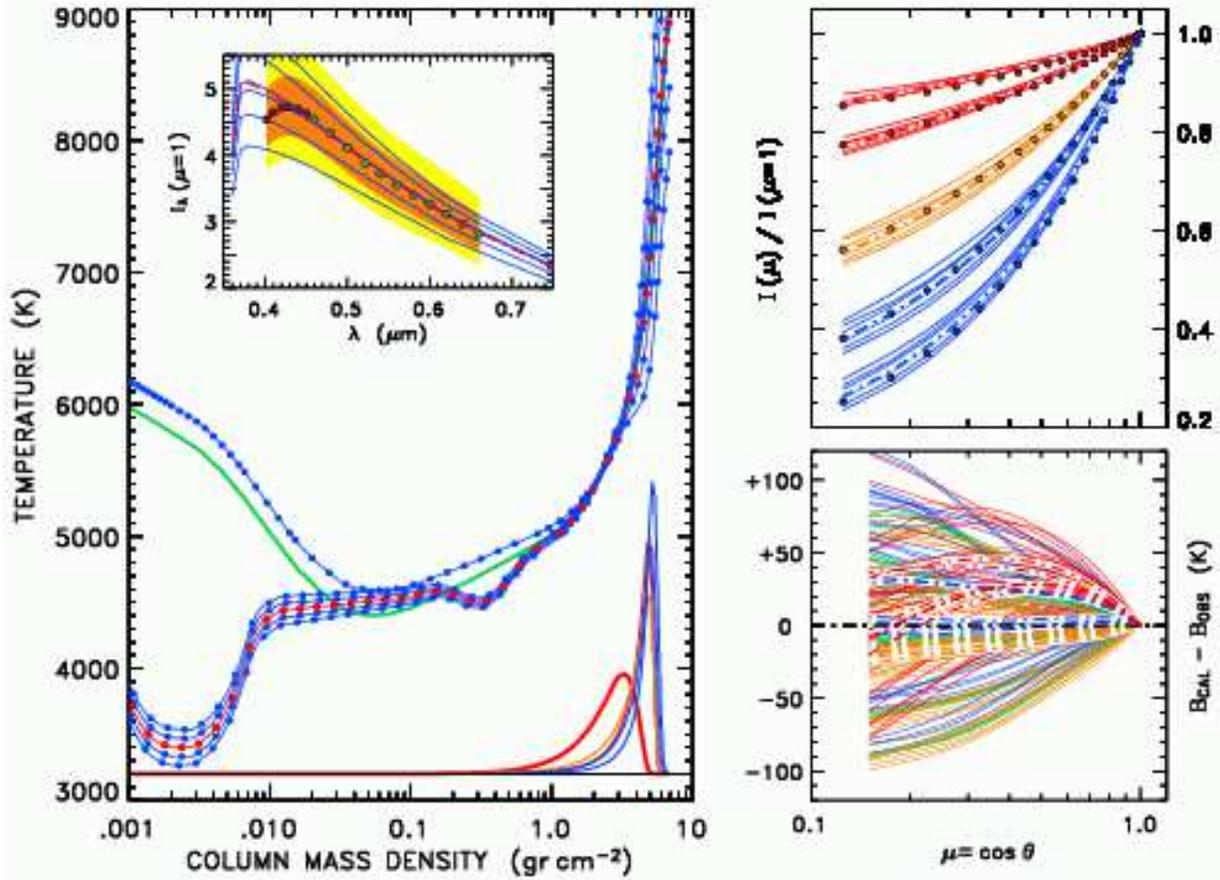}  
\caption[]{Continuum behavior for 6 component version
of proposed Double Dip model: small temperature depression in middle
photosphere improves thermal IR continuum
center-limb behavior without degrading good match of
base COmosphere model in visible and near IR.}
\end{figure}

\clearpage
\begin{figure}
\figurenum{17}
\vskip -30 mm
\caption[]{(a) Abundances for single component and 6 component realizations
of Double Dip model.  $\Delta{v}=1$ abundance distributions for both versions
of Double Dip is flatter with respect to $E_{\rm lower}$ than original
COmosphere model, although now clear separation
between $\Delta{v}=1$ and 2 systems is seen.

(b) Center-limb behavior for single component and 6 component realizations
of Double Dip model.
}
\end{figure}

\clearpage
%\begin{figure}
%\vskip -30mm
%\figurenum{17a}
\epsscale{1.0}
\plotone{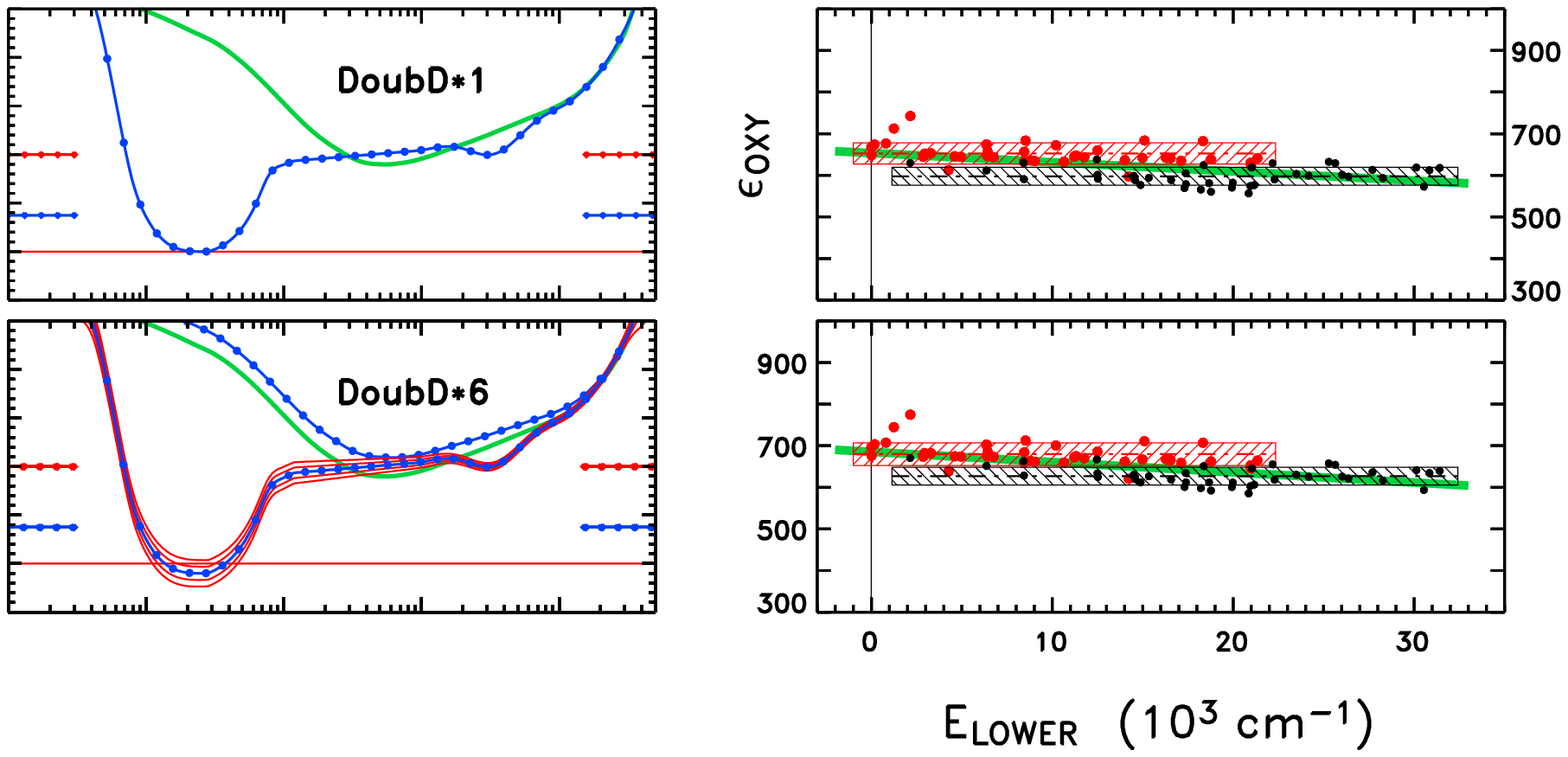}  
\centerline{Fig.~17a}
%\caption[]{}
%\end{figure}

\samepage
%\begin{figure}
%\vskip -40 mm
%\figurenum{17b}
\includegraphics[scale=0.75,angle=90]{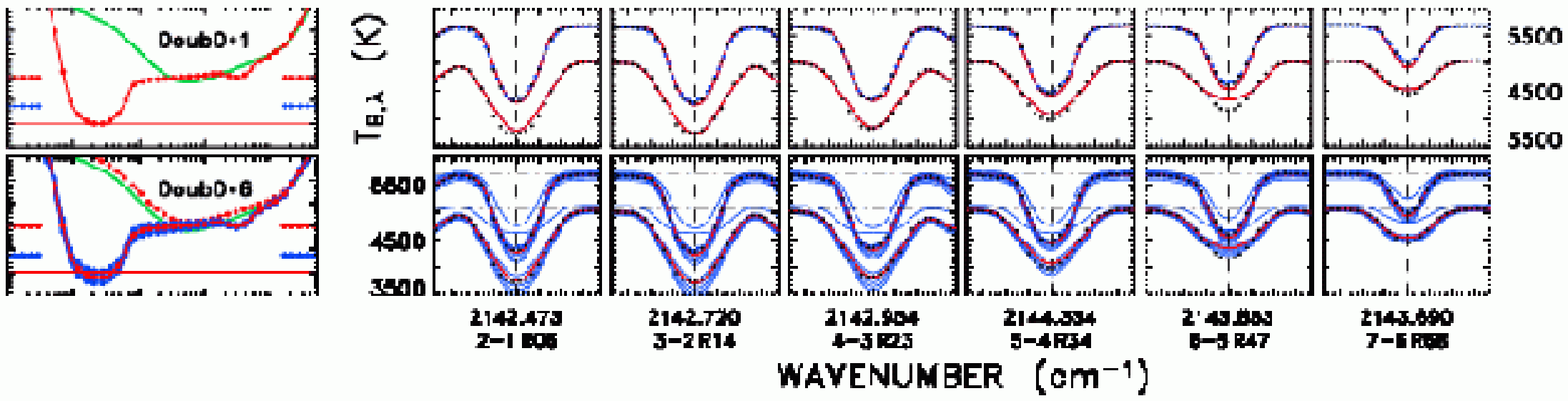}  
\centerline{Fig.~17b}
%\caption[]{}
%\end{figure}

\clearpage
\begin{figure}
\hskip -15mm
\figurenum{18}
\epsscale{1.0}
\plotone{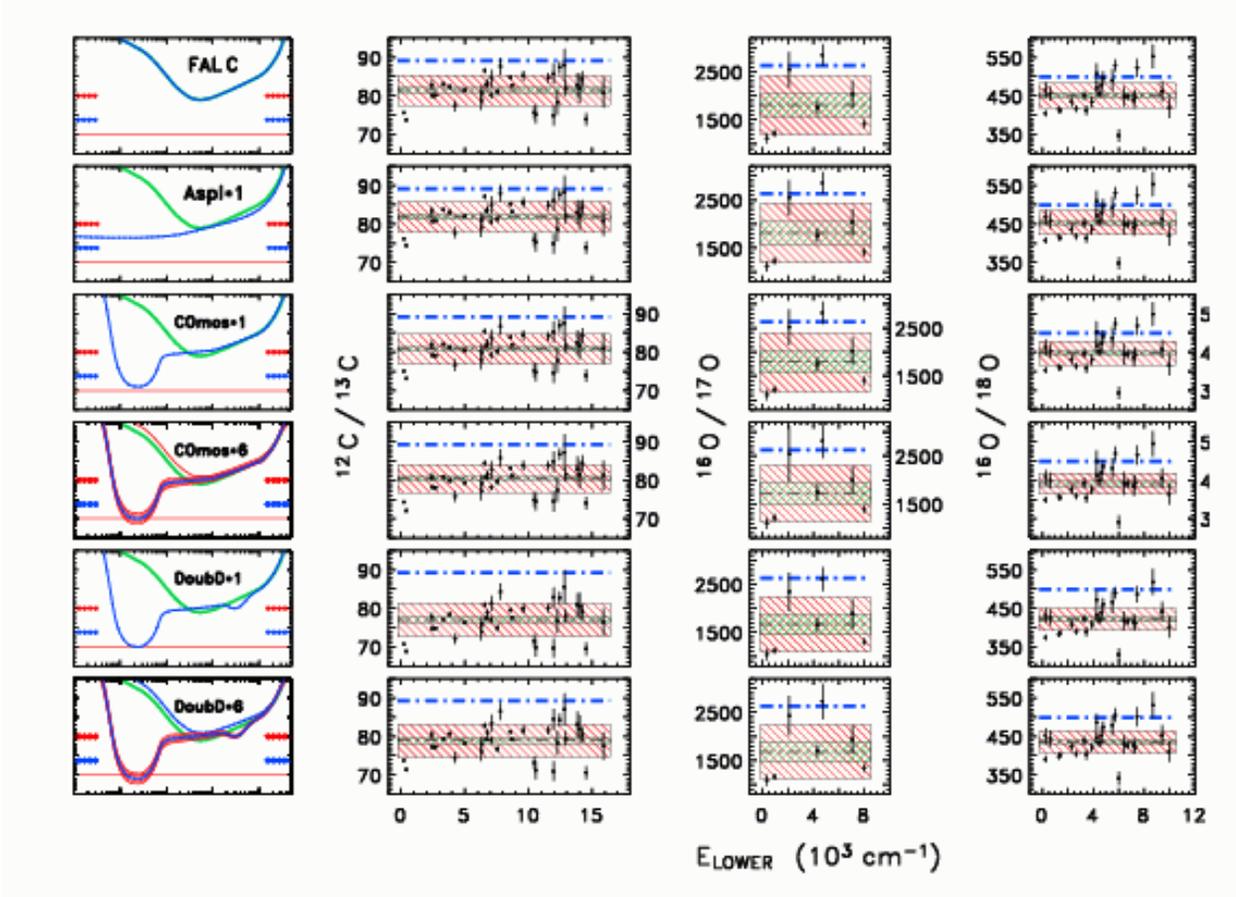}  
\caption[]{Isotopic abundance ratios for several atmospheric models.  
(Model dependent) $gf$-value corrections inferred
from $^{12}$C$^{16}$O sample were applied to isotopomer oscillator strengths.  
Derived isotopic ratios are minimally dependent on thermal structure
of model, and thus can be determined to better accuracy than absolute abundances.  
Outer (red) hatched bands indicate $\pm 1$
standard deviation for each sample, after applying 2\,$\sigma$ filter to eliminate outliers;
inner (green) hatched bands indicate $\pm 1$\ standard error of mean.  
Dot-dashed (blue) lines are terrestrial
reference values.}
\end{figure}

\clearpage
\begin{figure}
\figurenum{19}
\caption[]{Similar to Fig.~13, but now a montage of observed and synthesized lineshapes
of representative $^{13}$C$^{16}$O,
$^{12}$C$^{17}$O, and $^{12}$C$^{18}$O; based on
single component COmosphere model, best-fit $\epsilon_{\rm O}= 846$~ppm (with $\Delta= -5.5$
correction to $gf$-values), 
and $^{12}$C/$^{13}$C= 81, $^{16}$O/$^{17}$O= 1751, and $^{16}$O/$^{18}$O= 438.  
Isotopomers are coded, in brackets following transition 
designations, e.g., ``36''=~$^{13}$C$^{16}$O.  Again, 
relative intensity scales are graduated: this time, initial interval is $\pm$0.3\,\% around unit
level, and full range extends to only 6\,\% below it.  
Although the isotopic lines, in general, are very weak,
high S/N of ATMOS spectra allows them to be detected at high level of confidence.  
(Note, also, three $\Delta{v}=2$ $^{13}$C$^{16}$O transitions appended to $\Delta{v}=1$ sample.)}
\end{figure}

\clearpage
\begin{figure}
%\figurenum{19}
\hskip -5mm
%\vskip -10mm
\includegraphics[scale=0.90]{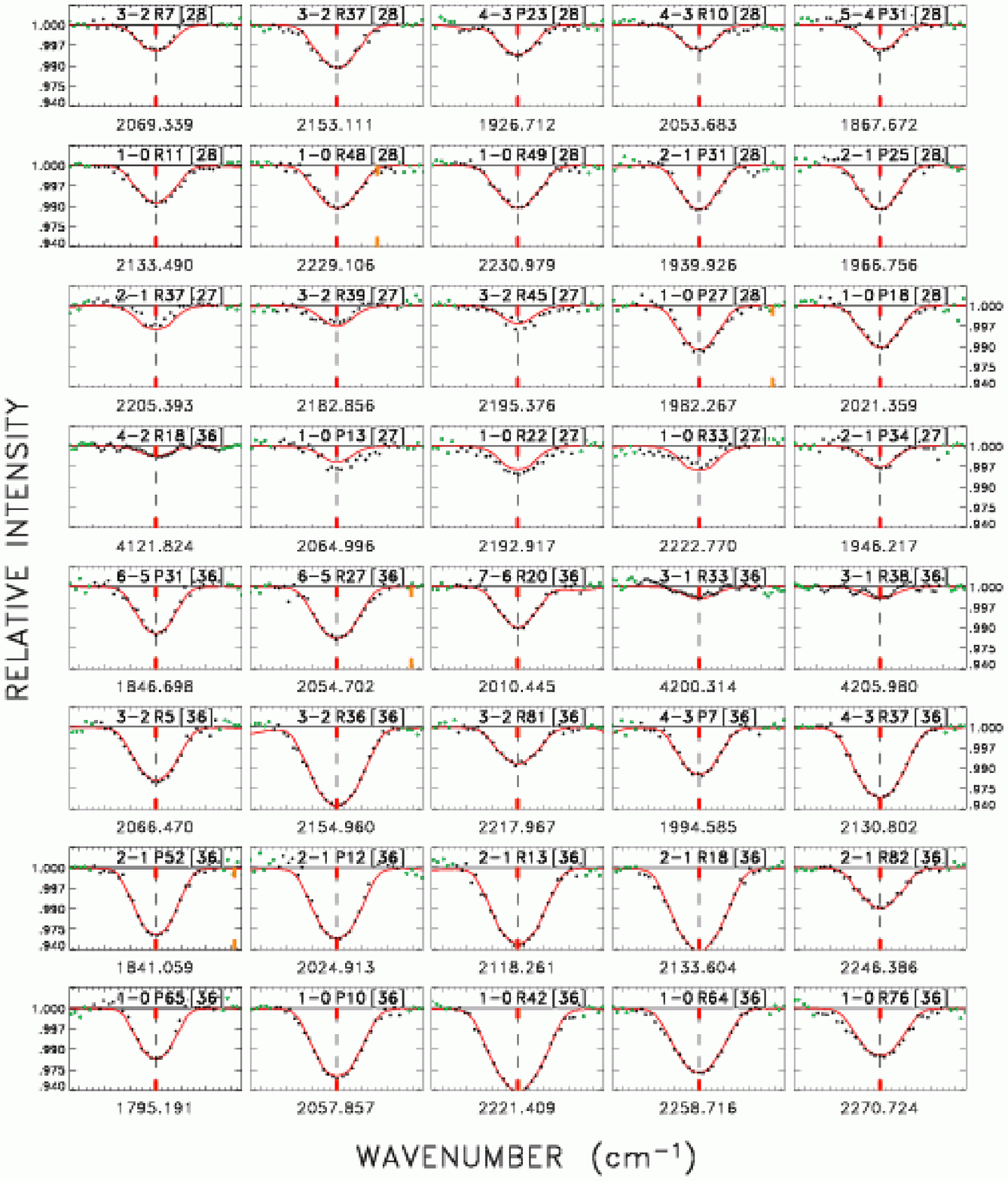}  
\centerline{Fig.~19}
%\caption[]{}
\end{figure}

\end{document}